\documentclass{JHEP3}
\usepackage{amsmath,feynmf}

\makeatletter
\let\oldref\old@ref
\makeatother

\allowdisplaybreaks[4]

\title{Non-renormalizability of $\theta$-expanded noncommutative QED}

\author{Raimar Wulkenhaar\thanks{Marie-Curie Fellow}  \\
Institut f\"ur Theoretische Physik, Universit\"at Wien,
Boltzmanngasse 5, A-1090 Wien, Austria\\
Email: \email{raimar@doppler.thp.univie.ac.at}}

\preprint{hep-th/0112248\\UWThPh-2001-56}


\abstract{
  Computing all divergent one-loop Green's functions of
  $\theta$-expanded noncommutative quantum electrodynamics up to first order
  in $\theta$, we show that this model is not renormalizable. The reason is a
  divergence in the electron four-point function which cannot be removed by
  field redefinitions. Ignoring this problem, we find however clear hints for
  new symmetries in massless $\theta$-expanded noncommutative QED: Four
  additional divergences which would be compatible with gauge and Lorentz
  symmetries and which are not reachable by field redefinitions are absent.
}

\begin{document}

\section{Introduction}
\label{Sec1}

The investigation of field theories on a very special type of noncommutative
spaces---the so-called noncommutative $\mathbb{R}^4$ involving a constant
antisymmetric tensor $\theta^{\mu\nu}$---became during the last two years an
industry, making it impossible to list all relevant contributions. Fortunately
one can refer to some reviews \cite{Seiberg:1999vs, Konechny:2000dp, 
Douglas:2001ba, Szabo:2001kg}, which reflect most of the achievements. 
Nevertheless, let us briefly recall a few selected milestones which lead
directly to the question addressed in this paper.

Many investigations on this subject are motivated by the discovery of
compactifications of M theory on noncommutative tori
\cite{Connes:1998cr} and the identification of these noncommutative
geometries as limiting cases of string theory \cite{Seiberg:1999vs}.
Quantum field theories on $\theta$-deformed space-time were shown to
be ultraviolet divergent \cite{Filk:1996dm} (motivated at this time by
the continuum version of the twisted Eguchi-Kawai model
\cite{Gonzalez-Arroyo:1983ac} which already possesses noncommutative
features). A more general result on divergences for spaces which are
noncommutative manifolds in the sense of \cite{Connes:1996gi} was
obtained in \cite{Varilly:1998gq}. The crucial question about the
physical relevance of field theories on noncommutative space-time is
whether these models are renormalizable. At the one-loop level,
renormalizability of $U(1)$-gauge theories on $\theta$-deformed spaces
was first proved in \cite{Martin:1999aq, Sheikh-Jabbari:1999iw,
  Krajewski:2000ja}. One-loop renormalizability of noncommutative
$U(n)$-gauge theories was proved in
\cite{Bonora:2000ga}\footnote{The two- and three-point functions of 
noncommutative $U(n)$ theory have already been computed in
  \cite{Armoni:2000xr}.}.  At higher loop order there was discovered a
new type of infrared divergences both in scalar field theories
\cite{Minwalla:2000px} and Yang-Mills theory
\cite{Hayakawa:1999zf,Matusis:2000jf}, which prevents a perturbative
renormalization of these models. A power-counting theorem for
noncommutative field theories which classifies the divergences was
established in \cite{Chepelev:2001hm}.

A very surprising result due to Seiberg and Witten
\cite{Seiberg:1999vs} was that noncommutative and commutative gauge
theories are related by a formal power series in $\theta$, the
so-called Seiberg-Witten map. The Seiberg-Witten map combines the
$\theta$-dependence through the $\star$-product with a cleverly chosen
$\theta$-dependence of the noncommutative gauge fields to produce an
action which is (commutatively) gauge-invariant\footnote{Expanding
  only the $\star$-product in $\theta$, gauge invariance of the
  truncated action is lost.} at \emph{any} order $n$ in $\theta$. In
this way the Seiberg-Witten map gives rise to a quantum field theory
of $N$-point gauge field Green's functions with up to $n$ factors of
$\theta$ which is free of any infrared divergences
\cite{Bichl:2001nf}\footnote{That work was inspired by
  \cite{Jurco:2000ja}.}. However, since $\theta$ has to be regarded as
an external field of power-counting dimension $-2$, the
renormalizability of such a theory (up to given order in $\theta$ but
any order in $\hbar$) is questionable. The fact that the
superficial divergences in the photon selfenergy are (to all orders in
$\theta$ and $\hbar$) field redefinitions \cite{Bichl:2001cq} gave
some hope that gauge theories on $\theta$-deformed space-time could
finally be renormalizable via the Seiberg-Witten map. However, it
became clear that one cannot proceed along this line without the
identification of new symmetries of the $\theta$-expanded action which
could restrict the structure of otherwise possible divergent
counterterms. The first candidates of such symmetries---Lorentz
rotation and dilatation---were shown to give no further information
beyond conformal symmetry of a Yang-Mills theory on commutative
space-time \cite{Bichl:2001yf}.  As a by-product, however, we have
obtained in \cite{Bichl:2001yf} a deeper understanding of the
Seiberg-Witten map: One has to distinguish between `observer Lorentz
transformations' (which transform $\theta^{\mu\nu}$ as a tensor) and
`particle Lorentz transformations' (which leave $\theta^{\mu\nu}$
invariant), see \cite{Colladay:1998fq}.  Demanding (A) observer
Lorentz invariance and (B) gauge invariance of the particle Lorentz
symmetry breaking of the physical action, the particle Lorentz
transformation of a field is the sum of its naive observer Lorentz
transformation and an additional part given by the Seiberg-Witten
differential equation (which is very conveniently derived in this
manner).

A brute-force approach to probe the existence of additional symmetries is to
compute one-loop Green's functions other than the selfenergy. For
$\theta$-deformed Maxwell theory (which has no $\theta$-independent
interactions) the computational effort becomes tremendous. In this paper we
therefore focus on $\theta$-deformed noncommutative QED \cite{Bichl:2001gu},
for which we are able to compute all divergent one-loop Green's functions up
to first order in $\theta$. The (already extremely lengthy) computation in
analytic regularization \cite{Speer:gj} is performed using a
\emph{Mathematica$^{TM}$} package \cite{Ertl}. The final result is simple and
(at least for some people) disappointing:
\begin{equation}
\text{\emph{Noncommutative QED cannot be renormalized by means of 
    Seiberg-Witten  expansion}.}  
\end{equation}
We provide some ideas how the Seiberg-Witten expansion can be used as a
computational technique to treat one-loop divergences of the full
($\theta$-unexpanded) noncommutative QED \cite{Hayakawa:1999zf}, but this
leads---similarly as UV/IR mixing---to problems at the two-loop level. This is
the end of the chapter on noncommutative quantum field theories treated by the
Seiberg-Witten map.

\section{Noncommutative $\mathbb{R}^4$}
\label{Sec2}

Our presentation of the noncommutative $\mathbb{R}^4$ is (with different
notations, however) based on \cite{Gracia-Bondia:1987} and the appendix of
\cite{Gracia-Bondia:2001ct}. Let $\mathcal{A}$ be the space of Schwartz class
functions\footnote{The Schwartz space $\mathcal{S}(\mathbb{R}^4)$ is
  the space of smooth complex-valued functions $a$ on $\mathbb{R}^4$
  such that for all multi-indices $\alpha,\beta$ there exist constants
  $C^{\alpha,\beta}$ with $| x^\alpha \partial^\beta a(x) | \leq
  C^{\alpha,\beta}$.} $a\equiv a(x)$ on $\mathbb{R}^4$, equipped with
the multiplication rule\footnote{It would be wrong in general to
  replace (\ref{star}) by 
  $(a\star b)(x)= \big(\exp(\frac{\mathrm{i}}{2}
  \theta^{\mu\nu} \frac{\partial}{\partial y^\mu}
  \frac{\partial}{\partial z^\nu}) a(y) b(z)\big)_{x=y=z}$, which
  for instance yields zero if $a(x)$ and $b(x)$ have disjoint support.}
\begin{align}
(a \star b)(x) &:= \int_{\mathbb{R}^4} \!\frac{d^4k}{(2\pi^4)}    
\int_{\mathbb{R}^4} \! d^4y \; a(x {+} \tfrac{1}{2} \theta{\cdot}k) 
\, b(x{+}y) \,\mathrm{e}^{\mathrm{i} k{\cdot} y} ~,
\label{star}
\end{align}
where $\theta \in M_4 \mathbb{R}$ is a real-valued antisymmetric
constant matrix, $\theta^{\mu\nu}=-\theta^{\nu\mu}$, and
$(\theta{\cdot}k)^\mu = \theta^{\mu\nu} k_\nu$, $k{\cdot}y = k_\mu
y^\mu$. Position space variables are denoted by $x,y,z$ and momentum
space variables by $k,l,p,q,r,s,t$. The multiplication (\ref{star}) is
associative but noncommutative. The $\star$-(anti)commutators are
defined by $[a,b]_\star=a\star b - b \star a$ and
$\{a,b\}_\star=a\star b + b \star a$.  There is an involution on
$\mathcal{A}$ given by complex conjugation $a^*(x)=\overline{a(x)}$
which satisfies $(a\star b)^* = b^* \star a^*$.  Partial derivatives
$\partial_\mu=\frac{\partial}{\partial x^\mu}$ are derivations with
respect to (\ref{star}), $\partial_\mu (a\star b) = \partial_\mu a
\star b + a \star \partial_\mu b $.  Since Schwartz class functions
are trace-class one can define an integral
\begin{align}
\int \! a = \int_{\mathbb{R}^4}\!  d^4x \, a(x)~,\qquad 
\int \! a\star b = \int \! b \star a~.  
\end{align}

An important concept is that of the multiplier algebra
\begin{align}
\mathcal{M} = \big\{ f: \mathbb{R}^4 \to \mathbb{C}~,~~
f\star a \in \mathcal{A} \text{ and } 
a\star f \in \mathcal{A} \text{ for all } a \in \mathcal{A} \big\} ~,
\label{MA}
\end{align}
where the product is given by (\ref{star}). The product of $f,g
\in \mathcal{M}$ is defined by associativity of (\ref{star}), 
$(f \star g) \star a = f \star (g \star a) \in \mathcal{A}$ 
for all $a \in \mathcal{A}$. The multiplier algebra
$\mathcal{M}$ contains, for example, the unit $1 \in \mathcal{M}$ and
the coordinate functions $x^\mu(y)\equiv y^\mu$, which both do not belong
to $\mathcal{A}$. The famous formula $[x^\mu,x^\nu]_\star
=\mathrm{i}\theta^{\mu\nu}$ is thus an identity in $\mathcal{M}$ and not in
$\mathcal{A}$. The multiplier algebra is the biggest compactification of
$\mathcal{A}$.  

Additionally we introduce the space $\mathcal{H}=\mathbb{C}^4 \otimes
L^2(\mathbb{R}^4)$ of square-integrable bispinors $\eta=\{\eta_s(x)\}_{s=1}^4$,
equipped with the sesquilinear inner product\footnote{We work in Minkowski
  space with metric $g^{\mu\nu}=\mathrm{diag}(1,-1,-1,-1)$. The inner product
  (\ref{inner}) is therefore not positive definite and $\mathcal{H}$ is not a
  Hilbert space.  The $\gamma$-matrices fulfill
  $\gamma^\mu\gamma^\nu+\gamma^\nu\gamma^\mu=2g^{\mu\nu} 1_{4\times4}$ and 
$\gamma^0 \big(\gamma^\mu\big)^\dag \gamma^0 = \gamma^\mu$.}
\begin{align}
\big\langle \xi,\eta \big\rangle_{\mathcal{H}} = \int d^4x
\sum_{s,s'=1}^4 \overline{\xi_s(x)} \big(\gamma^0\big)^{ss'} 
\eta_{s'}(x)~,
\label{inner}
\end{align}
where $\big(\gamma^0\big)^{ss'}$ are the matrix entries of the
$\gamma^0$-matrix. The multiplication
(\ref{star}) extends to an involutive action $\mathcal{A} \times \mathcal{H}
\to \mathcal{H}$ obtained by componentwise $\star$-multiplication,
\begin{align}
(a \star \eta)_s(x) &:= \int_{\mathbb{R}^4} \!\frac{d^4k}{(2\pi^4)}    
\int_{\mathbb{R}^4} \! d^4y \; a(x {+} \tfrac{1}{2} \theta{\cdot}k) 
\, \eta_s(x{+}y) \,\mathrm{e}^{\mathrm{i} k{\cdot} y} ~,\qquad 
\big\langle a \star \xi,\eta\big\rangle_{\mathcal{H}}= 
\big\langle \xi,a^*\star \eta\big\rangle_{\mathcal{H}}~.
\end{align}

We regard bosonic fields as distinguished elements $\phi_i \in
\mathcal{A}$ to which one assigns a power-counting dimension
$\mathrm{dim}(\phi_i) \in \mathbb{Z}$ and fermionic fields are
distinguished Grassmann-valued elements $\psi_i \in \mathcal{H}$ with
power-counting dimension $\mathrm{dim}(\psi_i) \in \mathbb{Z}/2$.  For
$\Gamma[\phi_i]$ being a (sufficiently regular) functional of the
fields $\phi_i$ (fermion fields now included) we define the functional
derivative
\begin{align}
\Big\langle \tilde{\phi}_j[\phi_k] , \frac{\delta}{\delta \phi_j}
\Big\rangle \Gamma[\phi_i] \equiv  
\Big\langle \tilde{\phi}_j[\phi_k] , \frac{\delta \Gamma[\phi_i]}{
\delta \phi_j} \Big\rangle 
&:= \lim_{\epsilon \to 0} \frac{1}{\epsilon} 
\Big( \Gamma\big[\phi_i + \epsilon \tilde{\phi}_i[\phi_k]\big]
-\Gamma[\phi_i] \Big)~,
\label{funcdev}
\end{align}
which replaces a field $\phi_j$ by the fuctional
$\tilde{\phi}_j[\phi_k]$ in a derivational manner. Summation over $j$
in (\ref{funcdev}) is self-understood. We use (\ref{funcdev}) to
define functional derivatives with respect to $\theta^{\rho\sigma}$ as
well. Allowing for an explicit $\theta$-dependence of the fields $U$
and $V$ we obtain from (\ref{star})
\begin{align}
\Big\langle \Theta^{\rho\sigma}, \frac{\delta (U \star V)}{\delta 
\theta^{\rho\sigma}} \Big\rangle = 
\Big\langle  \Theta^{\rho\sigma}, \frac{\delta U }{\delta 
\theta^{\rho\sigma}} \Big\rangle \star V + 
U \star \Big\langle  \Theta^{\rho\sigma}, \frac{\delta V}{\delta 
\theta^{\rho\sigma}} \Big\rangle + \frac{\mathrm{i}}{2}
\Theta^{\rho\sigma} \, (\partial_\rho U) \star (\partial_\sigma V)~,
\end{align}
where $\Theta^{\rho\sigma}$ has to be constant.

\section{Noncommutative quantum electrodynamics}
\label{Sec3}

We are going to study $\theta$-expanded noncommutative QED 
defined by the classical action (on noncommutative level, i.e.\ before
$\theta$-expansion) 
\begin{align}
\hat{\Sigma} &= \big\langle \hat{\psi} , \big(
\mathrm{i}\gamma^\mu \hat{D}_\mu -m \big) \hat{\psi}
\big\rangle_{\mathcal{H}}
-\frac{1}{4g^2} \int \hat{F}_{\mu\nu} \star \hat{F}^{\mu\nu} ~,
\label{YMD}
\\*
\hat{F}_{\mu\nu} &= \partial_\mu \hat{A}_\nu - 
\partial_\nu \hat{A}_\mu - \mathrm{i} [\hat{A}_\mu,\hat{A}_\nu]_\star
~,\qquad 
\hat{D}_\mu \hat{\psi} = \partial_\mu \hat{\psi} 
- \mathrm{i} \hat{A}_\mu \star \hat{\psi}~,
\nonumber
\end{align}
for the noncommutative gauge fields (photons)
$\hat{A}_\mu=(\hat{A}_\mu)^* \in \mathcal{A}$ and the noncommutative
fermion field (electrons) $\hat{\psi} \in \mathcal{H}$ to which we
assign the power-counting dimensions $\mathrm{dim}(\hat{A}_\mu)=1$ and 
$\mathrm{dim}(\hat{\psi})=\tfrac{3}{2}$.  Additionally we define
$\mathrm{dim}(\theta^{\rho\sigma})=-2$.  The action (\ref{YMD}) is
invariant under infinitesimal gauge transformations
$W^G_{\hat{\lambda}}$, observer Lorentz rotations $W^R_{\alpha\beta}$
and translations $W^T_\tau$, see \cite{Bichl:2001yf,Grimstrup:2001ja}:
\begin{align}
W^G_{\hat{\lambda}} &= \Big\langle (\partial_\mu \hat{\lambda} -
\mathrm{i} [\hat{A}_\mu,\hat{\lambda}]_\star),\frac{\delta}{\delta
  \hat{A}_\mu} \Big\rangle 
+  \Big\langle \mathrm{i} \hat{\lambda} \star \hat{\psi} ,
\frac{\delta}{\delta \hat{\psi}} \Big\rangle ~,
\qquad \hat{\lambda}=\hat{\lambda}^* \in \mathcal{A}~,
\\
W^R_{\alpha\beta} &= W^R_{\hat{A}+\hat{\psi};\alpha\beta} 
+ W^R_{\theta;\alpha\beta} ~,
\label{WR}
\\
W^R_{\hat{A}+\hat{\psi};\alpha\beta} &= \Big\langle 
\Big(\frac{1}{2} \{ x_\alpha, \partial_\beta \hat{A}_\mu\}_\star 
-\frac{1}{2} \{ x_\beta, \partial_\alpha \hat{A}_\mu\}_\star 
+ g_{\alpha\mu} \hat{A}_\beta 
- g_{\beta\mu} \hat{A}_\alpha \Big) ,  
\frac{\delta}{\delta \hat{A}_\mu} \Big\rangle 
\nonumber
\\
& + \Big\langle \Big( 
x_\alpha \star \partial_\beta \hat{\psi} 
- \mathrm{i} \theta_\alpha^{~\rho} \partial_\beta \partial_\rho 
\hat{\psi}
- x_\beta \star \partial_\alpha \hat{\psi} 
+ \mathrm{i} \theta_\beta^{~\rho} \partial_\alpha \partial_\rho 
\hat{\psi} +\frac{1}{4} 
[\gamma_\alpha,\gamma_\beta] \hat{\psi} \Big),
\frac{\delta}{\delta \hat{\psi}} \Big\rangle 
\tag{\oldref{WR}a}
\label{WRa}
\\
W^R_{\theta;\alpha\beta} &= \Big\langle 
(\delta_\alpha^\rho \theta_\beta^{~\sigma} 
+\delta_\alpha^\sigma \theta^\rho_{~\beta} 
-\delta_\beta^\rho \theta_\alpha^{~\sigma} 
-\delta_\beta^\sigma \theta^\rho_{~\alpha} ), 
\frac{\delta}{\delta \theta^{\rho\sigma}} \Big\rangle ~,
\tag{\oldref{WR}b}
\\
W^T_\tau &= 
\Big\langle \partial_\tau \hat{A}_\mu , 
\frac{\delta}{\delta \hat{A}_\mu} \Big\rangle 
+\Big\langle \partial_\tau \hat{\psi} , 
\frac{\delta}{\delta \hat{\psi}} \Big\rangle ~.
\end{align}
These Ward identity operators satisfy the following commutation relations:
\begin{align}
[W^G_{\hat{\lambda}_1}, W^G_{\hat{\lambda}_2}] &= 
W^G_{\hat{\lambda}_{12}}~,  
\nonumber
\\
[W^G_{\hat{\lambda}}, W^R_{\alpha\beta}] &= 
W^G_{\hat{\lambda}_{\alpha\beta}}~,  
& 
[W^G_{\hat{\lambda}}, W^T_{\alpha\beta}] &= W^G_{\hat{\lambda}_\tau}~,  
\nonumber
\\
[W^R_{\alpha\beta}, W^R_{\gamma\delta}] &= 
g_{\alpha\gamma} W^R_{\beta\delta} 
-g_{\beta\gamma} W^R_{\alpha\delta} 
-g_{\alpha\delta} W^R_{\beta\gamma} 
+ g_{\beta\delta} W^R_{\alpha\gamma} ~, \hspace*{-15em}
\nonumber
\\
[W^R_{\alpha\beta}, W^T_\tau] &= g_{\alpha\tau} W^T_\beta - 
g_{\beta\tau} W^T_\alpha ~,
&
[W^T_\tau, W^T_\sigma] &= 0~,
\end{align}
for certain $\hat{\lambda}_{12},\hat{\lambda}_{\alpha\beta},\hat{\lambda}_\tau
\in \mathcal{A}$ depending on $\hat{\lambda},\hat{\lambda}_1,\hat{\lambda}_2
\in \mathcal{A}$ as given in \cite{Bichl:2001yf}.

\section{Seiberg-Witten map}

There are two kinds of Lorentz transformations for a field theory on
$\theta$-deformed space-time \cite{Colladay:1998fq}: \emph{Observer Lorentz
  transformations} refer to passing to another reference frame; the physical
action has to be invariant under such a tranformation. \emph{Particle Lorentz
  transformations} refer to a repositioning of all particles in a given
reference frame in which the background field $\theta$ remains unchanged. In
general the physical action is not invariant under such a transformation.
Since the particle Lorentz symmetry breaking must in principle be observable,
it has to be gauge-invariant.

At first sight the particle Lorentz rotation is given by
$W^R_{\hat{A}+\hat{\psi};\alpha\beta}$ defined in (\ref{WRa}). However, that
transformation, applied to (\ref{YMD}), does not lead to a gauge-invariant
particle Lorentz symmetry breaking. This means that one rather has to split
(\ref{WR}) in the following way \cite{Bichl:2001yf}:
\begin{align}
W^R_{\alpha\beta} &= \tilde{W}^R_{\hat{A}+\hat{\psi};\alpha\beta} 
+ \tilde{W}^R_{\theta;\alpha\beta}~,
\label{WRT}
\\
\tilde{W}^R_{\hat{A}+\hat{\psi};\alpha\beta} &= 
W^R_{\hat{A}+\hat{\psi};\alpha\beta} 
- W^R_{\theta;\alpha\beta}(\theta^{\rho\sigma}) \Big(
\Big\langle \frac{d \hat{A}_\mu}{d \theta^{\rho\sigma}} , 
\frac{\delta}{\delta \hat{A}_\mu} \Big\rangle 
+\Big\langle \frac{d \hat{\psi}}{d \theta^{\rho\sigma}} , 
\frac{\delta}{\delta \hat{\psi}} \Big\rangle \Big)~,
\tag{\oldref{WRT}a}
\label{WRTa}
\\
\tilde{W}^R_{\theta;\alpha\beta} &= 
W^R_{\theta;\alpha\beta} 
+ W^R_{\theta;\alpha\beta}(\theta^{\rho\sigma}) \Big(
\Big\langle \frac{d \hat{A}_\mu}{d \theta^{\rho\sigma}} , 
\frac{\delta}{\delta \hat{A}_\mu} \Big\rangle 
+\Big\langle \frac{d \hat{\psi}}{d \theta^{\rho\sigma}} , 
\frac{\delta}{\delta \hat{\psi}} \Big\rangle \Big)~.
\tag{\oldref{WRT}b}
\end{align}
The transformation (\ref{WRTa}) is then a particle Lorentz rotation if 
$\tilde{W}^R_{\hat{A}+\hat{\psi}}$ applied to the action (\ref{YMD}) is
gauge-invariant. This condition is solved by
\cite{Bichl:2001yf,Grimstrup:2001ja} 
\begin{subequations}
\label{SW}
\begin{align}
\frac{d \hat{A}_\mu}{d \theta^{\rho\sigma}} 
&= -\frac{1}{8} \big\{ \hat{A}_\rho, 
\partial_\sigma \hat{A}_\mu + \hat{F}_{\sigma\mu} \big\}_\star 
+ \frac{1}{8} \big\{ \hat{A}_\sigma, 
\partial_\rho \hat{A}_\mu + \hat{F}_{\rho\mu} \big\}_\star +
\hat{\Omega}_{\rho\sigma\mu} ~,
\\
\frac{d \hat{\psi}}{d \theta^{\rho\sigma}} 
&= -\frac{1}{8} \hat{A}_\rho \star 
\big(\partial_\sigma \hat{\psi} + \hat{\bar{D}}_\sigma 
\hat{\psi} \big)  
+ \frac{1}{8} \hat{A}_\sigma \star 
\big(\partial_\rho \hat{\psi} + \hat{\bar{D}}_\rho
\hat{\psi} \big) + \hat{\Psi}_{\rho\sigma} ~,
\label{SWF}
\end{align}
\end{subequations}
where $\hat{\Omega}_{\rho\sigma\mu}$ and $\hat{\Psi}_{\rho\sigma}$
transform covariantly under gauge transformations:
\begin{subequations}
\label{OmegaPsi}
\begin{align}
W^G_{\hat{\lambda}} \hat{\Omega}_{\rho\sigma\mu} 
& = \mathrm{i} \big[
\hat{\lambda} ,  \hat{\Omega}_{\rho\sigma\mu}\big]_\star ~,
&
W^G_{\hat{\lambda}} \hat{\Psi}_{\rho\sigma} 
& =  \mathrm{i} \hat{\lambda} \star \hat{\Psi}_{\rho\sigma} ~.
\end{align}
Compatibility in (\ref{SW}) requires 
\begin{align}
\mathrm{dim}\big(\hat{\Omega}_{\rho\sigma\mu}\big) &=3~,
&
\mathrm{dim}\big(\hat{\Psi}_{\rho\sigma}\big) &= \tfrac{7}{2}~,
&
\hat{\Omega}_{\rho\sigma\mu}
=\big(\hat{\Omega}_{\rho\sigma\mu}\big)^*~.
\end{align}
\end{subequations}

The relations (\ref{SW}) can now be regarded as first-order (Seiberg-Witten
\cite{Seiberg:1999vs})  
differential equations for the noncommutative fields
$\hat{A}_\mu,\hat{\psi}$. As such they are solved by a power
series in $\theta$ and the initial values $A_\mu,\psi$ for 
$\hat{A}_\mu,\hat{\psi}$, respectively, at
$\theta=0$. Inserting this solution into the action (\ref{YMD}) one
obtaines an action $\Sigma[A_\mu,\psi,\theta]$ which at each order
$n$ in $\theta$ is invariant under commutative gauge transformations and
commutative observer Lorentz transformations \cite{Bichl:2001yf}:
\begin{subequations}
\begin{align}
W^G_{\lambda} &= \Big\langle (\partial_\mu \lambda -
\mathrm{i} [A_\mu,\lambda]),\frac{\delta}{\delta A_\mu} \Big\rangle 
+  \Big\langle \mathrm{i} \lambda \psi ,
\frac{\delta}{\delta \psi} \Big\rangle ~,
\\
W^R_{\alpha\beta} &= 
\Big\langle 
\Big( x_\alpha \partial_\beta A_\mu 
- x_\beta \partial_\alpha A_\mu 
+ g_{\alpha\mu} A_\beta 
- g_{\beta\mu} A_\alpha \Big) ,  
\frac{\delta}{\delta A_\mu} \Big\rangle 
\nonumber
\\
& 
+ \Big\langle \Big( 
x_\alpha \partial_\beta \psi 
- x_\beta \partial_\alpha \psi 
+\frac{1}{4} [\gamma_\alpha,\gamma_\beta] \psi \Big),
\frac{\delta}{\delta \psi} \Big\rangle 
\nonumber
\\
& 
+ \Big\langle 
(\delta_\alpha^\rho \theta_\beta^{~\sigma} 
+\delta_\alpha^\sigma \theta^\rho_{~\beta} 
-\delta_\beta^\rho \theta_\alpha^{~\sigma} 
-\delta_\beta^\sigma \theta^\rho_{~\alpha} ), 
\frac{\delta}{\delta \theta^{\rho\sigma}} \Big\rangle ~,
\\
W^T_\tau &= 
\Big\langle \partial_\tau A_\mu , 
\frac{\delta}{\delta A_\mu} \Big\rangle 
+\Big\langle \partial_\tau \psi , 
\frac{\delta}{\delta \psi} \Big\rangle ~.
\end{align}
\end{subequations}

\section{Noncommutative Yang-Mills-Dirac action to first order in $\theta$}

The solution of (\ref{SW}) is up to first order in $\theta$ given by 
\begin{subequations}
\label{SW1}
\begin{align}
\hat{A}_\mu &= A_\mu +\theta^{\alpha\beta} \big(
- \tfrac{1}{2} A_\alpha (\partial_\beta A_\mu + F_{\beta\mu}) 
\big) + \mathcal{O}(\theta^2) \,,
\label{A}
\\[1ex]
\hat{\psi} &= \psi + \theta^{\alpha\beta}\Big(
-\tfrac{1}{2} A_\alpha \partial_\beta \psi
+\tfrac{1}{4} \partial_\alpha A_\beta \psi
\nonumber
\\*
&+ \kappa_1 F_{\alpha\beta}\psi 
+ \kappa_2 F_{\nu\beta} \gamma^\nu_{~\alpha} \psi 
+ \kappa_3 F_{\nu\rho} 
\gamma^{\nu\rho}_{~~\alpha\beta} \psi 
+ \mathrm{i} \kappa_4 g^{\nu\rho} \gamma_{\alpha\beta} 
D_\nu D_\rho \psi
\nonumber
\\*
&+ \mathrm{i} \kappa_5 \gamma^\nu_{~\alpha} D_\beta D_\nu \psi
+ \kappa_6 m \gamma_\alpha D_\beta \psi
+ \kappa_7 m \gamma^\nu_{~\alpha\beta} D_\nu \psi
+ \mathrm{i} \kappa_8 m^2 \gamma_{\alpha\beta} \psi\Big)
+ \mathcal{O}(\theta^2) ~,
\label{psi}
\end{align}
\end{subequations}
where $D_\mu \psi = \partial_\mu \psi-\mathrm{i} A_\mu\psi$. The
$\kappa_i$ parametrize the solutions of (\ref{OmegaPsi})
for\footnote{It is convenient to shift $\kappa_1$ by $\frac{1}{8}$ in
  order to add in (\ref{psi}) the formally $\kappa$-independent term
  $\theta^{\alpha\beta} \partial_\alpha A_\beta \psi= \frac{1}{2}
  \theta^{\alpha\beta} F_{\alpha\beta} \psi$ to the
  $\hat{\Psi}$-independent part of (\ref{SWF}).}
$\hat{\Omega}_{\rho\mu\sigma}$ and $\hat{\Psi}_{\rho\sigma}$.  They
play the role of additional coupling constants which parametrize
(unphysical) \emph{field redefinitions}, see \cite{Bichl:2001cq} for
the model without fermions\footnote{That the ambiguity in the solution
  of the Seiberg-Witten differential equation is linked to field
  redefinitions was already noticed in \cite{Asakawa:1999cu}.}. Thus
we expect these coupling constants to be power series in Planck's
constant $\hbar$ in order to absorb possible divergences of the
effective quantum action\footnote{Note that the coefficients in front
  of the non-covariant field monomials in (\ref{SW1}) cannot be
  renormalized because they are fixed by gauge-invariance of the
  $\theta$-expanded action.}. It turns out that all possible
divergences in first order in $\theta$ are purely imaginary in
momentum space.  Hence we have written down only solutions of
(\ref{OmegaPsi}) which are purely imaginary in momentum space. In
particular, the $\kappa_i$ are real. We have introduced the completely
antisymmetric gamma matrices
\begin{subequations}
\begin{align}
\gamma^{\mu\nu} &= \tfrac{1}{2} [\gamma^\mu , \gamma^\nu]~, & 
\gamma^{\mu\nu\rho} &= \tfrac{1}{4} \{\gamma^\mu ,
[\gamma^\nu,\gamma^\rho]\} ~, & 
\gamma^{\mu\nu\rho\sigma} &= \tfrac{1}{8} [\gamma^\mu ,
\{\gamma^\nu,[\gamma^\rho,\gamma^\sigma]\}] ~,
\end{align}
fulfilling 
\begin{align}
\gamma^\mu \gamma^{\nu_1\dots \nu_n} -(-1)^n 
 \gamma^{\nu_1\dots \nu_n} \gamma^\mu = \sum_{j=1}^n 
(-1)^{j+1} \, 2 g^{\mu\nu_j} \, \gamma^{\nu_1\dots \nu_{j-1} 
\nu_{j+1} \dots \nu_j}~,\quad n=1,\dots,4~.
\end{align}
\end{subequations}
Note that $\hat{\Omega}_{\rho\sigma\mu} \in \mathcal{A}$ cannot
contain a term bilinear in $\hat{\psi}$ because---within our framework
presented in Section~\ref{Sec2}---there is no way to make an element
of $\mathcal{A}$ out of two elements of $\mathcal{H}$.  We shall see
(this is one of the main results of the paper) that the lack of such a
possibility makes the $\theta$-expanded noncommutative QED
unrenormalizable.  Therefore some readers may suggest to enlarge
somehow the framework of Section~\ref{Sec2} and to replace (\ref{A})
by
\begin{align}
\hat{A}_\mu(x) = A_\mu(x) 
+\theta^{\alpha\beta} \Big(
- \tfrac{1}{2} A_\alpha(x) & 
\big(\partial_\beta A_\mu(x) + F_{\beta\mu}(x) \big) 
\nonumber
\\*[-1ex]
&
+ \mathrm{i} 
g^2 \kappa_9 \sum_{s,s'=1}^4 \overline{\psi_s(x)} 
\big(\gamma^0 \gamma_{\mu\alpha\beta} \big)^{ss'} \psi_{s'}(x) 
\Big) + \mathcal{O}(\theta^2) \,.
\tag{\oldref{A}'}\label{A'}
\end{align}
To satisfy those readers we will indeed work with (\ref{A'}). We will see,
however, that (\ref{A'}) does not improve the result at all. Therefore, future
work on similar subjects can stay withing the mathematical framework of
Section~\ref{Sec2} from the very beginning.

Inserting (\ref{SW1}) into the noncommutative Yang-Mills-Dirac action
(\ref{YMD}) we obtain\footnote{We would like to draw the attention of the
  reader to the following typographical subtlety. From now on we will write
  down our formulae in terms of the \emph{adjoint spinor}
  $\bar{\psi}=\psi^\dag \gamma^0$ (a row of four elements of
  $L^2(\mathbb{R}^4)$).  There is no risk to confuse the adjoint spinor with
  the previous operation of complex conjugation $\overline{\psi_s(x)}$ (for
  one element of $L^2(\mathbb{R}^4)$).  The line symbolizing complex
  conjugation is longer. Additionally we restrict the number field we work
  with from $\mathbb{C}$ to $\mathbb{R}$ so that spinor $\psi$ and its adjoint
  $\bar{\psi}$ must be regarded as independent.}  to first order in $\theta$
\begin{align}
\Sigma_{c\ell} 
&= \int d^4 x\;\Big[ 
-\tfrac{1}{4 g^2} F_{\mu\nu} F^{\mu\nu} + 
\bar{\psi}\big(\gamma^\mu (\mathrm{i} \partial_\mu +
A_\mu)-m\big) \psi \nonumber
\nonumber 
\\
& + \theta^{\alpha\beta}\Big(
-\tfrac{1}{2 g^2} F_{\alpha\mu} F_{\beta\nu} F^{\mu\nu} 
+\tfrac{1}{8 g^2} F_{\alpha\beta} F_{\mu\nu} F^{\mu\nu} 
\nonumber
\\
& + 
(\tfrac{1}{4}{-} \kappa_2) \mathrm{i} 
\bar{\psi} \gamma^\mu 
(2 F_{\alpha\mu} D_\beta + \partial_\beta 
F_{\alpha\mu}) \psi 
+ (\kappa_1 {-}\tfrac{1}{2} \kappa_5) \mathrm{i} \bar{\psi} 
\gamma^\mu (2 F_{\alpha\beta} D_\mu 
+ \partial_\mu F_{\alpha\beta}) \psi 
\nonumber 
\\
& 
+ (\kappa_2 {+}2 \kappa_4) \mathrm{i} 
\bar{\psi} \gamma_\alpha  ( 2 F^\mu_{~\beta} D_\mu
+ \partial_\mu F^\mu_{~\beta}) \psi
+ \kappa_4 \bar{\psi} 
\gamma^\mu_{~\alpha\beta}
(- 2 D^\nu D_\nu D_\mu 
+ 2 \mathrm{i} F_\mu^{~\nu} D_\nu \psi 
+ \mathrm{i}\partial^\nu F_{\mu\nu})
\psi 
\nonumber 
\\
&
- (\tfrac{1}{2}\kappa_2 {+} 2 \kappa_3)\mathrm{i} 
\bar{\psi} \gamma^{\mu\nu}_{~~\alpha} \partial_\beta F_{\mu\nu} \psi
+ \mathrm{i} (2\kappa_3{+} \kappa_9) \bar{\psi}
\gamma^\nu_{~\alpha\beta} \partial^\mu F_{\mu\nu} \psi
\nonumber 
\\
& 
+(- 2 \kappa_1 {+}\kappa_6) m 
\bar{\psi} F_{\alpha\beta} \psi 
+(- 2 \kappa_3 {+} \kappa_7) m \bar{\psi} F_{\nu\rho} 
\gamma^{\nu\rho}_{~~\alpha\beta} \psi 
\nonumber 
\\
&
+(-2 \kappa_4 {+}2\kappa_7) \mathrm{i} m \bar{\psi} 
\gamma_{\alpha\beta} D^\nu D_\nu \psi 
+ (- \kappa_5 {+} \kappa_6 {+} 2\kappa_7) \mathrm{i}
m \bar{\psi} \gamma^\nu_{~\alpha} (2 D_\nu D_\beta 
+ \mathrm{i} F_{\nu\beta}) \psi 
\nonumber 
\\
& 
- (2\kappa_7 {+} 2\kappa_8) m^2 \bar{\psi} 
\gamma^\nu_{~\alpha\beta} D_\nu  \psi 
- 2 \mathrm{i} \kappa_8 m^3 \bar{\psi} \gamma_{\alpha\beta} \psi 
\nonumber
\\
& + \mathrm{i} g^2 \kappa_9 (\bar{\psi} \gamma^\mu \psi)
(\bar{\psi} \gamma_{\mu\alpha\beta}  \psi)
\Big)\Big] +\mathcal{O}(\theta^2)~.
\label{th-exp}
\end{align}
Note that $\{\kappa_1,\kappa_5,\kappa_6\}$ occur in (\ref{th-exp}) in the
combinations $2\kappa_1{-}\kappa_5$ and $\kappa_5{-}\kappa_6$ only. In the
massless case $m=0$ there is only the combination $2\kappa_1{-}\kappa_5$
suggesting not to eliminate $\kappa_6$. It is therefore no restriction
to put
\[
\kappa_5 =0~.
\]

In order to pass to quantum field theory we have to add a
BRST-invariant gauge-fixing term. This can be done before or after the
Seiberg-Witten map \cite{Bichl:2001nf}. We choose the more economical
way of a gauge-fixing of (\ref{th-exp}):
\begin{subequations}
\begin{align}
\Sigma_{gf} &= 
\int d^4x\;\Big( - \bar{c} \partial^\mu\partial_\mu c 
+ B \partial^\mu A_\mu + \frac{\alpha}{2} B^2 \Big)
=\int d^4x\; s \Big( \bar{c} \big(\partial^\mu A_\mu +
\frac{\alpha}{2} B \big) \Big)~,
\label{gf}
\\
s A_\mu &= \partial_\mu c ~,\quad
sc=0~,\quad
s\bar{c}=B~,\quad
sB=0~,\quad
s \psi = \mathrm{i} c \psi~,\quad
s \bar{\psi} = -\mathrm{i} \bar{\psi} c~.
\label{BRST}
\end{align}
\end{subequations}
Here one has to take into account that $\bar{c},c$ and $s$ are
Grassmann-valued.  Note that there are no interactions involving the
ghosts $c,\bar{c}$ or the multiplier field $B$. Thus, the nilpotent
BRST transformations (\ref{BRST}) are \emph{linear} transformations of
interacting fields. There is therefore no need to introduce sources
(antifields) for the BRST transformations, and the BRST invariance of
the total action $\Sigma=\Sigma_{cl}+\Sigma_{gf}$ can then be written
as follows:
\begin{align}
s \Sigma &= 
\Big\langle \partial_\mu c , 
\frac{\delta \Sigma}{\delta A_\mu} \Big\rangle 
+ \Big\langle {-} \mathrm{i} \bar{\psi} c , 
\frac{\delta \Sigma}{\delta \bar{\psi}} \Big\rangle 
+ \Big\langle \mathrm{i} c \psi , 
\frac{\delta \Sigma}{\delta \psi} \Big\rangle 
+ \Big\langle B , 
\frac{\delta \Sigma}{\delta \bar{c}} \Big\rangle =0~.
\label{Slavnov}
\end{align}

\section{Feynman rules}

We denote by $\Gamma[\phi_{i,c\ell}]$ the generating functional of
one-particle irreducible (1PI) Green's functions and by $Z_c[J_i]$ the
generating functional for connected Green's functions. Both are related by
Legendre transformation
\begin{align}
Z_c[J_i] &=\Gamma[\phi_{i,c\ell}] + \sum_i \int d^4x\; \phi_{i,c\ell}(x)\, 
J_i(x)~,\qquad J_i(x) = - \Big\langle \delta^4(x{-}y) , 
\frac{\delta \Gamma}{\delta \phi_{i,c\ell}(y)}\Big\rangle~.
\end{align}
Switching to momentum space\footnote{Our Fourier conventions are $f(x)=\int
  \frac{d^4k}{(2\pi)^4} \; \mathrm{e}^{-\mathrm{i} k\cdot x} \, \tilde{f}(k)$
  and $\tilde{f}(p)=\int d^4x\; \mathrm{e}^{\mathrm{i} k\cdot x} \, f(x)$.},
1PI Green's functions are obtained by functional derivation:
\begin{align}
(2\pi)^4 \delta^4(p_1+\dots+p_n) \Gamma_{\phi_1\dots\phi_n}(p_1,\dots,p_n)
= \frac{\delta^n \Gamma[\phi_{i,c\ell}]}{\delta 
\phi_{1,c\ell}(p_1) \dots \delta \phi_{n,c\ell}(p_n)}
\Big|_{\phi_{i,c\ell}=0}~. 
\end{align}
Functional derivation in momentum space is defined by
\begin{align}
\frac{\delta \phi_{i,c\ell}(p_i)}{\delta \phi_{j,c\ell}(p_j)}
= \delta^j_i \,(2\pi)^4 \delta^4(p_i{-}p_j)~. 
\label{1PI}
\end{align}
A parity factor of $-1$ has to be inserted for each commutation of a
Grassmann-valued derivative operator with a Grassmann-valued field.
Similarly, connected Green's functions are obtained by
\begin{align}
(2\pi)^4 \delta^4(p_1+\dots+p_n) \Delta^{\phi_1\dots\phi_n}(p_1,\dots,p_n)
= \frac{\delta^n Z_c[J_i]}{\delta 
J_1(p_1) \dots \delta J_n(p_n)} \Big|_{J_i=0}~. 
\end{align}
We introduce a bigrading $(\tau,\ell)$ for all Green's functions, with $\tau$
being the number of factors of $\theta$ and $\ell$ the number of loops. The
$(\ell{=}0)$-part corresponds to taking for $\Gamma$ the action
$\Sigma=\Sigma_{c\ell} + \Sigma_{gf}$.

\subsection{Propagators}
\begin{fmffile}{fmfgamma}
  
Feynman rules for propagators are obtained from the bilinear
$(\tau{=}0,\ell{=}0)$-part\footnote{We regard $\theta$-dependent terms which
    are bilinear in fields as vertices.} of $Z_c$. We only need the
  propagators for the sources of photons and electrons:
\begin{subequations}
\label{propagator}
\begin{align}
  &\parbox{30mm}{\begin{picture}(30,10)
\put(0,0){\begin{fmfgraph}(30,10)
\fmfleft{l}
\fmfright{r}
\fmf{fermion}{l,i}
\fmf{fermion}{i,r}
\end{fmfgraph}}
\put(5,8){\mbox{$-q$}}
\put(22,8){\mbox{$p$}}
\put(5,-1){\mbox{$\bar{\psi}$}}
\put(22,-1){\mbox{$\psi$}}
\end{picture}}
\quad
\Delta^{\bar{\psi}\psi}_{(0,0)}(q,p) = - \frac{\gamma^\mu p_\mu +m}{
p^2-m^2+\mathrm{i}\epsilon} ~,
\\[1ex]
&\parbox{30mm}{\begin{picture}(30,10)
\put(0,0){\begin{fmfgraph}(30,10)
\fmfleft{l}
\fmfright{r}
\fmf{photon}{i,l}
\fmf{photon}{i,r}
\fmf{phantom_arrow}{i,l}
\fmf{phantom_arrow}{i,r}
\end{fmfgraph}}
\put(5,8){\mbox{$p$}}
\put(22,8){\mbox{$q$}}
\put(5,-1){\mbox{$A_\mu$}}
\put(22,-1){\mbox{$A_\nu$}}
\end{picture}}
\quad
\Delta^{AA}_{(0,0)\mu\nu}(p,q) = \frac{g^2}{p^2+\mathrm{i}\epsilon} 
\Big(g_{\mu\nu}-\big(1{-}\frac{\alpha}{g^2}\big) 
\frac{p_\mu p_\nu}{p^2+\mathrm{i}\epsilon}
\Big)~.
\end{align}
\end{subequations}
The ghost propagator and the mixed $A$-$B$ propagator are not required because
there are no vertices involving $B$ and ghosts.

\subsection{Vertices independent of $\theta$}

Feynman rules for vertices are obtained from the interaction $(\ell{=}0)$-part
of $\Gamma$, i.e.\ from the interaction part of the total action $\Sigma$. The
only vertex which is independent of $\theta$ is the standard QED vertex:
\begin{align}
&\parbox{30mm}{\begin{picture}(30,20)
\put(0,0){\begin{fmfgraph}(30,20)
\fmfleft{l1,l2}
\fmfright{r}
\fmf{fermion}{i,l1}
\fmf{fermion}{l2,i}
\fmf{photon}{r,i}
\fmf{phantom_arrow}{r,i}
\end{fmfgraph}}
\put(22,12){\mbox{$p$}}
\put(10,2){\mbox{$-q$}}
\put(10,17){\mbox{$r$}}
\put(2,3){\mbox{$\bar{\psi}$}}
\put(2,15){\mbox{$\psi$}}
\put(22,5){\mbox{$A_\mu$}}
\end{picture}}
\quad
\Gamma_{A\bar{\psi}\psi}^{(0,0)\mu}(p,q,r) = \gamma^\mu~.
\label{Aff}
\end{align}

The free part of the
action $\Sigma$ leads to the following Green's functions, which by definition
do not give rise to Feynman rules:
\begin{align}
\Gamma^{(0,0)\mu\nu}_{AA}(p,q) 
&= - \frac{1}{g^2} \big( p^2 g^{\mu\nu} - p^\mu p^\nu \big) ~,
&
\Gamma^{(0,0)}_{\bar{\psi}\psi}(q,p) 
&= p_\mu \gamma^\mu-m~,
\nonumber
\\
\Gamma^{(0,0)\mu}_{AB}(p,q) &= - \mathrm{i} p^\mu ~, 
&
\Gamma^{(0,0)}_{BB}(p,q) &= \alpha ~, 
&
\Gamma^{(0,0)}_{\bar{c}c}(q,p) 
&= p^2~.
\end{align}

\subsection{Vertices linear in $\theta$}

At first order in $\theta$ we have the following graphs:
\begin{align}
&\parbox{20mm}{\begin{picture}(20,10)
\put(0,0){\begin{fmfgraph}(20,10)
\fmfleft{l}
\fmfright{r}
\fmf{fermion}{l,i}
\fmf{fermion}{i,r}
\fmfdot{i}
\end{fmfgraph}}
\put(3,8){\mbox{$-q$}}
\put(15,8){\mbox{$p$}}
\put(3,-1){\mbox{$\bar{\psi}$}}
\put(15,-1){\mbox{$\psi$}}
\end{picture}}
&&
\parbox{30mm}{\begin{picture}(30,20)
\put(0,0){\begin{fmfgraph}(30,20)
\fmfleft{l1,l2}
\fmfright{r}
\fmf{fermion}{i,l1}
\fmf{fermion}{l2,i}
\fmf{photon}{r,i}
\fmf{phantom_arrow}{r,i}
\fmfdot{i}
\end{fmfgraph}}
\put(22,12){\mbox{$p$}}
\put(10,2){\mbox{$-q$}}
\put(10,17){\mbox{$r$}}
\put(2,3){\mbox{$\bar{\psi}$}}
\put(2,15){\mbox{$\psi$}}
\put(22,5){\mbox{$A_\mu$}}
\end{picture}}
&&
\parbox{30mm}{\begin{picture}(30,20)
\put(0,0){\begin{fmfgraph}(30,20)
\fmfleft{l1,l2}
\fmfright{r1,r2}
\fmf{fermion}{i,l1}
\fmf{fermion}{l2,i}
\fmf{photon}{r1,i}
\fmf{phantom_arrow,tension=0}{r1,i}
\fmf{photon}{r2,i}
\fmf{phantom_arrow,tension=0}{r2,i}
\fmfdot{i}
\end{fmfgraph}}
\put(18,2){\mbox{$q$}}
\put(19,17){\mbox{$p$}}
\put(10,2){\mbox{$-r$}}
\put(10,17){\mbox{$s$}}
\put(2,3){\mbox{$\bar{\psi}$}}
\put(2,15){\mbox{$\psi$}}
\put(22,6){\mbox{$A_\nu$}}
\put(22,13){\mbox{$A_\mu$}}
\end{picture}}
&&
\parbox{30mm}{\begin{picture}(30,20)
\put(0,0){\begin{fmfgraph}(30,20)
\fmfleft{l1,l2}
\fmftop{t1,t2,r3,t3}
\fmfbottom{b1,b2,r1,b3}
\fmfright{r2}
\fmf{fermion,tension=2}{i,l1}
\fmf{fermion,tension=2}{l2,i}
\fmf{photon}{r1,i}
\fmf{phantom_arrow,tension=0}{r1,i}
\fmf{photon}{r2,i}
\fmf{phantom_arrow,tension=0}{r2,i}
\fmf{photon}{r3,i}
\fmf{phantom_arrow,tension=0}{r3,i}
\fmfdot{i}
\end{fmfgraph}}
\put(20,2){\mbox{$r$}}
\put(16,19){\mbox{$p$}}
\put(5,0){\mbox{$-s$}}
\put(7,17){\mbox{$t$}}
\put(2,3){\mbox{$\bar{\psi}$}}
\put(2,15){\mbox{$\psi$}}
\put(22,6){\mbox{$A_\nu$}}
\put(25,12){\mbox{$q$}}
\put(21,18){\mbox{$A_\mu$}}
\put(12,2){\mbox{$A_\rho$}}
\end{picture}}
\nonumber
\\[2ex]
&\Gamma^{(1,0)}_{\bar{\psi}\psi}(p) &&
\Gamma^{(1,0)\mu}_{A\bar{\psi}\psi}(p,q,r) &&
\Gamma^{(1,0)\mu\nu}_{AA\bar{\psi}\psi}(p,q,r,s) &&
\Gamma^{(1,0)\mu\nu\rho}_{AAA\bar{\psi}\psi}(p,q,r,s,t) 
\nonumber
\\[2ex]
&
\parbox{30mm}{\begin{picture}(30,20)
\put(0,0){\begin{fmfgraph}(30,20)
\fmfleft{l1,l2}
\fmfright{r1,r2}
\fmf{fermion}{i1,l1}
\fmf{fermion}{l2,i1}
\fmf{fermion}{r1,i2}
\fmf{fermion}{i2,r2}
\fmf{dots,tension=3}{i1,i2}
\fmfdot{i1}
\end{fmfgraph}}
\put(19,2){\mbox{$q$}}
\put(18,17){\mbox{$-p$}}
\put(9,2){\mbox{$-r$}}
\put(9,17){\mbox{$s$}}
\put(2,3){\mbox{$\bar{\psi}$}}
\put(2,15){\mbox{$\psi$}}
\put(24,6){\mbox{$\psi$}}
\put(24,13){\mbox{$\bar{\psi}$}}
\end{picture}}
&&
\parbox{30mm}{\begin{picture}(30,20)
\put(0,0){\begin{fmfgraph}(30,20)
\fmfleft{l1,l2}
\fmfright{r}
\fmf{photon}{l1,i}
\fmf{phantom_arrow}{l1,i}
\fmf{photon}{l2,i}
\fmf{phantom_arrow}{l2,i}
\fmf{photon}{r,i}
\fmf{phantom_arrow,tension=3}{r,i}
\fmfdot{i}
\end{fmfgraph}}
\put(22,12){\mbox{$p$}}
\put(10,2){\mbox{$q$}}
\put(10,17){\mbox{$r$}}
\put(2,5){\mbox{$A_\nu$}}
\put(2,14){\mbox{$A_\rho$}}
\put(22,5){\mbox{$A_\mu$}}
\end{picture}} \hspace*{-5em}
\nonumber
\\[2ex]
&\Gamma^{(1,0)}_{\bar{\psi}\psi;\bar{\psi}\psi}(p,q;r,s) &&
\Gamma^{(1,0)\mu\nu\rho}_{AAA;1+2}(p,q,r) \equiv 
\Gamma^{(1,0)\mu\nu\rho}_{AAA;1}(p,q,r) 
+\Gamma^{(1,0)\mu\nu\rho}_{AAA;2}(p,q,r) \hspace*{-25em}
\label{vertexfunctions}
\end{align}
where
\begin{subequations}
\begin{align}
\Gamma_{\bar{\psi}\psi}^{(1,0)}(q,p) &= 
2 \mathrm{i} \theta_{\alpha\beta} 
\Big(- \kappa_8 m^3 \gamma^{\alpha\beta} 
+ (\kappa_4{-}\kappa_7) m p^2 \gamma^{\alpha\beta} 
- m (\kappa_6 {+} 2 \kappa_7) 
p_\mu p^\beta \gamma^{\mu\alpha} 
\nonumber
\\[-1ex]
&\qquad\qquad
+ (\kappa_7{+}\kappa_8) m^2 p_\mu \gamma^{\mu\alpha\beta} 
- \kappa_4 p_\mu p^2 \gamma^{\mu\alpha\beta} \Big)~,
\\[1ex]
\Gamma_{A\bar{\psi}\psi}^{(1,0)\mu}(p,q,r) &= 
\mathrm{i} \theta_{\alpha\beta} \Big(
(2 \kappa_2 {-} \tfrac{1}{2}) p^\alpha r^\beta \gamma^\mu 
-(2 \kappa_2 {-} \tfrac{1}{2}) p^\nu r^\beta \gamma_\nu g^{\mu\alpha} 
+ ( \tfrac{1}{4} {+} 2 \kappa_1 {-} \kappa_2 ) 
p^\nu p^\beta \gamma_\nu g^{\mu\alpha}
\nonumber
\\
&
+(4 \kappa_1) \gamma_\nu r^\nu p^\beta g^{\mu\alpha} 
+ (2 \kappa_6 {-} 4 \kappa_1) m g^{\mu\alpha} p^\beta
+ (\kappa_2 {+} 2\kappa_4) (p^2 +2pr) \gamma^\beta g^{\mu\alpha} 
\nonumber
\\
&
- (\kappa_2 {+} 2\kappa_4) (p^\mu +2 r^\mu) r^\alpha \gamma^\beta  
- (\kappa_2 {+} 4\kappa_3) \gamma^{\mu\nu\alpha} p_\nu p^\beta 
\nonumber
\\
&
- ( \kappa_6 {+} 2 \kappa_7) m (p^\beta + 2 r^\beta)
\gamma^{\mu\alpha} 
+ (\kappa_6 {+} 2 \kappa_7) m (p_\nu +2 r_\nu) 
\gamma^{\nu\beta}  g^{\mu\alpha}
\nonumber
\\
&
+ (2 \kappa_4 {-} 2 \kappa_7) m (p^\mu +2 r^\mu) \gamma^{\alpha\beta}  
+ (2 \kappa_7 {-} 4\kappa_3) m p_\nu \gamma^{\mu\nu\alpha\beta} 
\nonumber
\\
&
+ (2 \kappa_3  {+} \kappa_9 {-} \kappa_4) p^\mu p_\nu \gamma^{\nu\alpha\beta}
- (2 \kappa_3 {+} \kappa_9 {-} \kappa_4) p^2 \gamma^{\mu\alpha\beta}
- 2 \kappa_4 (p^2 +pr + r^2) \gamma^{\mu\alpha\beta} 
\nonumber
\\
&
- 2 \kappa_4 (p^\mu r_\nu + r^\mu p_\nu + 2 r^\mu r_\nu) 
\gamma^{\nu\alpha\beta} 
+ (2 \kappa_7 {+} 2 \kappa_8) m^2 \gamma^{\mu\alpha\beta}\Big)~,
\\[1ex]
\Gamma_{AA\bar{\psi}\psi}^{(1,0)\mu\nu}(p,q,r,&s) = 
\mathrm{i} \theta_{\alpha\beta} \Big(
  (\tfrac{1}{2} {-}2\kappa_2)  (\gamma^\mu g^{\nu\alpha} p^\beta
+  \gamma^\nu  g^{\mu\alpha} q^\beta)
\nonumber
\\
&
+ (\tfrac{1}{2} {-}2\kappa_2) \gamma_\rho  
(p^\rho  g^{\alpha\mu} g^{\beta\nu} 
- q^\rho  g^{\alpha\mu} g^{\beta\nu})
+ 4 \kappa_1 (g^{\nu\alpha} q^\beta \gamma^\mu
+ g^{\mu\alpha} p^\beta \gamma^\nu)
\nonumber
\\
&
+ (2\kappa_2 {+} 4\kappa_4) g^{\mu\nu} \gamma^\alpha  (p^\beta +
q^\beta)
+ (2\kappa_2 {+} 4\kappa_4) \gamma^\beta (g^{\mu\alpha} p^\nu + 
g^{\nu\alpha} q^\mu) 
\nonumber
\\
&- 2 \kappa_4 \gamma^{\mu\alpha\beta} (p^\nu +q^\nu + 2 s^\nu)
- 2 \kappa_4 \gamma^{\nu\alpha\beta} (p^\mu +q^\mu + 2 s^\mu)
\nonumber
\\
&
- 2 \kappa_4 g^{\mu\nu} \gamma^{\rho\alpha\beta} (p_\rho + q_\rho + 2
s_\rho)
+ (4\kappa_7 {+} 2\kappa_6) m (
\gamma^{\mu\beta} g^{\nu\alpha}+\gamma^{\nu\beta} g^{\mu\alpha})
\nonumber
\\
&
+ (4 \kappa_4{-} 4\kappa_7) m g^{\mu\nu} \gamma^{\alpha\beta} \Big)~,
\\[1ex]
\Gamma_{AAA\bar{\psi}\psi}^{(1,0)\mu\nu\rho}(p,q,r&,s,t) = 
- 4 \mathrm{i} \theta_{\alpha\beta} \kappa_4 
\big(g^{\mu\nu} \gamma^{\rho\alpha\beta}
+g^{\nu\rho} \gamma^{\mu\alpha\beta}
+g^{\rho\mu} \gamma^{\nu\alpha\beta}\big)~,
\\[1ex]
\Gamma^{(1,0)}_{\bar{\psi}\psi ;\bar{\psi}\psi}(p,q;r,&s) =  
\mathrm{i} g^2 \theta_{\alpha\beta} \kappa_9 \gamma_\mu \otimes 
\gamma^{\mu\alpha\beta}~,
\label{4f}
\\[1ex]
\Gamma^{(1,0)\mu\nu\rho}_{AAA;1}(p,q,&r) 
= \frac{1}{g^2}\big(1+\lambda_1\big) 
\mathrm{i} \theta_{\alpha\beta} \Big(
g^{\alpha\mu}g^{\beta\nu}\big( (qr)p^{\rho} - (pr)q^{\rho} 
\big)
+g^{\alpha\nu}g^{\beta\rho}\big( (rp)q^{\mu} - (qp)r^{\mu} \big)
\nonumber
\\*
&
+g^{\alpha\rho}g^{\beta\mu}\big( (pq)r^{\nu} - (rq)p^{\nu} \big) 
\nonumber
\\
&+ g^{\alpha\mu} \big( 
( g^{\nu\rho} (pq) - p^\nu q^\rho ) r^{\beta}
+( g^{\nu\rho} (rp) - r^\nu p^\rho ) q^{\beta}\big) 
\nonumber
\\
&+ g^{\alpha\nu} \big( 
( g^{\rho\mu} (qr) - q^\rho r^\mu ) p^{\beta}
+( g^{\rho\mu} (pq) - p^\rho q^\mu ) r^{\beta}\big) 
\nonumber
\\*
&+ g^{\alpha\rho} \big( 
( g^{\mu\nu} (rp) - r^\mu p^\nu ) q^{\beta}
+( g^{\mu\nu} (qr) - q^\mu r^\nu ) p^{\beta}\big) 
\nonumber
\\
&+ g^{\mu\nu}\big(p^{\rho} q^{\alpha} r^{\beta} 
+ q^{\rho} p^{\alpha} r^{\beta} \big)
 + g^{\nu\rho}\big(q^{\mu} r^{\alpha} p^{\beta} 
+ r^{\mu} q^{\alpha} p^{\beta} \big)
\nonumber
\\*[-0.5ex]
& + g^{\rho\mu}\big(r^{\nu} p^{\alpha} q^{\beta} 
+ p^{\nu} r^{\alpha} q^{\beta} \big)\Big)~,
\label{tau1}
\\[1ex]
\Gamma^{(1,0)\mu\nu\rho}_{AAA;2}(p,q,&r) 
= \frac{1}{g^2}\big(1+\lambda_2\big) \mathrm{i} 
\theta_{\alpha\beta} \Big(
-g^{\alpha\mu} ( g^{\nu\rho} (rq) - r^\nu q^\rho ) p^{\beta}
- g^{\alpha\nu}  ( g^{\rho\mu} (pr) - p^\rho r^\mu ) q^{\beta}
\nonumber
\\*
& \qquad 
- g^{\alpha\rho} ( g^{\mu\nu} (qp) - q^\mu p^\nu ) r^{\beta} \Big)~.
\label{tau2}
\end{align}
\end{subequations}
The last two terms (\ref{tau1}),(\ref{tau2}) occur as a sum with
$\lambda_1=\lambda_2=0$, we split them artificially in order to discuss
possible extensions 
\begin{align}
\text{YM} = -\frac{1}{4 g^2}  \int \Big( \hat{F}_{\mu\nu}
\star \hat{F}^{\mu\nu} + 2 \lambda_1 \theta^{\alpha\beta} 
\hat{F}_{\alpha\mu} \star
\hat{F}_{\beta\nu} \star \hat{F}^{\mu\nu} 
-  \frac{1}{2} \lambda_2 \theta^{\alpha\beta} \hat{F}_{\alpha\beta} 
\star \hat{F}_{\mu\nu} \star \hat{F}^{\mu\nu} \Big)
\label{YM-ext}
\end{align}
to the bosonic action. These Feynman rules are subject to momentum
conservation $p{+}q=0$, $p{+}q{+}r=0$, $p{+}q{+}r{+}s=0$ and
$p{+}q{+}r{+}s{+}t=0$, respectively.  
The strange fifth graph in (\ref{vertexfunctions})
symbolizes a single vertex. The dotted line permits
momentum exchange but does not affect the spinor structure. 

\subsection{Concatenation of propagators and vertices}

To an 1PI Feynman graph one associates an integral by concatenation of
propagators (inner lines only) and vertices, matching momenta and
Lorentz indices and preserving the sense of the arrows. Concatenation
from left to right of fermion propagators and vertices is always in
opposite sense of the arrow, because the adjoint spinor symbolized by
an outgoing arrow is on the left. To any closed loop with loop
momentum $k_i$ one associates an integration operator
$\frac{\hbar}{\mathrm{i}} \int \frac{d^4 k_i}{(2\pi)^4}$. To each
closed electron line one associates the operator $-\mathrm{tr}$, where
$\mathrm{tr}$ denotes the trace over the $\gamma$-matrices in the
loop. If the graph has an $S$-fold symmetry the integral has to be
devided by $S$.

The integration over all internal loop momenta of a subgraph with $N_A$
external photon lines and $N_\psi$ external electron lines and with 
a total number of $T$ factors of $\theta$ is expected to be ultraviolet
divergent if 
\begin{align}
4 + 2T - N_A - \frac{3}{2} N_\psi \geq 0~.  
\label{PC}
\end{align}
Note that $N_\psi$ is always even. The problem is to make sense of these
meaningless integrals in a way preserving locality.

\section{One-loop Feynman graphs independent of $\theta$}

First we compute all divergent one-loop graphs built out of the single
vertex (\ref{Aff}) which is independent of $\theta$. From (\ref{PC})
we see that the problematic graphs are those with
\begin{align}
(N_\psi,N_A) \in \big\{\;(2,0)\,,~(2,1)\,,~ (0,2)\,,~(0,3)\,,~(0,4)\;
\big\}~.  
\end{align}
We employ analytic regularization in terms of a complex variable
$\varepsilon$, $|\varepsilon| \to 0$, see Appendix~\ref{appa} for details. The
advantage is that with analytic regularization we are on the safe side with
respect to the algebra of $\gamma$-matrices. Moreover, we can---in the
divergent part---arbitrarily shift the integration momentum and naively
eliminate common factors in the numerator and denominator (verified for all
integrals to evaluate).

For the electron selfenergy $(N_\psi,N_A) =(2,0)$ we obtain
\begin{align}
\Gamma^{(0,1)}_{\bar{\psi}\psi}(q,p) 
&= \quad \parbox{40mm}{\begin{picture}(40,20)
\put(0,6){\begin{fmfgraph}(40,10)
\fmfbottom{l,r}
\fmftop{t}
\fmf{fermion}{i1,l}
\fmf{fermion,tension=0.5}{i2,i1}
\fmf{fermion}{r,i2}
\fmffreeze
\fmf{photon,tension=0.5,right=0.5}{t,i1}
\fmf{phantom_arrow,tension=0.5,right=0.5}{t,i1}
\fmf{photon,tension=0.5,left=0.5}{t,i2}
\fmf{phantom_arrow,tension=0.5,left=0.5}{t,i2}
\end{fmfgraph}}
\put(35,3){\mbox{$p$}}
\put(1,15){\mbox{$-k{+}\frac{p}{2}$}}
\put(7,10){\mbox{$\mu$}}
\put(31,10){\mbox{$\nu$}}
\put(28,15){\mbox{$k{-}\frac{p}{2}$}}
\put(16,3){\mbox{$k{+}\frac{p}{2}$}}
\end{picture}}
\nonumber
\\*
&= \Gamma^{(0,0)\mu}_{A\bar{\psi}\psi}({-}k{+}\tfrac{p}{2},
{-}p,k{+}\tfrac{p}{2})
\Delta^{\bar{\psi}\psi}_{(0,0)}({-}k{-}\tfrac{p}{2},k{+}\tfrac{p}{2})
\Gamma^{(0,0)\nu}_{A\bar{\psi}\psi}(k{-}\tfrac{p}{2},{-}k{-}\tfrac{p}{2},p)
\nonumber
\\*
&\hspace*{23em} \times 
\Delta^{AA}_{(0,0)\mu\nu}({-}k{+}\tfrac{p}{2}, k{-}\tfrac{p}{2}) 
\nonumber 
\\*
&= \frac{\hbar g^2}{(4\pi)^2\varepsilon} \Big(
\frac{\alpha}{g^2} \gamma^\rho p_\rho
-\big(3{+}\frac{\alpha}{g^2}\big)m \Big) 
+ \mathcal{O}(1)
\nonumber
\\*
&=  \frac{\hbar g^2}{(4\pi)^2\varepsilon} \Big(
\frac{\alpha}{2 g^2 } N_\psi 
+ 3 m \frac{\partial}{\partial m} \Big) 
\Gamma^{(0,1)}_{\bar{\psi}\psi}(p)+ \mathcal{O}(1)~.
\label{fs0}
\end{align}

Next, the one-loop QED vertex correction $(N_\psi,N_A)=(2,1)$ is computed to
\begin{align}
\Gamma^{(0,1)\mu}_{A\bar{\psi}\psi}(p,q,r) 
&= \quad \parbox{40mm}{\begin{picture}(40,35)
\put(0,6){\begin{fmfgraph}(40,30)
\fmfright{r1,r2}
\fmfleft{l}
\fmf{fermion,tension=2}{r1,i1}
\fmf{fermion}{i1,i3,i2}
\fmf{fermion,tension=2}{i2,r2}
\fmf{photon,tension=1}{l,i3}
\fmf{phantom_arrow,tension=1}{l,i3}
\fmffreeze
\fmf{photon}{i4,i1}
\fmf{phantom_arrow}{i4,i1}
\fmf{photon}{i4,i2}
\fmf{phantom_arrow}{i4,i2}
\end{fmfgraph}}
\put(30,5){\mbox{$r$}}
\put(30,15){\mbox{$k,\sigma$}}
\put(30,27){\mbox{$-k,\rho$}}
\put(16,11){\mbox{$k{+}r$}}
\put(4,17){\mbox{$p,\mu$}}
\put(10,28){\mbox{$k{+}p{+}r$}}
\end{picture}}
\nonumber
\\
& = 
\Delta^{AA}_{(0,0)\sigma\rho}(k,{-}k) 
\Gamma^{(0,0)\rho}_{A\bar{\psi}\psi}({-}k,{-}p{-}r,k{+}p{+}r)
\Delta^{\bar{\psi}\psi}_{(0,0)}(-k{-}p{-}r,k{+}p{+}r) 
\nonumber
\\ 
& \qquad\qquad \times 
\Gamma^{(0,0)\mu}_{A\bar{\psi}\psi}(p,{-}k{-}p{-}r,k{+}r)
\Delta^{\bar{\psi}\psi}_{(0,0)}({-}k{-}r,k{+}r) 
 \Gamma^{(0,0)\sigma}_{A\bar{\psi}\psi}(k,{-}k{-}r,r)
\nonumber
\\
&= \frac{\hbar \alpha}{(4\pi)^2\varepsilon} \gamma^\mu
+ \mathcal{O}(1)
\nonumber
\\*
&= \frac{\hbar g^2}{(4\pi)^2\varepsilon} 
\Big( \frac{\alpha}{2 g^2 } N_{\psi} + 0 \,N_A \Big)
\Gamma^{(0,0)\sigma}_{A\bar{\psi}\psi}(p,q,r)+ \mathcal{O}(1)~.
\label{ffp0}
\end{align}

For the photon selfenergy $(N_\psi,N_A)=(0,2)$ we obtain
\begin{align}
\Gamma_{AA}^{(0,1)\mu\nu}(p,q) 
&=~-~\parbox{40mm}{\begin{picture}(40,30)
\put(0,0){\begin{fmfgraph}(40,30)
\fmfleft{l}
\fmfright{r}
\fmf{photon,tension=1}{l,i1}
\fmf{phantom_arrow,tension=1}{l,i1}
\fmf{photon,tension=1}{r,i2}
\fmf{phantom_arrow,tension=1}{r,i2}
\fmf{fermion,left=1}{i1,i2}
\fmf{fermion,left=1}{i2,i1}
\end{fmfgraph}}
\put(6,17){\mbox{$p,\mu$}}
\put(31,17){\mbox{$-p,\nu$}}
\put(17,4){\mbox{$k{-}\frac{p}{2}$}}
\put(17,24){\mbox{$k{+}\frac{p}{2}$}}
\end{picture}}
\nonumber
\\
&= -\mathrm{tr} \big(
\Gamma^{(0,0)\mu}_{A\bar{\psi}\psi}(p,{-}k{-}\tfrac{p}{2}, 
k{-}\tfrac{p}{2}) 
\Delta^{\bar{\psi}\psi}_{(0,0)}(-k{+}\tfrac{p}{2},k{-}\tfrac{p}{2})
\Gamma^{(0,0)\nu}_{A\bar{\psi}\psi}(-p,{-}k{+}\tfrac{p}{2}, 
k{+}\tfrac{p}{2})  
\nonumber
\\*
& \hspace*{20em} \times 
\Delta^{\bar{\psi}\psi}_{(0,0)}(-k{-}\tfrac{p}{2},
k{+}\tfrac{p}{2})\big)
\nonumber
\\
& 
= \frac{\hbar g^2}{(4\pi)^2 \varepsilon} \Big( -\frac{4}{3 g^2}  
(p^2 g^{\mu\nu} - p^\mu p^\nu) \Big) + \mathcal{O}(1)
\nonumber
\\*
&= \frac{\hbar g^2}{(4\pi)^2 \varepsilon} 
\Big( 0\, N_A -\frac{4}{3} g^2 \frac{\partial}{\partial g^2} \Big)
\Gamma^{AA}_{(0,0)}(p)+ \mathcal{O}(1)~.
\label{ps0}
\end{align}

The photon three- and four-point functions $(N_\psi,N_A)=(0,3)$ and 
$(N_\psi,N_A)=(0,4)$ are actually convergent due to gauge
invariance and its preservation in analytic regularization. For
instance, in the three-point function 
\begin{align}
\Gamma_{AAA}^{(0,1)\mu\nu\rho}(p,q,r) 
&= \left\{ ~- \hspace*{-1em} 
\parbox{40mm}{\begin{picture}(40,30)
\put(0,0){\begin{fmfgraph}(40,30)
\fmfsurround{l1,l2,l3}
\fmf{photon,tension=2}{l1,i1}
\fmf{phantom_arrow,tension=1}{l1,i1}
\fmf{photon,tension=2}{l2,i2}
\fmf{phantom_arrow,tension=1}{l2,i2}
\fmf{photon,tension=2}{l3,i3}
\fmf{phantom_arrow,tension=1}{l3,i3}
\fmf{fermion,left=0.7}{i1,i3}
\fmf{fermion,left=0.4}{i3,i2}
\fmf{fermion,left=0.7}{i2,i1}
\end{fmfgraph}}
\put(35,17){\mbox{$p,\mu$}}
\put(12,1){\mbox{$q,\nu$}}
\put(6,22){\mbox{$r,\rho$}}
\put(19,26){\mbox{$k{+}r$}}
\put(8,14){\mbox{$k$}}
\put(26,3){\mbox{$k{+}p{+}r$}}
\end{picture}} \quad \right\}
+ \Big\{ ~(q,\nu) \leftrightarrow (r,\rho)~\Big\}~
= \mathcal{O}(1) \label{ppp0}
\end{align}
the divergent contributions of both graphs cancel (Furry's theorem).

In order to absorb the divergences (\ref{fs0}), (\ref{ffp0}) and
(\ref{ps0}) the fields and parameters of the model must depend on
$\hbar$ in the following way:
\begin{subequations}
\label{hbarren}
\begin{align}
g &= \Big(1+ \frac{2}{3} \frac{\hbar g_0^2}{(4\pi)^2 \varepsilon} +
\mathcal{O}(\hbar^2) \Big) g_0 ~,
\\
m &= \Big(1- 3  \frac{\hbar g_0^2}{(4\pi)^2 \varepsilon} +
\mathcal{O}(\hbar^2) \Big) m_0 ~,
\\
\bar{\psi} &= \Big(1 - \frac{\alpha}{2 g_0^2} 
\frac{\hbar g_0^2}{(4\pi)^2 \varepsilon} +
\mathcal{O}(\hbar^2) \Big) \bar{\psi}_0 ~,
\\
\psi &= \Big(1 - \frac{\alpha}{2 g_0^2} 
\frac{\hbar g_0^2}{(4\pi)^2 \varepsilon} +
\mathcal{O}(\hbar^2) \Big) \psi_0 ~,
\\
A_\mu &= (A_\mu)_0 + \mathcal{O}(\hbar^2) ~.
\end{align}
\end{subequations}

\section{One-loop Feynman graphs linear in $\theta$}

In this section we compute the divergent one-loop Green's functions involving
a vertex linear in $\theta$.  Assuming a regularization scheme preserving
gauge invariance (such as analytic regularization), we expect at order 1 in
$\theta$ the following Hermitean counter\-terms of dimension\footnote{The
  power-counting dimensions $\text{dim}$ are $\text{dim}(d^4x)=-4$ and
  $\text{dim}(\delta(x{-}y))=4$. \label{foot}} $0$ which are purely imaginary
in momentum space:
\begin{subequations}
\label{CT}
\begin{align}
B_1 &= \int d^4x \, \theta^{\alpha\beta}
F_{\alpha\mu} F_{\beta\nu} F^{\mu\nu} ~, 
&
B_2 &= \int d^4x \, \theta^{\alpha\beta}
F_{\alpha\beta} F_{\mu\nu} F^{\mu\nu} ~,
\label{B}
\\*[1ex]
\Phi_0 &= \int d^4x\, \mathrm{i} \theta^{\alpha\beta} 
\bar{\psi} \gamma^\mu (2 F_{\alpha\mu} D_\beta + \partial_\beta 
F_{\alpha\mu}) \psi~,
\\
\Phi_1 &= 
\int d^4x\, \mathrm{i} \theta^{\alpha\beta} \bar{\psi} \gamma^\mu 
(2 F_{\alpha\beta} D_\mu + \partial_\mu F_{\alpha\beta}) \psi~,
&
\Phi_2 &= \int d^4x\, \mathrm{i} \theta^{\alpha\beta} 
\bar{\psi} \gamma_\alpha
(2 F_{\mu\beta} D^\mu + \partial^\mu F_{\mu\beta}) \psi~,
\nonumber
\\*
\Phi_3 &= \int d^4x\,\theta^{\alpha\beta} \bar{\psi} 
\gamma^{\mu\alpha\beta} (-2 D^\nu D_\nu D_\mu + 2 \mathrm{i} 
F_\mu^{~\nu} D_\nu + \mathrm{i}
\partial^\nu F_{\mu\nu}) \psi~, \hspace*{-10em} 
\nonumber
\\*
\Phi_4 &= \int d^4x\, \mathrm{i} \theta^{\alpha\beta} 
\bar{\psi} \gamma_{\mu\nu\alpha}
\partial_\beta F^{\mu\nu} \psi~,
&
\Phi_5 &= \int d^4x\,\mathrm{i} \theta^{\alpha\beta} \bar{\psi} 
\gamma_{\mu\alpha\beta}\partial_\nu F^{\nu\mu} \psi~,
\\[1ex]
\tilde{\Phi}_1 &= \int d^4x\,\mathrm{i} \theta^{\alpha\beta} 
(\bar{\psi} \gamma^\mu \psi)
(\bar{\psi} \gamma_{\mu\alpha\beta} \psi)~,
\\[1ex]
M_1 &= \int d^4 x \, \mathrm{i} \theta^{\alpha\beta} m^3 \bar{\psi} 
\gamma_{\alpha\beta} \psi~,
&
M_2 &= \int d^4 x \, \theta^{\alpha\beta} m^2 \bar{\psi} 
\gamma_{\mu\alpha\beta} D^\mu \psi~,
\nonumber
\\*
M_3 &= \int d^4 x \, \mathrm{i} \theta^{\alpha\beta} m \bar{\psi} 
\gamma^\mu_{~\alpha} (2 D_\mu D_\beta + \mathrm{i} F_{\mu\beta}) 
\psi~,
&
M_4 &= \int d^4 x \, \mathrm{i} \theta^{\alpha\beta} m \bar{\psi} 
\gamma_{\alpha\beta} D^\nu D_\nu \psi~,
\nonumber
\\*
M_5 &= \int d^4 x \, \theta^{\alpha\beta} m \bar{\psi} 
\gamma_{\mu\nu\alpha\beta} F^{\mu\nu} \psi~,
&
M_6 &= \int d^4 x \, \theta^{\alpha\beta} m \bar{\psi} 
F_{\alpha\beta} \psi~.
\label{M}
\end{align}
\end{subequations}
Naive comparison of (\ref{th-exp}) with (\ref{CT}) shows immediately that the
noncommutative Yang-Mills-Dirac action (\ref{YMD}) expanded to first order in
$\theta$ cannot expected to be renormalizable. Indeed, for the absorption of
the five divergences corresponding to $\Phi_1\dots\Phi_5$ we only have the
field redefinition parameters $\kappa_1,\kappa_2,\kappa_3,\kappa_4$ at
disposal, because $\kappa_9$ (if included) is already fixed by the divergence
corresponding to $\tilde{\Phi}_1$. Even if we allow for a renormalization of
$\theta$ to deal with $\Phi_0$, there is at least one missing parameter to
achieve renormalizability in the massless case without a symmetry. The massive
case is worse: there are only three parameters $\kappa_6,\kappa_7,\kappa_8$ to
absorb the six divergences corresponding to $M_1\dots M_6$. Finally, for the
bosonic divergences corresponding to $B_1,B_2$ there is no parameter at all
available (for $\lambda_1=\lambda_2=0$ in (\ref{YM-ext})). Thus, unless there
are symmetries, $\theta$-expanded noncommutative QED cannot be
renormalizable\footnote{This refers to renormalizability of the
  $\theta$-expansion of (\ref{YMD}). One can always add to (\ref{YMD})
  on unexpanded level 
  $\theta$-dependent terms as in (\ref{YM-ext}) in order to compensate
  the divergences of the type (\ref{CT}). However, such a
  theory looses all predictivity and hence would be regarded as
  `non-renormalizable'.}. 

For simplicity we choose from now on the
Feynman gauge $\alpha=g_0^2=g^2+\mathcal{O}(\hbar )$. 

\subsection{Two electrons, no photon}
 
We begin with the electron selfenergy, allowing for a renormalization
of $\theta$:
\begin{align}
&\Gamma^{(1,1)}_{\bar{\psi}\psi}(q,p) 
\nonumber
\\*
& = ~
\parbox{40mm}{\begin{picture}(40,20)
\put(0,6){\begin{fmfgraph}(40,10)
\fmfkeep{FF1}
\fmfbottom{l,r}
\fmftop{t}
\fmf{fermion}{i1,l}
\fmf{fermion,tension=0.5}{i2,i1}
\fmf{fermion}{r,i2}
\fmffreeze
\fmf{photon,tension=0.5,right=0.5}{t,i1}
\fmf{phantom_arrow,tension=0.5,right=0.5}{t,i1}
\fmf{photon,tension=0.5,left=0.5}{t,i2}
\fmf{phantom_arrow,tension=0.5,left=0.5}{t,i2}
\fmfdot{i1}
\end{fmfgraph}}
\put(35,3){\mbox{$p$}}
\put(7,15){\mbox{$-k$}}
\put(7,10){\mbox{$\mu$}}
\put(31,10){\mbox{$\nu$}}
\put(28,15){\mbox{$k$}}
\put(14,3){\mbox{$k{+}p$}}
\end{picture}}
{}~ +~
\parbox{40mm}{\begin{picture}(40,20)
\put(0,6){\begin{fmfgraph}(40,10)
\fmfkeep{FF2}
\fmfbottom{l,r}
\fmftop{t}
\fmf{fermion}{i1,l}
\fmf{fermion,tension=0.5}{i2,i1}
\fmf{fermion}{r,i2}
\fmffreeze
\fmf{photon,tension=0.5,right=0.5}{t,i1}
\fmf{phantom_arrow,tension=0.5,right=0.5}{t,i1}
\fmf{photon,tension=0.5,left=0.5}{t,i2}
\fmf{phantom_arrow,tension=0.5,left=0.5}{t,i2}
\fmfdot{i2}
\end{fmfgraph}}
\put(35,3){\mbox{$p$}}
\put(7,15){\mbox{$-k$}}
\put(7,10){\mbox{$\mu$}}
\put(31,10){\mbox{$\nu$}}
\put(28,15){\mbox{$k$}}
\put(14,3){\mbox{$k{+}p$}}
\end{picture}}
{}~+~ 
\parbox{40mm}{\begin{picture}(40,20)
\put(0,6){\begin{fmfgraph}(40,10)
\fmfkeep{FF3}
\fmfbottom{l,r}
\fmftop{t}
\fmf{fermion}{i1,l}
\fmf{fermion,tension=1}{i2,i3,i1}
\fmf{fermion}{r,i2}
\fmffreeze
\fmf{photon,tension=0.5,right=0.5}{t,i1}
\fmf{phantom_arrow,tension=0.5,right=0.5}{t,i1}
\fmf{photon,tension=0.5,left=0.5}{t,i2}
\fmf{phantom_arrow,tension=0.5,left=0.5}{t,i2}
\fmfdot{i3}
\end{fmfgraph}}
\put(35,3){\mbox{$p$}}
\put(7,15){\mbox{$-k$}}
\put(7,10){\mbox{$\mu$}}
\put(31,10){\mbox{$\nu$}}
\put(28,15){\mbox{$k$}}
\put(15,3){\mbox{$k{+}p$}}
\end{picture}}
\nonumber
\\
& +~
\parbox{40mm}{\begin{picture}(40,20)
\put(0,3){\begin{fmfgraph}(40,15)
\fmfkeep{FF4}
\fmfbottom{l,r}
\fmftop{t}
\fmf{fermion}{i1,l}
\fmf{dots,tension=3.5}{i1,i2}
\fmf{fermion}{r,i2}
\fmffreeze
\fmf{fermion,tension=0.5,right=0.8}{i2,t}
\fmf{fermion,tension=0.5,right=0.8}{t,i1}
\fmfdot{i1}
\end{fmfgraph}}
\put(35,2){\mbox{$p$}}
\put(28,13){\mbox{$k$}}
\end{picture}}
{}~+~
\parbox{40mm}{\begin{picture}(40,20)
\put(0,3){\begin{fmfgraph}(40,15)
\fmfkeep{FF5}
\fmfbottom{l,r}
\fmftop{t}
\fmf{fermion}{i1,l}
\fmf{dots,tension=3.5}{i1,i2}
\fmf{fermion}{r,i2}
\fmffreeze
\fmf{fermion,tension=0.5,right=0.8}{i2,t}
\fmf{fermion,tension=0.5,right=0.8}{t,i1}
\fmfdot{i2}
\end{fmfgraph}}
\put(35,2){\mbox{$p$}}
\put(28,13){\mbox{$k$}}
\end{picture}}
\nonumber
\\
& + \dfrac{1}{2} ~
\parbox{40mm}{\begin{picture}(40,20)
\put(0,6){\begin{fmfgraph}(40,15)
\fmfkeep{FF6}
\fmfbottom{l,r}
\fmftop{t}
\fmf{fermion}{i,l}
\fmf{fermion}{r,i}
\fmffreeze
\fmf{photon,tension=0.5,right=0.8}{t,i}
\fmf{phantom_arrow,tension=0.5,right=0.8}{t,i}
\fmf{photon,tension=0.5,left=0.8}{t,i}
\fmf{phantom_arrow,tension=0.5,left=0.8}{t,i}
\fmfdot{i}
\end{fmfgraph}}
\put(35,3){\mbox{$p$}}
\put(10,15){\mbox{$k$}}
\put(12,10){\mbox{$\mu$}}
\put(26,10){\mbox{$\nu$}}
\put(28,15){\mbox{$-k$}}
\end{picture}}
{}~-~
\parbox{40mm}{\begin{picture}(40,20)
\put(0,3){\begin{fmfgraph}(40,15)
\fmfkeep{FF7}
\fmfbottom{l,r}
\fmftop{t}
\fmf{fermion}{i1,l}
\fmf{fermion}{r,i1}
\fmffreeze
\fmf{dots,tension=3}{i1,i2}
\fmf{fermion,tension=0.5,right=1}{i2,t}
\fmf{fermion,tension=0.5,right=1}{t,i2}
\fmfdot{i2}
\end{fmfgraph}}
\put(35,2){\mbox{$p$}}
\put(27,13){\mbox{$k$}}
\end{picture}}
{}~-~
\parbox{40mm}{\begin{picture}(40,20)
\put(0,3){\begin{fmfgraph}(40,15)
\fmfkeep{FF8}
\fmfbottom{l,r}
\fmftop{t}
\fmf{fermion}{i1,l}
\fmf{fermion}{r,i1}
\fmffreeze
\fmf{dots,tension=3}{i1,i2}
\fmf{fermion,tension=0.5,right=1}{i2,t}
\fmf{fermion,tension=0.5,right=1}{t,i2}
\fmfdot{i1}
\end{fmfgraph}}
\put(35,2){\mbox{$p$}}
\put(27,13){\mbox{$k$}}
\end{picture}}
\nonumber
\\
& = \frac{\hbar g^2}{(4\pi)^2\varepsilon} 
\Big(\frac{1}{2} N_\psi + 0\, N_A + 3 m \frac{\partial}{\partial m} 
+ \tau \theta^{\alpha\beta} \frac{\partial}{\partial
  \theta^{\alpha\beta}}
\nonumber
\\
& \qquad 
+ \Big(\frac{1}{48} - \frac{1}{6} \kappa_1 + \frac{1}{2} \kappa_2 +
\frac{2}{3} \kappa_3 + (-\tau - \frac{1}{2}) \kappa_4 
+ \frac{1}{3} \kappa_9 \Big) \frac{\partial}{\partial \kappa_4}
\nonumber
\\
& \qquad 
+ \Big(\frac{1}{6} + \frac{2}{3} \kappa_1 - \kappa_2 -
\frac{2}{3} \kappa_3 - 3 \kappa_4 + (-\tau - 3) \kappa_6 + 2 \kappa_7
- \frac{1}{3} \kappa_9 
\Big) \frac{\partial}{\partial \kappa_6}
\nonumber
\\
& \qquad 
+ \Big(- \frac{1}{2} \kappa_2 - \kappa_3 - \frac{3}{2} \kappa_4 
+ \frac{1}{2} \kappa_6 + (-\tau - 3) \kappa_7 + \frac{1}{2} \kappa_9 
\Big) \frac{\partial}{\partial \kappa_7}
\nonumber
\\
& \qquad 
+ \Big(\frac{1}{16} - \frac{1}{2} \kappa_1 + \kappa_2 +
\kappa_3 + \kappa_4 + \kappa_6 + 2 \kappa_7 + (-\tau - 6) \kappa_8
+ \frac{1}{2} \kappa_9 \Big) \frac{\partial}{\partial \kappa_8}\Big) 
\Big(\Gamma^{(1,0)}_{\bar{\psi}\psi}(p)\Big) 
\nonumber
\\
& +\mathcal{O}(1)\,.
\label{fs1}
\end{align}
This is indeed a renormalization of $\kappa_4,\kappa_6,\kappa_7,
\kappa_8$! 

\subsection{Four electrons, no photon}

Next we consider one-loop corrections to the electron four-point
function $\Gamma_{\bar{\psi}\psi;\bar{\psi}\psi}(p,q;r,s)$ in first order in
$\theta$. Since 
$\Gamma_{\bar{\psi}\psi;\bar{\psi}\psi}(p,q;r,s)$ is independent of
external momenta and no further label distinguishes the two electron
pairs in $\Gamma_{\bar{\psi}\psi;\bar{\psi}\psi}(p,q;r,s)$, we can
consider $\Gamma_{\bar{\psi}\psi;\bar{\psi}\psi}(p,q;r,s)$ in the
asymmetric form where the electrons labelled by $p,q$ attach to
$\gamma^\mu$ and those labelled by $r,s$ attach to
$\gamma_{\mu\alpha\beta}$ in (\ref{4f}). This reduces the number of
graphs to compute: we simply choose the external momenta of the
divergent part of the one-loop graphs according to this
prescription. All but the last two graphs below can then be arranged
in pairs which contain the following concatenation of Feynman rules:
\[
\gamma^\rho (-\gamma^\tau (s_\tau{+}k_\tau)-m) \gamma^\sigma 
+ \gamma^\sigma (-\gamma^\tau (-r_\tau{-}k_\tau)-m) \gamma^\rho
= 2 \gamma^{\rho\sigma\tau} k_\tau + \text{terms independent of $k$}.
\]
Thus, these subgraphs correspond to the external momenta
$(r,s)$. We therefore have
\begin{align}
&\Gamma^{(1,1)}_{\bar{\psi}\psi;\bar{\psi}\psi} (p,q;r,s) 
\nonumber
\\
& = ~
\parbox{50mm}{\begin{picture}(50,28)
\put(10,1){\begin{fmfgraph}(40,25)
\fmfright{r1,r2}
\fmfleft{l1,l2}
\fmf{fermion,tension=2}{r1,i1}
\fmf{fermion,tension=2}{r2,i2}
\fmf{fermion,tension=1}{i1,i3}
\fmf{fermion,tension=1}{i2,i4}
\fmf{fermion,tension=2}{i3,l1}
\fmf{fermion,tension=2}{i4,l2}
\fmf{photon,tension=1}{i5,i1}
\fmf{phantom_arrow,tension=0}{i5,i1}
\fmf{photon,tension=1}{i5,i2}
\fmf{phantom_arrow,tension=0}{i5,i2}
\fmf{photon,tension=1}{i6,i3}
\fmf{phantom_arrow,tension=0}{i6,i3}
\fmf{photon,tension=1}{i6,i4}
\fmf{phantom_arrow,tension=0}{i6,i4}
\fmfdot{i2}
\end{fmfgraph}}
\put(18,0){\mbox{$-r$}}
\put(18,24){\mbox{$-p$}}
\put(39,0){\mbox{$s$}}
\put(38,24){\mbox{$q$}}
\put(26,1){\mbox{$s{+}k$}}
\put(26,23){\mbox{$q{-}k$}}
\put(40,17){\mbox{$-k,\nu$}}
\put(40,9){\mbox{$k,\sigma$}}
\put(3,17){\mbox{$k{+}r{+}s,\mu$}}
\put(0,9){\mbox{${-}k{-}r{-}s,\rho$}}
\end{picture}}
{}~+~ 
\parbox{50mm}{\begin{picture}(40,28)
\put(10,1){\begin{fmfgraph}(40,25)
\fmfright{r1,r2}
\fmfleft{l1,l2}
\fmf{fermion,tension=3}{r1,i1}
\fmf{fermion,tension=3}{r2,i2}
\fmf{fermion,tension=1}{i1,i3}
\fmf{fermion,tension=1}{i2,i4}
\fmf{fermion,tension=3}{i3,l1}
\fmf{fermion,tension=3}{i4,l2}
\fmf{photon,tension=0.5}{i4,i1}
\fmf{phantom,tension=0.5}{i2,i3}
\fmffreeze
\fmf{photon}{i2,i7}
\fmf{photon,tension=2,rubout=4}{i7,i8}
\fmf{photon}{i8,i3}
\fmf{phantom_arrow}{i5,i1}
\fmf{phantom_arrow}{i5,i4}
\fmf{phantom_arrow}{i6,i3}
\fmf{phantom_arrow}{i6,i2}
\fmfdot{i2}
\end{fmfgraph}}
\put(18,0){\mbox{$-r$}}
\put(18,25){\mbox{$-p$}}
\put(39,0){\mbox{$s$}}
\put(39,25){\mbox{$q$}}
\put(25,0.5){\mbox{$-r{-}k$}}
\put(26,24){\mbox{$q{-}k$}}
\put(36,8){\mbox{${-}k{-}r{-s},\rho$}}
\put(36,17){\mbox{$-k,\nu$}}
\put(7,17){\mbox{$k{+}r{+}s,\mu$}}
\put(18,8){\mbox{$k,\sigma$}} 
\end{picture}}
\nonumber
\\[1ex]
& +~{} 
\parbox{50mm}{\begin{picture}(50,28)
\put(10,1){\begin{fmfgraph}(40,25)
\fmfright{r1,r2}
\fmfleft{l1,l2}
\fmf{fermion,tension=2}{r1,i1}
\fmf{fermion,tension=2}{r2,i2}
\fmf{fermion,tension=1}{i1,i3}
\fmf{fermion,tension=1}{i2,i4}
\fmf{fermion,tension=2}{i3,l1}
\fmf{fermion,tension=2}{i4,l2}
\fmf{photon,tension=1}{i5,i1}
\fmf{phantom_arrow,tension=0}{i5,i1}
\fmf{photon,tension=1}{i5,i2}
\fmf{phantom_arrow,tension=0}{i5,i2}
\fmf{photon,tension=1}{i6,i3}
\fmf{phantom_arrow,tension=0}{i6,i3}
\fmf{photon,tension=1}{i6,i4}
\fmf{phantom_arrow,tension=0}{i6,i4}
\fmfdot{i4}
\end{fmfgraph}}
\put(18,0){\mbox{$-r$}}
\put(18,24){\mbox{$-p$}}
\put(39,0){\mbox{$s$}}
\put(38,24){\mbox{$q$}}
\put(26,1){\mbox{$s{+}k$}}
\put(26,23){\mbox{$q{-}k$}}
\put(40,17){\mbox{$-k,\nu$}}
\put(40,9){\mbox{$k,\sigma$}}
\put(3,17){\mbox{$k{+}r{+}s,\mu$}}
\put(0,9){\mbox{${-}k{-}r{-}s,\rho$}}
\end{picture}}
{}~+~ 
\parbox{50mm}{\begin{picture}(40,28)
\put(10,1){\begin{fmfgraph}(40,25)
\fmfright{r1,r2}
\fmfleft{l1,l2}
\fmf{fermion,tension=3}{r1,i1}
\fmf{fermion,tension=3}{r2,i2}
\fmf{fermion,tension=1}{i1,i3}
\fmf{fermion,tension=1}{i2,i4}
\fmf{fermion,tension=3}{i3,l1}
\fmf{fermion,tension=3}{i4,l2}
\fmf{photon,tension=0.5}{i4,i1}
\fmf{phantom,tension=0.5}{i2,i3}
\fmffreeze
\fmf{photon}{i2,i7}
\fmf{photon,tension=2,rubout=4}{i7,i8}
\fmf{photon}{i8,i3}
\fmf{phantom_arrow}{i5,i1}
\fmf{phantom_arrow}{i5,i4}
\fmf{phantom_arrow}{i6,i3}
\fmf{phantom_arrow}{i6,i2}
\fmfdot{i4}
\end{fmfgraph}}
\put(18,0){\mbox{$-r$}}
\put(18,25){\mbox{$-p$}}
\put(39,0){\mbox{$s$}}
\put(39,25){\mbox{$q$}}
\put(25,0.5){\mbox{$-r{-}k$}}
\put(26,24){\mbox{$q{-}k$}}
\put(36,8){\mbox{${-}k{-}r{-s},\rho$}}
\put(36,17){\mbox{$-k,\nu$}}
\put(7,17){\mbox{$k{+}r{+}s,\mu$}}
\put(18,8){\mbox{$k,\sigma$}} 
\end{picture}}
\nonumber
\\[1ex]
& +~{} 
\parbox{50mm}{\begin{picture}(50,28)
\put(10,1){\begin{fmfgraph}(40,25)
\fmfright{r1,r2}
\fmfleft{l1,l2}
\fmf{fermion,tension=2}{r1,i1}
\fmf{fermion,tension=2}{r2,i2}
\fmf{fermion,tension=1}{i1,i3}
\fmf{phantom,tension=1}{i2,i4}
\fmf{fermion,tension=2}{i3,l1}
\fmf{fermion,tension=2}{i4,l2}
\fmf{photon,tension=1}{i5,i1}
\fmf{phantom_arrow,tension=0}{i5,i1}
\fmf{photon,tension=1}{i5,i2}
\fmf{phantom_arrow,tension=0}{i5,i2}
\fmf{photon,tension=1}{i6,i3}
\fmf{phantom_arrow,tension=0}{i6,i3}
\fmf{photon,tension=1}{i6,i4}
\fmf{phantom_arrow,tension=0}{i6,i4}
\fmffreeze
\fmf{fermion}{i2,i7,i4}
\fmfdot{i7}
\end{fmfgraph}}
\put(18,0){\mbox{$-r$}}
\put(18,24){\mbox{$-p$}}
\put(39,0){\mbox{$s$}}
\put(38,24){\mbox{$q$}}
\put(26,1){\mbox{$s{+}k$}}
\put(26,23.5){\mbox{$q{-}k$}}
\put(40,17){\mbox{$-k,\nu$}}
\put(40,9){\mbox{$k,\sigma$}}
\put(3,17){\mbox{$k{+}r{+}s,\mu$}}
\put(0,9){\mbox{${-}k{-}r{-}s,\rho$}}
\end{picture}}
{}~+~ 
\parbox{50mm}{\begin{picture}(40,28)
\put(10,1){\begin{fmfgraph}(40,25)
\fmfright{r1,r2}
\fmfleft{l1,l2}
\fmf{fermion,tension=3}{r1,i1}
\fmf{fermion,tension=3}{r2,i2}
\fmf{fermion,tension=1}{i1,i3}
\fmf{phantom,tension=1}{i2,i4}
\fmf{fermion,tension=3}{i3,l1}
\fmf{fermion,tension=3}{i4,l2}
\fmf{photon,tension=0.5}{i4,i1}
\fmf{phantom,tension=0.5}{i2,i3}
\fmffreeze
\fmf{photon}{i2,i7}
\fmf{photon,tension=2,rubout=4}{i7,i8}
\fmf{photon}{i8,i3}
\fmf{phantom_arrow}{i5,i1}
\fmf{phantom_arrow}{i5,i4}
\fmf{phantom_arrow}{i6,i3}
\fmf{phantom_arrow}{i6,i2}
\fmffreeze
\fmf{fermion}{i2,i9,i4}
\fmfdot{i9}
\end{fmfgraph}}
\put(18,0){\mbox{$-r$}}
\put(18,25){\mbox{$-p$}}
\put(39,0){\mbox{$s$}}
\put(39,25){\mbox{$q$}}
\put(25,0.5){\mbox{$-r{-}k$}}
\put(26,24.5){\mbox{$q{-}k$}}
\put(36,8){\mbox{${-}k{-}r{-s},\rho$}}
\put(36,17){\mbox{$-k,\nu$}}
\put(7,17){\mbox{$k{+}r{+}s,\mu$}}
\put(18,8){\mbox{$k,\sigma$}} 
\end{picture}}
\nonumber
\\[1ex]
& +~ \dfrac{1}{2}~{} 
\parbox{50mm}{\begin{picture}(40,28)
\put(10,0){\begin{fmfgraph}(40,25)
\fmfright{r1,r2}
\fmfleft{l1,l2}
\fmf{fermion,tension=3}{r2,i1}
\fmf{fermion,tension=3}{r1,i2}
\fmf{fermion,tension=1}{i2,i3}
\fmf{fermion,tension=3}{i1,l2}
\fmf{fermion,tension=3}{i3,l1}
\fmf{photon,tension=0.5}{i4,i1}
\fmf{phantom_arrow,tension=0.5}{i4,i1}
\fmf{photon,tension=0.5}{i4,i2}
\fmf{phantom_arrow,tension=0.5}{i4,i2}
\fmf{photon,tension=0.5}{i5,i1}
\fmf{phantom_arrow,tension=0.5}{i5,i1}
\fmf{photon,tension=0.5}{i5,i3}
\fmf{phantom_arrow,tension=0.5}{i5,i3}
\fmfdot{i1}
\end{fmfgraph}}
\put(13,3){\mbox{$-r$}}
\put(13,21){\mbox{$-p$}}
\put(42,3){\mbox{$s$}}
\put(42,21){\mbox{$q$}}
\put(34,16){\mbox{$-k,\nu$}}
\put(9,16){\mbox{$k{+}r{+}s,\mu$}}
\put(38,8){\mbox{$k,\sigma$}}
\put(2,8){\mbox{${-}k{-}r{-}s,\rho$}}
\put(26,0){\mbox{$s{+}k$}}
\end{picture}}
{}~+ \dfrac{1}{2} 
\parbox{40mm}{\begin{picture}(40,28)
\put(0,0){\begin{fmfgraph}(40,25)
\fmfright{r1,r2}
\fmfleft{l1,l2}
\fmf{fermion,tension=3}{r2,i1}
\fmf{fermion,tension=3}{r1,i2}
\fmf{fermion,tension=1}{i2,i3}
\fmf{fermion,tension=3}{i1,l2}
\fmf{fermion,tension=3}{i3,l1}
\fmf{photon,tension=0.5}{i4,i1}
\fmf{phantom_arrow,tension=0.5}{i4,i1}
\fmf{photon,tension=0.5}{i4,i2}
\fmf{phantom_arrow,tension=0.5}{i4,i2}
\fmf{photon,tension=0.5}{i5,i1}
\fmf{phantom_arrow,tension=0.5}{i5,i1}
\fmf{photon,tension=0.5}{i5,i3}
\fmf{phantom_arrow,tension=0.5}{i5,i3}
\fmfdot{i1}
\end{fmfgraph}}
\put(3,3){\mbox{$-r$}}
\put(3,21){\mbox{$-p$}}
\put(32,3){\mbox{$s$}}
\put(32,21){\mbox{$q$}}
\put(24,16){\mbox{$k{+}r{+}s,\mu$}}
\put(5,16){\mbox{${-}k,\nu$}}
\put(28,8){\mbox{${-}k{-}r{-}s,\rho$}}
\put(5,8){\mbox{$k,\sigma$}}
\put(14,0){\mbox{$-r{-}k$}}
\end{picture}}
\nonumber
\\[1ex]
& ~+ ~
\parbox{40mm}{\begin{picture}(40,28)
\put(0,0){\begin{fmfgraph}(40,25)
\fmfright{r1,r2}
\fmfleft{l1,l2}
\fmf{phantom}{r1,i1,l1}
\fmf{phantom}{r2,i2,l2}
\fmf{dots,tension=5}{i1,i2}
\fmffreeze
\fmf{fermion,tension=1}{r2,i3}
\fmf{fermion,tension=1}{i3,i2}
\fmf{fermion,tension=1}{r1,i4}
\fmf{fermion,tension=1}{i4,i1}
\fmf{fermion,tension=1}{i1,l1}
\fmf{fermion,tension=1}{i2,l2}
\fmffreeze
\fmfforce{((0.8w,0.5h))}{i5}
\fmf{photon,right=0.5}{i5,i3}
\fmf{phantom_arrow,right=0.5}{i5,i3}
\fmf{photon,left=0.5}{i5,i4}
\fmf{phantom_arrow,left=0.5}{i5,i4}
\fmfdot{i2}
\end{fmfgraph}}
\put(3,3){\mbox{$-r$}}
\put(2,20){\mbox{$-p$}}
\put(35,2){\mbox{$s$}}
\put(34,21){\mbox{$q$}}
\put(19,2){\mbox{$s{+}k$}}
\put(19,20){\mbox{$q{-}k$}}
\put(33,15){\mbox{$-k$}}
\put(34,7){\mbox{$k$}}
\end{picture}}
{} ~+ ~
\parbox{40mm}{\begin{picture}(40,28)
\put(0,0){\begin{fmfgraph}(40,25)
\fmfright{r1,r2}
\fmfleft{l1,l2}
\fmf{phantom}{r1,i1,l1}
\fmf{phantom}{r2,i2,l2}
\fmf{dots,tension=5}{i1,i2}
\fmffreeze
\fmf{fermion,tension=1}{r2,i3}
\fmf{fermion,tension=1}{i3,i2}
\fmf{fermion,tension=1}{i1,i4}
\fmf{fermion,tension=1}{i4,l1}
\fmf{fermion,tension=1}{r1,i1}
\fmf{fermion,tension=1}{i2,l2}
\fmffreeze
\fmfforce{((0.3w,0.5h))}{i5}
\fmfforce{((0.5w,0.85h))}{i6}
\fmf{photon,left=0.3}{i6,i3}
\fmf{phantom_arrow,left=0.3}{i6,i3}
\fmf{photon,left=0.4,rubout=4}{i5,i6}
\fmf{photon,right=0.5}{i5,i4}
\fmf{phantom_arrow,right=0.5}{i5,i4}
\fmfdot{i2}
\end{fmfgraph}}
\put(2,3){\mbox{$-r$}}
\put(2,20){\mbox{$-p$}}
\put(32,3){\mbox{$s$}}
\put(33,20){\mbox{$q$}}
\put(25,15){\mbox{$q{-}k$}}
\put(12.5,3){\mbox{$-r{-}k$}}
\put(7,10){\mbox{$k$}}
\put(22,23){\mbox{$-k$}}
\end{picture}}
\nonumber
\\[1ex]
& ~+ ~
\parbox{40mm}{\begin{picture}(40,28)
\put(0,0){\begin{fmfgraph}(40,25)
\fmfright{r1,r2}
\fmfleft{l1,l2}
\fmf{phantom}{r1,i1,l1}
\fmf{phantom}{r2,i2,l2}
\fmf{dots,tension=5}{i1,i2}
\fmffreeze
\fmf{fermion,tension=1}{i3,l2}
\fmf{fermion,tension=1}{i2,i3}
\fmf{fermion,tension=1}{i4,l1}
\fmf{fermion,tension=1}{i1,i4}
\fmf{fermion,tension=1}{r1,i1}
\fmf{fermion,tension=1}{r2,i2}
\fmffreeze
\fmfforce{((0.2w,0.5h))}{i5}
\fmf{photon,left=0.5}{i5,i3}
\fmf{phantom_arrow,left=0.5}{i5,i3}
\fmf{photon,right=0.5}{i5,i4}
\fmf{phantom_arrow,right=0.5}{i5,i4}
\fmfdot{i2}
\end{fmfgraph}}
\put(1,2){\mbox{$-r$}}
\put(1,21){\mbox{$-p$}}
\put(33,3){\mbox{$s$}}
\put(33,20){\mbox{$q$}}
\put(14,4){\mbox{$s{+}k$}}
\put(14,20){\mbox{$q{-}k$}}
\put(1,15){\mbox{$-k$}}
\put(4,7){\mbox{$k$}}
\end{picture}}
{} ~+ ~
\parbox{40mm}{\begin{picture}(40,28)
\put(0,0){\begin{fmfgraph}(40,25)
\fmfright{r1,r2}
\fmfleft{l1,l2}
\fmf{phantom}{r1,i1,l1}
\fmf{phantom}{r2,i2,l2}
\fmf{dots,tension=5}{i1,i2}
\fmffreeze
\fmf{fermion,tension=1}{i3,l2}
\fmf{fermion,tension=1}{i2,i3}
\fmf{fermion,tension=1}{i4,i1}
\fmf{fermion,tension=1}{r1,i4}
\fmf{fermion,tension=1}{i1,l1}
\fmf{fermion,tension=1}{r2,i2}
\fmffreeze
\fmfforce{((0.7w,0.5h))}{i5}
\fmfforce{((0.5w,0.85h))}{i6}
\fmf{photon,right=0.3}{i6,i3}
\fmf{phantom_arrow,right=0.3}{i6,i3}
\fmf{photon,right=0.4,rubout=4}{i5,i6}
\fmf{photon,left=0.5}{i5,i4}
\fmf{phantom_arrow,left=0.5}{i5,i4}
\fmfdot{i2}
\end{fmfgraph}}
\put(2,3){\mbox{$-r$}}
\put(2,20){\mbox{$-p$}}
\put(32,4){\mbox{$s$}}
\put(33,20){\mbox{$q$}}
\put(7,15){\mbox{$q{-}k$}}
\put(15,3.5){\mbox{$-r{-}k$}}
\put(31,10){\mbox{$k$}}
\put(12,24){\mbox{$-k$}}
\end{picture}}
\nonumber
\\[1ex]
&
{}~+ 
\parbox{40mm}{\begin{picture}(40,28)
\put(0,0){\begin{fmfgraph}(40,25)
\fmfright{r1,r2}
\fmfleft{l1,l2}
\fmf{fermion,tension=1}{r2,i3}
\fmf{fermion,tension=1}{i3,i1}
\fmf{fermion}{r1,i2}
\fmf{fermion,tension=1}{i1,i4}
\fmf{fermion,tension=1}{i4,l2}
\fmf{fermion}{i2,l1}
\fmf{dots,tension=4}{i1,i2}
\fmffreeze
\fmfforce{((0.5w,0.95h))}{i5}
\fmf{photon,left=0.5}{i5,i3}
\fmf{phantom_arrow,left=0.5}{i5,i3}
\fmf{photon,right=0.5}{i5,i4}
\fmf{phantom_arrow,right=0.5}{i5,i4}
\fmfdot{i2}
\end{fmfgraph}}
\put(3,3){\mbox{$-r$}}
\put(2,19){\mbox{$-p$}}
\put(32,3){\mbox{$s$}}
\put(33,19){\mbox{$q$}}
\put(23,11){\mbox{$q{+}k$}}
\put(5,11){\mbox{$-p{+}k$}}
\put(9,24){\mbox{$-k$}}
\put(26,24){\mbox{$k$}}
\end{picture}}
{}~+ ~
\parbox{40mm}{\begin{picture}(40,28)
\put(0,0){\begin{fmfgraph}(40,25)
\fmfright{r1,r2}
\fmfleft{l1,l2}
\fmf{fermion,tension=1}{r1,i3}
\fmf{fermion,tension=1}{i3,i1}
\fmf{fermion}{r2,i2}
\fmf{fermion,tension=1}{i1,i4}
\fmf{fermion,tension=1}{i4,l1}
\fmf{fermion}{i2,l2}
\fmf{dots,tension=4}{i1,i2}
\fmffreeze
\fmfforce{((0.5w,0.05h))}{i5}
\fmf{photon,right=0.5}{i5,i3}
\fmf{phantom_arrow,right=0.5}{i5,i3}
\fmf{photon,left=0.5}{i5,i4}
\fmf{phantom_arrow,left=0.5}{i5,i4}
\fmfdot{i1}
\end{fmfgraph}}
\put(3,5){\mbox{$-r$}}
\put(2,21){\mbox{$-p$}}
\put(32,5){\mbox{$s$}}
\put(33,21){\mbox{$q$}}
\put(26,11){\mbox{$s{+}k$}}
\put(4,11){\mbox{$-r{+}k$}}
\put(8,0){\mbox{$-k$}}
\put(27,0){\mbox{$k$}}
\end{picture}}
\nonumber
\\[1ex]
& 
= \frac{\hbar g^2}{(4\pi)^2 \varepsilon} \Big(
\frac{1}{2} N_\psi 
- \frac{4}{3} g^2 \frac{\partial}{\partial g^2} 
+\tau \theta^{\alpha\beta} \frac{\partial}{\partial \theta^{\alpha\beta}} 
+ \Big( \frac{3}{4} + (-\tau + \frac{4}{3}) \kappa_9\Big)
\frac{\partial}{\partial \kappa_9} \Big) 
\Gamma^{(1,0)}_{\bar{\psi}\psi;\bar{\psi}\psi}(p,q;r,s)
\nonumber
\\
&+\mathcal{O}(1)~.
\label{ffff1}
\end{align}
The divergence which is expressed by the factor $\frac{3}{4}$ in (\ref{ffff1})
is problematic. We had mentioned in the discussion around (\ref{A'}) that the
mathematical framework of Section~\ref{Sec2} imposes $\kappa_9=0$ to all
orders in $\hbar$. But this contradicts (\ref{ffff1}) which enforces an
$\hbar$-renormalization of $\kappa_9$. That was the reason why we have
included $\kappa_9$. Nevertheless it will turn out that the inclusion of
$\kappa_9$ does not help. In order to make this transparent we introduce a
switch $\zeta$ which according to (\ref{ffff1}) should take the value
$\zeta{=}1$ and for mathematical reasons the value $\zeta{=}0$ (leaving us
with an unrenormalizable divergence in the electron four-point function). We
therefore write
\begin{align}
\Gamma^{(1,1)}_{\bar{\psi}\psi;\bar{\psi}\psi}(p,q;r,s) 
& = \frac{\hbar g^2}{(4\pi)^2 \varepsilon} \Big(
\frac{1}{2} N_\psi 
- \frac{4}{3} g^2 \frac{\partial}{\partial g^2} 
+\tau \theta^{\alpha\beta} \frac{\partial}{\partial \theta^{\alpha\beta}} 
\nonumber
\\*
& \qquad 
+ \Big( \frac{3}{4} \zeta + (-\tau + \frac{4}{3}) \kappa_9\Big)
\frac{\partial}{\partial \kappa_9} \Big) 
\Gamma^{(1,0)}_{\bar{\psi}\psi;\bar{\psi}\psi}(p,q;r,s) 
+ \mathcal{O}(1)~.
\tag{\oldref{ffff1}'}
\end{align}

\subsection{Two electrons, one photon}

Now we turn to the computation of the electron-photon vertex to first order in
$\theta$:
\begin{subequations}
\label{ffp1}
\begin{align}
&\Gamma^{(1,1)\mu}_{A\bar{\psi}\psi}(p,q,r) 
\nonumber
\\*
&= 
\parbox{40mm}{\begin{picture}(40,36)
\put(0,6){\begin{fmfgraph}(40,30)
\fmfkeep{AFFA}
\fmfright{r1,r2}
\fmfleft{l}
\fmf{fermion,tension=2}{r1,i1}
\fmf{fermion}{i1,i3,i2}
\fmf{fermion,tension=2}{i2,r2}
\fmf{photon,tension=1}{l,i3}
\fmf{phantom_arrow,tension=1}{l,i3}
\fmffreeze
\fmf{photon}{i4,i1}
\fmf{phantom_arrow}{i4,i1}
\fmf{photon}{i4,i2}
\fmf{phantom_arrow}{i4,i2}
\fmfdot{i2}
\end{fmfgraph}}
\put(30,5){\mbox{$r$}}
\put(30,15){\mbox{$k,\sigma$}}
\put(30,27){\mbox{$-k,\rho$}}
\put(16,11){\mbox{$k{+}r$}}
\put(4,17){\mbox{$p,\mu$}}
\put(10,28){\mbox{$k{+}p{+}r$}}
\end{picture}}
{}~ + ~
\parbox{40mm}{\begin{picture}(40,36)
\put(0,6){\begin{fmfgraph}(40,30)
\fmfkeep{AFFB}
\fmfright{r1,r2}
\fmfleft{l}
\fmf{fermion,tension=2}{r1,i1}
\fmf{fermion}{i1,i3,i2}
\fmf{fermion,tension=2}{i2,r2}
\fmf{photon,tension=1}{l,i3}
\fmf{phantom_arrow,tension=1}{l,i3}
\fmffreeze
\fmf{photon}{i4,i1}
\fmf{phantom_arrow}{i4,i1}
\fmf{photon}{i4,i2}
\fmf{phantom_arrow}{i4,i2}
\fmfdot{i1}
\end{fmfgraph}}
\put(30,5){\mbox{$r$}}
\put(30,15){\mbox{$k,\sigma$}}
\put(30,27){\mbox{$-k,\rho$}}
\put(16,11){\mbox{$k{+}r$}}
\put(4,17){\mbox{$p,\mu$}}
\put(10,28){\mbox{$k{+}p{+}r$}}
\end{picture}}
{}~ + ~
\parbox{40mm}{\begin{picture}(40,36)
\put(0,6){\begin{fmfgraph}(40,30)
\fmfkeep{AFFC}
\fmfright{r1,r2}
\fmfleft{l}
\fmf{fermion,tension=2}{r1,i1}
\fmf{fermion}{i1,i3,i2}
\fmf{fermion,tension=2}{i2,r2}
\fmf{photon,tension=1}{l,i3}
\fmf{phantom_arrow,tension=1}{l,i3}
\fmffreeze
\fmf{photon}{i4,i1}
\fmf{phantom_arrow}{i4,i1}
\fmf{photon}{i4,i2}
\fmf{phantom_arrow}{i4,i2}
\fmfdot{i3}
\end{fmfgraph}}
\put(30,5){\mbox{$r$}}
\put(30,15){\mbox{$k,\sigma$}}
\put(30,27){\mbox{$-k,\rho$}}
\put(16,11){\mbox{$k{+}r$}}
\put(4,17){\mbox{$p,\mu$}}
\put(10,28){\mbox{$k{+}p{+}r$}}
\end{picture}}
\nonumber
\\
& + ~ 
\parbox{40mm}{\begin{picture}(40,30)
\put(0,0){\begin{fmfgraph}(40,30)
\fmfkeep{AFFD}
\fmfright{r1,r2}
\fmfleft{l}
\fmf{fermion,tension=2}{r1,i1}
\fmf{photon}{i4,i1}
\fmf{phantom_arrow,tension=0}{i4,i1}
\fmf{photon}{i4,i3}
\fmf{phantom_arrow,tension=0}{i4,i3}
\fmf{photon}{i5,i2}
\fmf{phantom_arrow,tension=0}{i5,i2}
\fmf{photon}{i5,i3}
\fmf{phantom_arrow,tension=0}{i5,i3}
\fmf{fermion,tension=2}{i2,r2}
\fmf{photon,tension=1}{l,i3}
\fmf{phantom_arrow,tension=1}{l,i3}
\fmffreeze
\fmf{fermion}{i1,i2}
\fmfdot{i3}
\end{fmfgraph}}
\put(33,3){\mbox{$r$}}
\put(32,15){\mbox{$-k$}}
\put(7,6){\mbox{$k{+}r,\nu$}}
\put(12,1){\mbox{$-k{-}r,\rho$}}
\put(2,11){\mbox{$p,\mu$}}
\put(10,26){\mbox{$k{+}p{+}r,\tau$}}
\put(-4,20){\mbox{$-k{-}p{-}r,\sigma$}}
\end{picture}}
{}~ + ~ 
\parbox{40mm}{\begin{picture}(40,30)
\put(0,2){\begin{fmfgraph}(40,30)
\fmfkeep{AFFE}
\fmfright{r2,r1}
\fmfleft{l}
\fmf{phantom}{i2,r2}
\fmf{fermion}{i2,r1}
\fmf{photon,tension=2}{l,i2}
\fmf{phantom_arrow,tension=2}{l,i2}
\fmffreeze
\fmf{fermion,tension=2}{r2,i1}
\fmf{fermion}{i1,i2}
\fmffreeze
\fmfforce{(0.4w,0)}{i3}
\fmf{photon,left=0.5}{i3,i2}
\fmf{phantom_arrow,left=0.5}{i3,i2}
\fmf{photon,right=0.5}{i3,i1}
\fmf{phantom_arrow,right=0.5}{i3,i1}
\fmfdot{i2}
\end{fmfgraph}}
\put(32,7){\mbox{$r$}}
\put(26,0){\mbox{$k,\sigma$}}
\put(-1,11){\mbox{$-k,\rho$}}
\put(4,19){\mbox{$p,\mu$}}
\put(22,12){\mbox{$k{+}r$}}
\end{picture}}
{}~ + ~ 
\parbox{40mm}{\begin{picture}(40,30)
\put(0,0){\begin{fmfgraph}(40,30)
\fmfkeep{AFFF}
\fmfright{r1,r2}
\fmfleft{l}
\fmf{phantom}{r2,i2}
\fmf{fermion}{r1,i2}
\fmf{photon,tension=2}{l,i2}
\fmf{phantom_arrow,tension=2}{l,i2}
\fmffreeze
\fmf{fermion,tension=2}{i1,r2}
\fmf{fermion}{i2,i1}
\fmffreeze
\fmfforce{(0.4w,h)}{i3}
\fmf{photon,left=0.5}{i3,i1}
\fmf{phantom_arrow,left=0.5}{i3,i1}
\fmf{photon,right=0.5}{i3,i2}
\fmf{phantom_arrow,right=0.5}{i3,i2}
\fmfdot{i2}
\end{fmfgraph}}
\put(32,4){\mbox{$r$}}
\put(24,31){\mbox{$-k,\rho$}}
\put(2,20){\mbox{$k,\sigma$}}
\put(4,11){\mbox{$p,\mu$}}
\put(22,18){\mbox{$k{+}p{+}r$}}
\end{picture}}
\nonumber
\\
& + ~ 
\parbox{40mm}{\begin{picture}(40,38)
\put(0,6){\begin{fmfgraph}(40,30)
\fmfkeep{AFFG}
\fmfright{r1,r2}
\fmfleft{l}
\fmf{fermion,tension=2}{r1,i1}
\fmf{fermion}{i1,i3}
\fmf{phantom}{i3,i2}
\fmf{fermion,tension=2}{i2,r2}
\fmf{photon,tension=1}{l,i3}
\fmf{phantom_arrow,tension=1}{l,i3}
\fmffreeze
\fmf{fermion}{i3,i5,i2}
\fmf{photon}{i4,i1}
\fmf{phantom_arrow}{i4,i1}
\fmf{photon}{i4,i2}
\fmf{phantom_arrow}{i4,i2}
\fmfdot{i5}
\end{fmfgraph}}
\put(30,5){\mbox{$r$}}
\put(30,15){\mbox{$k,\sigma$}}
\put(30,27){\mbox{$-k,\rho$}}
\put(16,11){\mbox{$k{+}r$}}
\put(4,17){\mbox{$p,\mu$}}
\put(10,28){\mbox{$k{+}p{+}r$}}
\end{picture}}
+ ~
\parbox{40mm}{\begin{picture}(40,38)
\put(0,6){\begin{fmfgraph}(40,30)
\fmfkeep{AFFH}
\fmfright{r1,r2}
\fmfleft{l}
\fmf{fermion,tension=2}{r1,i1}
\fmf{fermion}{i3,i2}
\fmf{phantom}{i1,i3}
\fmf{fermion,tension=2}{i2,r2}
\fmf{photon,tension=1}{l,i3}
\fmf{phantom_arrow,tension=1}{l,i3}
\fmffreeze
\fmf{fermion}{i1,i5,i3}
\fmf{photon}{i4,i1}
\fmf{phantom_arrow}{i4,i1}
\fmf{photon}{i4,i2}
\fmf{phantom_arrow}{i4,i2}
\fmfdot{i5}
\end{fmfgraph}}
\put(30,5){\mbox{$r$}}
\put(30,15){\mbox{$k,\sigma$}}
\put(30,27){\mbox{$-k,\rho$}}
\put(16,11){\mbox{$k{+}r$}}
\put(4,17){\mbox{$p,\mu$}}
\put(10,28){\mbox{$k{+}p{+}r$}}
\end{picture}}
+ \dfrac{1}{2} ~
\parbox{40mm}{\begin{picture}(40,38)
\put(0,6){\begin{fmfgraph}(40,30)
\fmfkeep{AFFI}
\fmfright{r1,r2}
\fmfleft{l}
\fmfforce{((0.3w,h))}{i2}
\fmf{fermion,tension=1}{r1,i1}
\fmf{fermion,tension=1}{i1,r2}
\fmf{photon,tension=1}{l,i1}
\fmf{phantom_arrow,tension=1}{l,i1}
\fmffreeze
\fmf{photon,left=0.65}{i2,i1}
\fmf{photon,right=0.65}{i2,i1}
\fmf{phantom_arrow,left=0.65}{i2,i1}
\fmf{phantom_arrow,right=0.65}{i2,i1}
\fmfdot{i1}
\end{fmfgraph}}
\put(30,14){\mbox{$r$}}
\put(4,17){\mbox{$p,\mu$}}
\put(3,30){\mbox{$-k$}}
\put(21,32){\mbox{$k$}}
\end{picture}}
\nonumber
\\
& {}+ ~
\parbox{40mm}{\begin{picture}(40,28)
\put(0,6){\begin{fmfgraph}(40,25)
\fmfkeep{AFFJ}
\fmfright{r1,r2}
\fmfleft{l}
\fmf{fermion,tension=1}{r1,i1}
\fmf{fermion,tension=1}{i2,r2}
\fmf{dots,tension=3}{i1,i2}
\fmf{photon,tension=1}{l,i3}
\fmf{phantom_arrow,tension=1}{l,i3}
\fmf{fermion,left=0.9}{i1,i3}
\fmf{fermion,left=0.9}{i3,i2}
\fmfdot{i1}
\end{fmfgraph}}
\put(30,14){\mbox{$r$}}
\put(14,8){\mbox{$k{-}\frac{p}{2}$}}
\put(4,15){\mbox{$p,\mu$}}
\put(14,27){\mbox{$k{+}\frac{p}{2}$}}
\end{picture}}
{}~+ ~
\parbox{40mm}{\begin{picture}(40,28)
\put(0,6){\begin{fmfgraph}(40,25)
\fmfkeep{AFFK}
\fmfright{r1,r2}
\fmfleft{l}
\fmf{fermion,tension=1}{r1,i1}
\fmf{fermion,tension=1}{i2,r2}
\fmf{dots,tension=3}{i1,i2}
\fmf{photon,tension=1}{l,i3}
\fmf{phantom_arrow,tension=1}{l,i3}
\fmf{fermion,left=0.9}{i1,i3}
\fmf{fermion,left=0.9}{i3,i2}
\fmfdot{i2}
\end{fmfgraph}}
\put(30,14){\mbox{$r$}}
\put(14,8){\mbox{$k{-}\frac{p}{2}$}}
\put(4,15){\mbox{$p,\mu$}}
\put(14,27){\mbox{$k{+}\frac{p}{2}$}}
\end{picture}}
\nonumber
\\[-1ex]
& - ~
\parbox{40mm}{\begin{picture}(40,28)
\put(0,6){\begin{fmfgraph}(40,25)
\fmfkeep{AFFL}
\fmfright{r1,r2}
\fmfleft{l}
\fmf{fermion,tension=1}{r1,i1}
\fmf{fermion,tension=1}{i1,r2}
\fmf{dots,tension=4.5}{i1,i2}
\fmf{photon,tension=1}{l,i3}
\fmf{phantom_arrow,tension=1}{l,i3}
\fmf{fermion,left=1}{i2,i3}
\fmf{fermion,left=1}{i3,i2}
\fmfdot{i2}
\end{fmfgraph}}
\put(31,14){\mbox{$r$}}
\put(13,8){\mbox{$k{-}\frac{p}{2}$}}
\put(2,15){\mbox{$p,\mu$}}
\put(13,26){\mbox{$k{+}\frac{p}{2}$}}
\end{picture}}
{}~- ~
\parbox{40mm}{\begin{picture}(40,28)
\put(0,6){\begin{fmfgraph}(40,25)
\fmfkeep{AFFM}
\fmfright{r1,r2}
\fmfleft{l}
\fmf{fermion,tension=1}{r1,i1}
\fmf{fermion,tension=1}{i1,r2}
\fmf{dots,tension=4.5}{i1,i2}
\fmf{photon,tension=1}{l,i3}
\fmf{phantom_arrow,tension=1}{l,i3}
\fmf{fermion,left=1}{i2,i3}
\fmf{fermion,left=1}{i3,i2}
\fmfdot{i1}
\end{fmfgraph}}
\put(31,14){\mbox{$r$}}
\put(13,8){\mbox{$k{-}\frac{p}{2}$}}
\put(2,15){\mbox{$p,\mu$}}
\put(13,26){\mbox{$k{+}\frac{p}{2}$}}
\end{picture}}
\nonumber
\\
& = \frac{\hbar g^2}{(4\pi)^2 \varepsilon} \Big(\frac{1}{2} N_\psi 
+ 0\, N_A + 3 m \frac{\partial}{\partial m} 
+ \tau \theta^{\alpha\beta} \frac{\partial}{\partial \theta^{\alpha\beta}} 
\nonumber
\\
& \qquad 
+ \Big( -\frac{13}{48} -\frac{1}{2} \lambda_1 
+\frac{1}{6} \lambda_2 
+ (-\tau - \frac{1}{6}) \kappa_1 + \frac{1}{2} \kappa_2 - \frac{4}{3}
\kappa_3 - \frac{3}{2} \kappa_4 - \frac{2}{3} \kappa_9 \Big) 
\frac{\partial}{\partial \kappa_1}
\nonumber
\\
& \qquad 
+ \Big( -\frac{1}{12} + \lambda_1  -\frac{1}{3} \lambda_2 
+ \frac{2}{3} \kappa_1 -\tau \kappa_2 -\frac{2}{3} \kappa_3 
+ 3 \kappa_4 - \frac{1}{3} \kappa_9\Big) \frac{\partial}{\partial
  \kappa_2}
\nonumber
\\
& \qquad 
+ \Big( \frac{13}{32} +\frac{5}{24} \lambda_2 
- \frac{1}{4} \kappa_1 -\frac{1}{4} \kappa_2 
-\tau \kappa_3 
+ \frac{3}{4} \kappa_4\Big) \frac{\partial}{\partial
  \kappa_3}
\nonumber
\\
& \qquad 
+ \Big(\frac{1}{48} - \frac{1}{6} \kappa_1 + \frac{1}{2} \kappa_2 +
\frac{2}{3} \kappa_3 + (-\tau - \frac{1}{2}) \kappa_4 
+ \frac{1}{3} \kappa_9 \Big) \frac{\partial}{\partial \kappa_4}
\nonumber
\\
& \qquad 
+ \Big(\frac{1}{6} + \frac{2}{3} \kappa_1 - \kappa_2 -
\frac{2}{3} \kappa_3 - 3 \kappa_4 + (-\tau - 3) \kappa_6 + 2 \kappa_7
- \frac{1}{3} \kappa_9 \Big) \frac{\partial}{\partial \kappa_6}
\nonumber
\\
& \qquad 
+ \Big(- \frac{1}{2} \kappa_2 - \kappa_3 - \frac{3}{2} \kappa_4 
+ \frac{1}{2} \kappa_6 + (-\tau - 3) \kappa_7 + \frac{1}{2} \kappa_9\Big) 
\frac{\partial}{\partial \kappa_7}
\nonumber
\\
& \qquad 
+ \Big(\frac{1}{16} - \frac{1}{2} \kappa_1 + \kappa_2 +
\kappa_3 + \kappa_4 + \kappa_6 + 2 \kappa_7 +(-\tau - 6) \kappa_8 
+ \frac{1}{2} \kappa_9
\Big) \frac{\partial}{\partial \kappa_8}
\nonumber
\\
& \qquad 
+\Big(\frac{3}{4}\zeta + (-\tau +\frac{4}{3}) \kappa_9 \Big)
\frac{\partial}{\partial \kappa_9}\Big) 
\Big(\Gamma^{(1,0)\mu}_{A\bar{\psi}\psi}(p,q,r)\Big)
\label{ffp1a}
\\
& + \frac{\hbar g^2}{(4\pi)^2 \varepsilon} \mathrm{i}
\theta^{\alpha\beta} \Big(
\Big(3 +5 \lambda_1 -\frac{10}{3}\lambda_2 \Big) m p_\beta g^\mu_\alpha 
+\Big(\frac{3}{2}+\frac{5}{6}\lambda_2 \Big) 
m p_\nu \gamma^{\mu\nu}_{~~\alpha\beta}
\nonumber
\\
&\qquad
+ \Big(-2 \lambda_1 +\frac{2}{3} \lambda_2 + \frac{1}{4} \tau \Big)
\big((2 g^\mu_\nu p_\alpha - 2 g^\mu_\alpha p_\nu) r_\beta
- g^\mu_\alpha p_\nu p_\beta\big)\gamma^\nu 
\nonumber
\\
&\qquad 
+\Big(\frac{3}{4}\zeta -\frac{1}{2} \lambda_1 
+\frac{1}{6} \lambda_2 \Big)
(p^2 g^{\mu\nu} - p^\mu p^\nu) \gamma_{\nu\alpha\beta} \Big)
+ \mathcal{O}(1)~.
\label{ffp1b}
\end{align}
\end{subequations}

There are several observations: 
\begin{itemize}
\item The renormalizations of
  $\kappa_4,\kappa_6,\kappa_7,\kappa_8,\kappa_9$ are the same in
  (\ref{fs1}), (\ref{ffff1}) and (\ref{ffp1a}), which together with
  the transversality of the additional terms in (\ref{ffp1b}) verifies
  the Ward identity (see the next Section).

\item The field redefinitions parametrized by $\kappa_i$ always give
  rise to renormalizations of $\kappa_j$, never to a renormalization
  of the ``physical''  counterterms required by (\ref{ffp1b}).
  
\item We obtain divergences in (\ref{ffp1b}) which have no counterpart
  in the original action (\ref{th-exp}), in particular, they cannot be
  field redefinitions. In the massive case not all of them can be
  eliminated.
  
\item In the massless case $m=0$ both remaining divergences in
  (\ref{ffp1b}) are absent for $6\lambda_1 -2 \lambda_2 = 9\zeta +
  \mathcal{O}(\hbar)$ and $\tau=12\zeta+ \mathcal{O}(\hbar)$.

\end{itemize}
The last remark means that for $\zeta=0$ (mathematical framework) in the
massless case no $\theta$-renormalization and no extension terms
(\ref{YM-ext}) to the bosonic action are required. Thus we have the choice of
a serious problem in (\ref{ffff1}) or in (\ref{ffp1}). It is clear that we
prefer to make (\ref{ffp1}) as nice as possible.

\subsection{No electrons, two photons}

It remains to compute the divergent one-loop graphs without external
fermion lines to first order in $\theta$. The exciting question is whether
these divergences are compatible with $\lambda_1=\lambda_2=0$ for
$\tau=12\zeta=0$. For the photon two-point function we obtain
\begin{align}
&\Gamma_{AA}^{(1,1)\mu\nu}(p,q) 
\nonumber
\\*
&= ~ 
- ~ \parbox{40mm}{\begin{picture}(40,30)
\put(0,0){\begin{fmfgraph}(40,30)
\fmfleft{l}
\fmfright{r}
\fmf{photon,tension=1}{l,i1}
\fmf{phantom_arrow,tension=1}{l,i1}
\fmf{photon,tension=1}{r,i2}
\fmf{phantom_arrow,tension=1}{r,i2}
\fmf{fermion,left=1}{i1,i2}
\fmf{fermion,left=1}{i2,i1}
\fmfdot{i1}
\end{fmfgraph}}
\put(6,17){\mbox{$p,\mu$}}
\put(31,17){\mbox{$-p,\nu$}}
\put(17,3){\mbox{$k{-}\frac{p}{2}$}}
\put(17,24){\mbox{$k{+}\frac{p}{2}$}}
\end{picture}}
{}~-~
\parbox{40mm}{\begin{picture}(40,30)
\put(0,0){\begin{fmfgraph}(40,30)
\fmfleft{l}
\fmfright{r}
\fmf{photon,tension=1}{l,i1}
\fmf{phantom_arrow,tension=1}{l,i1}
\fmf{photon,tension=1}{r,i2}
\fmf{phantom_arrow,tension=1}{r,i2}
\fmf{fermion,left=1}{i1,i2}
\fmf{fermion,left=1}{i2,i1}
\fmfdot{i2}
\end{fmfgraph}}
\put(6,17){\mbox{$p,\mu$}}
\put(31,17){\mbox{$-p,\nu$}}
\put(17,3){\mbox{$k{-}\frac{p}{2}$}}
\put(17,24){\mbox{$k{+}\frac{p}{2}$}}
\end{picture}}
\nonumber
\\
& ~-~ \parbox{40mm}{\begin{picture}(40,30)
\put(0,0){\begin{fmfgraph}(40,30)
\fmfleft{l}
\fmfright{r}
\fmftop{i4}
\fmf{photon,tension=1}{l,i1}
\fmf{phantom_arrow,tension=1}{l,i1}
\fmf{photon,tension=1}{r,i2}
\fmf{phantom_arrow,tension=1}{r,i2}
\fmf{phantom,left=1}{i1,i2}
\fmf{fermion,left=1}{i2,i1}
\fmffreeze
\fmf{fermion,left=0.4}{i1,i3}
\fmf{fermion,left=0.4}{i3,i2}
\fmf{phantom,tension=1.7}{i3,i4}
\fmfdot{i3}
\end{fmfgraph}}
\put(6,17){\mbox{$p,\mu$}}
\put(31,17){\mbox{$-p,\nu$}}
\put(17,3){\mbox{$k{-}\frac{p}{2}$}}
\put(17,24){\mbox{$k{+}\frac{p}{2}$}}
\end{picture}}
{}~-~
\parbox{40mm}{\begin{picture}(40,30)
\put(0,0){\begin{fmfgraph}(40,30)
\fmfleft{l}
\fmfright{r}
\fmfbottom{i4}
\fmf{photon,tension=1}{l,i1}
\fmf{phantom_arrow,tension=1}{l,i1}
\fmf{photon,tension=1}{r,i2}
\fmf{phantom_arrow,tension=1}{r,i2}
\fmf{fermion,left=1}{i1,i2}
\fmf{phantom,left=1}{i2,i1}
\fmffreeze
\fmf{fermion,left=0.4}{i2,i3}
\fmf{fermion,left=0.4}{i3,i1}
\fmf{phantom,tension=1.7}{i3,i4}
\fmfdot{i3}
\end{fmfgraph}}
\put(6,17){\mbox{$p,\mu$}}
\put(31,17){\mbox{$-p,\nu$}}
\put(17,3){\mbox{$k{-}\frac{p}{2}$}}
\put(17,24){\mbox{$k{+}\frac{p}{2}$}}
\end{picture}}
{}~-~
\parbox{40mm}{\begin{picture}(40,30)
\put(0,5){\begin{fmfgraph}(40,30)
\fmfbottom{l,r}
\fmf{photon,tension=1}{l,i1}
\fmf{phantom_arrow,tension=1}{l,i1}
\fmf{photon,tension=1}{r,i1}
\fmf{phantom_arrow,tension=1}{r,i1}
\fmf{fermion,right=1}{i1,i1}
\fmfdot{i1}
\end{fmfgraph}}
\put(6,10){\mbox{$p,\mu$}}
\put(27,10){\mbox{$-p,\nu$}}
\put(16,22){\mbox{$k$}}
\end{picture}}
\nonumber
\\*
&= \mathcal{O}(1)~. \label{ps1}
\end{align}
In (\ref{ps1}) the divergent parts of the first, second and fifth
graphs are identically zero and those of the third and fourth graphs
are antisymmetric in $(p,\mu)\leftrightarrow (-p,\nu)$ so that their
sum is zero. The result of (\ref{ps1}) was clear from the beginning
because there is no gauge-invariant purely bosonic action part to
first order in $\theta$. 

\subsection{No electrons, three photons}
\label{Sec85}

Now we compute the photon three-point function to first order in
$\theta$: 
\begin{align}
&\Gamma_{AAA}^{(1,1)\mu\nu\rho}(p,q,r) 
\nonumber
\\*
&= \left\{\left\{ ~- \hspace*{-1em} 
\parbox{40mm}{\begin{picture}(40,30)
\put(0,0){\begin{fmfgraph}(40,30)
\fmfsurround{l1,l2,l3}
\fmf{photon,tension=2}{l1,i1}
\fmf{phantom_arrow,tension=1}{l1,i1}
\fmf{photon,tension=2}{l2,i2}
\fmf{phantom_arrow,tension=1}{l2,i2}
\fmf{photon,tension=2}{l3,i3}
\fmf{phantom_arrow,tension=1}{l3,i3}
\fmf{fermion,left=0.7}{i1,i3}
\fmf{fermion,left=0.4}{i3,i2}
\fmf{fermion,left=0.7}{i2,i1}
\fmfdot{i1}
\end{fmfgraph}}
\put(35,17){\mbox{$p,\mu$}}
\put(12,1){\mbox{$q,\nu$}}
\put(6,22){\mbox{$r,\rho$}}
\put(19,26){\mbox{$k{+}r$}}
\put(8,14){\mbox{$k$}}
\put(26,3){\mbox{$k{+}p{+}r$}}
\end{picture}} \quad \right\}
+ \Big\{ ~(q,\nu) \leftrightarrow (r,\rho)~\Big\}\right\}
\nonumber
\\*
& + \Big\{\, (p,\mu) \to (q,\nu) \to (r,\rho) \to (p,\mu) \,\Big\} 
+ \Big\{\, (p,\mu) \to (r,\rho) \to (q,\nu) \to (p,\mu) \,\Big\} 
\nonumber
\\
&+ \left\{\left\{ ~- \hspace*{-1em} 
\parbox{40mm}{\begin{picture}(40,30)
\put(0,0){\begin{fmfgraph}(40,30)
\fmfsurround{l1,l2,l3}
\fmf{photon,tension=2}{l1,i1}
\fmf{phantom_arrow,tension=1}{l1,i1}
\fmf{photon,tension=2}{l2,i2}
\fmf{phantom_arrow,tension=1}{l2,i2}
\fmf{photon,tension=2}{l3,i3}
\fmf{phantom_arrow,tension=1}{l3,i3}
\fmf{fermion,left=0.7}{i1,i3}
\fmf{fermion,left=0.4}{i3,i2}
\fmf{phantom,left=0.7}{i2,i1}
\fmffreeze
\fmfforce{(0.6w,0.78h)}{i4}
\fmf{fermion,left=0.3}{i2,i4}
\fmf{fermion,left=0.3}{i4,i1}
\fmfdot{i4}
\end{fmfgraph}}
\put(35,17){\mbox{$p,\mu$}}
\put(12,1){\mbox{$q,\nu$}}
\put(6,22){\mbox{$r,\rho$}}
\put(19,26){\mbox{$k{+}r$}}
\put(8,14){\mbox{$k$}}
\put(26,3){\mbox{$k{+}p{+}r$}}
\end{picture}} \quad \right\}
+ \Big\{ ~(q,\nu) \leftrightarrow (r,\rho)~\Big\}\right\}
\nonumber
\\*
& + \Big\{\, (p,\mu) \to (q,\nu) \to (r,\rho) \to (p,\mu) \,\Big\} 
+ \Big\{\, (p,\mu) \to (r,\rho) \to (q,\nu) \to (p,\mu) \,\Big\} 
\nonumber
\\
&+ \left\{ ~- \hspace*{-1em} 
\parbox{40mm}{\begin{picture}(40,30)
\put(0,0){\begin{fmfgraph}(40,30)
\fmfsurround{l1,l2,l3}
\fmf{photon,tension=2}{l1,i2}
\fmf{phantom_arrow,tension=2}{l1,i2}
\fmf{photon,tension=2}{l3,i1}
\fmf{phantom_arrow,tension=2}{l3,i1}
\fmf{photon,tension=2}{l2,i1}
\fmf{phantom_arrow,tension=2}{l2,i1}
\fmf{fermion,left=1}{i1,i2}
\fmf{fermion,left=1}{i2,i1}
\fmfdot{i1}
\end{fmfgraph}}
\put(35,17){\mbox{$p,\mu$}}
\put(12,1){\mbox{$q,\nu$}}
\put(4,22){\mbox{$r,\rho$}}
\put(19,26){\mbox{$k{-}\frac{p}{2}$}}
\put(26,3){\mbox{$k{+}\frac{p}{2}$}}
\end{picture}} \quad \right\}
+ \Big\{ ~(p,\mu) \leftrightarrow (r,\rho)~\Big\}
+ \Big\{ ~(p,\mu) \leftrightarrow (q,\nu)~\Big\}
\nonumber
\\
& - \hspace*{-1em} 
\parbox{40mm}{\begin{picture}(40,30)
\put(0,0){\begin{fmfgraph}(40,30)
\fmfsurround{l1,l2,l3}
\fmf{photon,tension=2}{l1,i1}
\fmf{phantom_arrow,tension=2}{l1,i1}
\fmf{photon,tension=2}{l3,i1}
\fmf{phantom_arrow,tension=2}{l3,i1}
\fmf{photon,tension=2}{l2,i1}
\fmf{phantom_arrow,tension=2}{l2,i1}
\fmffreeze
\fmf{fermion,left}{i1,i1}
\fmfdot{i1}
\end{fmfgraph}}
\put(35,17){\mbox{$p,\mu$}}
\put(12,1){\mbox{$q,\nu$}}
\put(4,22){\mbox{$r,\rho$}}
\put(28,4){\mbox{$k$}}
\end{picture}} 
\nonumber
\\
& 
= \frac{\hbar g^2}{(4\pi)^2 \varepsilon} \Big(
0 \,N_A -\frac{4}{3} g^2 \frac{\partial}{\partial g^2} 
+ 12 \zeta \theta^{\alpha\beta} \frac{\partial}{\partial
  \theta^{\alpha\beta}} 
+ \big(-\frac{4}{3} \lambda_1 -12 \zeta(1{+}\lambda_1)\big) 
\frac{\partial}{\partial \lambda_1}
\nonumber
\\
& 
\hspace*{10em} + \big(-\frac{4}{3} \lambda_2 -12 \zeta(1{+}\lambda_2)\big) 
\frac{\partial}{\partial \lambda_2}
\Big) \Gamma^{(1,0)\mu\nu\rho}_{AAA;1+2}(p,q,r) + \mathcal{O}(1)~.
\label{ppp1}
\end{align}
We confirm indeed that for $\zeta =0$ the choice $\lambda_1=\lambda_2=0$ in
(\ref{YM-ext}), which corresponds to the $\theta$-expansion of the unmodified
noncommutative Yang-Mills-Dirac action (\ref{YMD}), is stable at one-loop
level. A deeper understanding of this result is missing. It points however to
a link between the bosonic action and an effective fermion action as in
\cite{Langmann:2001cv}. Before entering the discussion we have to prove that
this nice result (ignoring the $\kappa_9$-problem) is not going to change by
the remaining divergent Green's functions.

\section{Ward identities}

The action $\Sigma=\Sigma_{c\ell}+\Sigma_{gf}$ given in (\ref{YMD}) and
(\ref{gf}) is invariant under Abelian BRST transformations (\ref{Slavnov}).
Switching to momentum space, functional derivation with respect to $c(p)$ and
restriction to the physical sector $\{B,\bar{c},c\}=0$ leads to the following
form of the Ward identity:
\begin{align}
\bigg[p_\mu \frac{\delta \Gamma}{\delta A_\mu(p)} 
+ \int \frac{d^4q}{(2\pi)^4}\frac{d^4r}{(2\pi)^4}\;
(2\pi)^4\delta(p{+}q{+}r) \,\Big(
\bar{\psi}(q) \frac{\overrightarrow{\delta} \Gamma}{\delta
  \bar{\psi}(-r)} 
-\frac{\Gamma \overleftarrow{\delta}}{\delta
  \psi(-q)} \psi(r) \Big) \bigg]_{\{B,\bar{c},c\}=0} \,=0~.
\label{WIM}
\end{align}
From (\ref{WIM}) we derive the following identities:
\begin{subequations}
\begin{align}
0 &= \bigg[p_\mu \frac{\delta^n}{\delta A_{\nu_1}(q_1) \dots 
\delta A_{\nu_n}(q_n) } \frac{\delta \Gamma}{\delta A_\mu(p)}
\bigg]_{\{A,\bar{\psi},\psi,B,\bar{c},c\}=0} ~,
\\
0 &= \bigg[p_\mu \frac{\overrightarrow{\delta}}{\delta \bar{\psi}(q)} 
\frac{\delta^{n+1} \Gamma}{\delta A_\mu(p) \delta A_{\nu_1}(p_1)
\dots \delta A_{\nu_n}(p_n) } 
\frac{\overleftarrow{\delta}}{\delta \psi(r)} 
\nonumber
\\
& 
\qquad + \frac{\overrightarrow{\delta}}{\delta \bar{\psi}(p{+}q)} 
\frac{\delta^n \Gamma}{\delta A_{\nu_1}(p_1)
\dots \delta A_{\nu_n}(p_n) }  
\frac{\overleftarrow{\delta}}{\delta \psi(r)} 
\nonumber
\\
& 
\qquad - \frac{\overrightarrow{\delta}}{\delta \bar{\psi}(q)} 
\frac{\delta^n \Gamma}{\delta A_{\nu_1}(p_1)
\dots \delta A_{\nu_n}(p_n) }  
\frac{\overleftarrow{\delta}}{\delta \psi(p{+}r)}
\bigg]_{\{A,\bar{\psi},\psi,B,\bar{c},c\}=0} ~,
\end{align}
\end{subequations}
which means
\begin{subequations}
\label{WI}
\begin{align}
p_\mu \Gamma^{(\tau,\ell)\mu\nu_1\dots\nu_n}_{A\dots A}(p,q_1,\dots,q_n) &=
0~, 
\label{WI0}
\\
p_\mu \Gamma^{(\tau,\ell)\mu}_{A\bar{\psi}\psi}(p,q,r) &= 
\Gamma^{(\tau,\ell)}_{\bar{\psi}\psi}(q,p{+}r)   
-\Gamma^{(\tau,\ell)}_{\bar{\psi}\psi}(p{+}q,r) ~,
\label{WI1}
\\
p_\mu \Gamma^{(\tau,\ell)\mu\nu}_{AA\bar{\psi}\psi}(p,q,r,s) &= 
\Gamma^{(\tau,\ell)\nu}_{A\bar{\psi}\psi}(q,r,p{+}s)   
-\Gamma^{(\tau,\ell)\nu}_{A\bar{\psi}\psi}(q,p{+}r,s) ~,
\label{WI2}
\\
p_\mu \Gamma^{(\tau,\ell)\mu\nu\rho}_{AAA\bar{\psi}\psi}(p,q,r,s,t) &= 
\Gamma^{(\tau,\ell)\nu\rho}_{AA\bar{\psi}\psi}(q,r,s,p{+}t)   
-\Gamma^{(\tau,\ell)\nu\rho}_{AA\bar{\psi}\psi}(q,r,p{+}s,t) ~.
\label{WI3}
\end{align}
\end{subequations}

On tree-level $(\tau,0)$ the identities (\ref{WI}) are easy to verify.
Let us first investigate (\ref{WI1}) for $(\tau,\ell)=(1,1)$. 
We perform the manipulations directly on the
integrals encoded in the Feynman graphs. Denoting by the subscript
$\frac{1}{\varepsilon}$ the divergent part in $\varepsilon$ in
analytic regularization, let us consider
\begin{align}
& 
p_\mu \left( 
\parbox{40mm}{\begin{picture}(40,36)
\put(0,6){\begin{fmfgraph}(40,30)
\fmfright{r1,r2}
\fmfleft{l}
\fmf{fermion,tension=2}{r1,i1}
\fmf{fermion}{i1,i3,i2}
\fmf{fermion,tension=2}{i2,r2}
\fmf{photon,tension=1}{l,i3}
\fmf{phantom_arrow,tension=1}{l,i3}
\fmffreeze
\fmf{photon}{i4,i1}
\fmf{phantom_arrow}{i4,i1}
\fmf{photon}{i4,i2}
\fmf{phantom_arrow}{i4,i2}
\fmfdot{i2}
\end{fmfgraph}}
\put(30,5){\mbox{$r$}}
\put(31,15){\mbox{$k,\sigma$}}
\put(31,27){\mbox{$-k,\rho$}}
\put(16,11){\mbox{$k{+}r$}}
\put(4,17){\mbox{$p,\mu$}}
\put(10,28){\mbox{$k{+}p{+}r$}}
\end{picture}} \right)_{\!\!\tfrac{1}{\varepsilon}}
\nonumber
\\
&= \Big(\int \!\!\frac{d^4 k}{(2\pi)^4} \, 
\Gamma^{(1,0)\rho}_{A\bar{\psi}\psi}(-k,-p{-}r,k{+}p{+}r) 
\nonumber
\\
& \qquad
\times \frac{-\gamma^\alpha(k{+}p{+}r)_\alpha - m}{(k{+}p{+}r)^2-m^2
  +\mathrm{i} \epsilon} p_\mu \gamma^\mu 
\frac{-\gamma^\beta(k{+}r)_\beta - m}{(k{+}r)^2-m^2
  +\mathrm{i} \epsilon} \gamma^\sigma \Delta^{AA}_{(0,0)\sigma\rho}(k,{-}k) 
\Big)_{\!\!\tfrac{1}{\varepsilon}}
\nonumber
\\
&= \Big(\int \!\!\frac{d^4 k}{(2\pi)^4} \, 
\Gamma^{(1,0)\rho}_{A\bar{\psi}\psi}(-k,-p{-}r,k{+}p{+}r) 
\nonumber
\\
& \qquad
\times \Big(
\frac{(-\gamma^\alpha(k{+}p{+}r)_\alpha - m)(
\gamma^\mu(k{+}p{+}r)_\mu-m)}{(k{+}p{+}r)^2-m^2
  +\mathrm{i} \epsilon} \,
\frac{-\gamma^\beta(k{+}r)_\beta - m}{(k{+}r)^2-m^2
  +\mathrm{i} \epsilon} 
\nonumber
\\
& \qquad\qquad 
-\frac{-\gamma^\alpha(k{+}p{+}r)_\alpha - m}{(k{+}p{+}r)^2-m^2
  +\mathrm{i} \epsilon} \,
\frac{(\gamma^\mu(k{+}r)_\mu - m)(
-\gamma^\beta(k{+}r)_\beta-m)}{(k{+}r)^2-m^2
  +\mathrm{i} \epsilon} \Big)
\gamma^\sigma \Delta^{AA}_{(0,0)\sigma\rho}(k,{-}k) 
\Big)_{\!\!\tfrac{1}{\varepsilon}}
\nonumber
\\
&= - \Big(\int\!\! \frac{d^4 k}{(2\pi)^4} \, 
\Gamma^{(1,0)\rho}_{A\bar{\psi}\psi}(-k,-p{-}r,k{+}p{+}r) 
\frac{-\gamma^\beta(k{+}r)_\beta - m}{(k{+}r)^2-m^2
  +\mathrm{i} \epsilon} \gamma^\sigma \Delta^{AA}_{(0,0)\sigma\rho}(k,{-}k) 
\Big)_{\!\!\tfrac{1}{\varepsilon}}
\nonumber
\\*
& + \Big(\int \!\! \frac{d^4 k}{(2\pi)^4} \, 
\Gamma^{(1,0)\rho}_{A\bar{\psi}\psi}(-k,-p{-}r,k{+}p{+}r) 
\frac{-\gamma^\alpha(k{+}p{+}r)_\alpha - m}{(k{+}p{+}r)^2-m^2
  +\mathrm{i} \epsilon}  \gamma^\sigma \Delta^{AA}_{(0,0)\sigma\rho}(k,{-}k) 
\Big)_{\!\!\tfrac{1}{\varepsilon}}~.
\nonumber
\end{align}
We have used the following property of analytic regularization, see
Appendix~\ref{appa}:  
\begin{align}
\Big(\int \frac{d^4 k}{(2\pi)^4} \, &
\frac{(g^{\mu\nu} (k{+}p)_\mu(k{+}p)_\nu - m^2) k_{\rho_1}\dots
  k_{\rho_n}}{((k{+}p)^2-m^2+\mathrm{i}\epsilon)
\prod_{i=1}^s ((k{+}p_i)^2-m^2_i+\mathrm{i}\epsilon)}
\Big)_{\!\!\tfrac{1}{\varepsilon}}
\nonumber
\\
& \hspace*{10em} = \Big(\int \frac{d^4 k}{(2\pi)^4} \, 
\frac{ k_{\rho_1}\dots  k_{\rho_n}}{
\prod_{i=1}^s ((k{+}p_i)^2-m^2_i+\mathrm{i}\epsilon)}
\Big)_{\!\!\tfrac{1}{\varepsilon}}~.
\label{anreg}
\end{align}
Using the tree-level Ward identity (\ref{WI1}),
\begin{align*}
\Gamma^{(1,0)\rho}_{A\bar{\psi}\psi}(-k,-p{-}r,k{+}p{+}r) 
&= 
p_\mu \Gamma^{(1,0)\mu\rho}_{AA\bar{\psi}\psi}(p,-k,-p{-}r,k{+}r) 
+\Gamma^{(1,0)\rho}_{A\bar{\psi}\psi}(-k,-r,k{+}r) ~,
\end{align*}
we conclude
\begin{subequations}
\label{W3}
\begin{align}
&  p_\mu \left( 
\parbox{40mm}{\begin{picture}(40,36)
\put(0,6){\begin{fmfgraph}(40,30)
\fmfright{r1,r2}
\fmfleft{l}
\fmf{fermion,tension=2}{r1,i1}
\fmf{fermion}{i1,i3,i2}
\fmf{fermion,tension=2}{i2,r2}
\fmf{photon,tension=1}{l,i3}
\fmf{phantom_arrow,tension=1}{l,i3}
\fmffreeze
\fmf{photon}{i4,i1}
\fmf{phantom_arrow}{i4,i1}
\fmf{photon}{i4,i2}
\fmf{phantom_arrow}{i4,i2}
\fmfdot{i2}
\end{fmfgraph}}
\put(30,5){\mbox{$r$}}
\put(31,15){\mbox{$k,\sigma$}}
\put(31,27){\mbox{$-k,\rho$}}
\put(17,11){\mbox{$k{+}r$}}
\put(4,17){\mbox{$p,\mu$}}
\put(11,28){\mbox{$k{+}p{+}r$}}
\end{picture}} 
\quad + \quad 
\parbox{40mm}{\begin{picture}(40,30)
\put(1,2){\begin{fmfgraph}(40,30)
\fmfright{r2,r1}
\fmfleft{l}
\fmf{phantom}{i2,r2}
\fmf{fermion}{i2,r1}
\fmf{photon,tension=2}{l,i2}
\fmf{phantom_arrow,tension=2}{l,i2}
\fmffreeze
\fmf{fermion,tension=2}{r2,i1}
\fmf{fermion}{i1,i2}
\fmffreeze
\fmfforce{(0.4w,0)}{i3}
\fmf{photon,left=0.5}{i3,i2}
\fmf{phantom_arrow,left=0.5}{i3,i2}
\fmf{photon,right=0.5}{i3,i1}
\fmf{phantom_arrow,right=0.5}{i3,i1}
\fmfdot{i2}
\end{fmfgraph}}
\put(32,7){\mbox{$r$}}
\put(28,0){\mbox{$k,\sigma$}}
\put(0,11){\mbox{$-k,\rho$}}
\put(4,19){\mbox{$p,\mu$}}
\put(23,12){\mbox{$k{+}r$}}
\end{picture}}
\right)_{\!\!\tfrac{1}{\varepsilon}}
\nonumber
\\*
& \qquad\qquad = \left(
\parbox{40mm}{\begin{picture}(40,20)
\put(1,6){\begin{fmfgraph}(40,10)
\fmfbottom{l,r}
\fmftop{t}
\fmf{fermion}{i1,l}
\fmf{fermion,tension=0.5}{i2,i1}
\fmf{fermion}{r,i2}
\fmffreeze
\fmf{photon,tension=0.5,right=0.5}{t,i1}
\fmf{phantom_arrow,tension=0.5,right=0.5}{t,i1}
\fmf{photon,tension=0.5,left=0.5}{t,i2}
\fmf{phantom_arrow,tension=0.5,left=0.5}{t,i2}
\fmfdot{i1}
\end{fmfgraph}}
\put(35,3){\mbox{$p{+}r$}}
\put(8,15){\mbox{$-k$}}
\put(8,10){\mbox{$\rho$}}
\put(32,10){\mbox{$\sigma$}}
\put(30,15){\mbox{$k$}}
\put(15,3){\mbox{$k{+}p{+}r$}}
\end{picture}}
\quad - \quad 
\parbox{41mm}{\begin{picture}(40,20)
\put(1,6){\begin{fmfgraph}(40,10)
\fmfbottom{l,r}
\fmftop{t}
\fmf{fermion}{i1,l}
\fmf{fermion,tension=0.5}{i2,i1}
\fmf{fermion}{r,i2}
\fmffreeze
\fmf{photon,tension=0.5,right=0.5}{t,i1}
\fmf{phantom_arrow,tension=0.5,right=0.5}{t,i1}
\fmf{photon,tension=0.5,left=0.5}{t,i2}
\fmf{phantom_arrow,tension=0.5,left=0.5}{t,i2}
\fmfdot{i1}
\end{fmfgraph}}
\put(35,3){\mbox{$r$}}
\put(8,15){\mbox{$-k$}}
\put(8,10){\mbox{$\rho$}}
\put(32,10){\mbox{$\sigma$}}
\put(30,15){\mbox{$k$}}
\put(17,3){\mbox{$k{+}r$}}
\end{picture}}
\right)_{\!\!\tfrac{1}{\varepsilon}} ~.
\label{W31}
\end{align}

In the same way one proves
\begin{align}
&  p_\mu \left( 
\parbox{40mm}{\begin{picture}(40,36)
\put(0,6){\begin{fmfgraph}(40,30)
\fmfright{r1,r2}
\fmfleft{l}
\fmf{fermion,tension=2}{r1,i1}
\fmf{fermion}{i1,i3,i2}
\fmf{fermion,tension=2}{i2,r2}
\fmf{photon,tension=1}{l,i3}
\fmf{phantom_arrow,tension=1}{l,i3}
\fmffreeze
\fmf{photon}{i4,i1}
\fmf{phantom_arrow}{i4,i1}
\fmf{photon}{i4,i2}
\fmf{phantom_arrow}{i4,i2}
\fmfdot{i1}
\end{fmfgraph}}
\put(30,5){\mbox{$r$}}
\put(31,15){\mbox{$k,\sigma$}}
\put(31,27){\mbox{$-k,\rho$}}
\put(17,11){\mbox{$k{+}r$}}
\put(4,17){\mbox{$p,\mu$}}
\put(11,28){\mbox{$k{+}p{+}r$}}
\end{picture}} 
\quad + \quad 
\parbox{40mm}{\begin{picture}(40,30)
\put(0,0){\begin{fmfgraph}(40,30)
\fmfright{r1,r2}
\fmfleft{l}
\fmf{phantom}{r2,i2}
\fmf{fermion}{r1,i2}
\fmf{photon,tension=2}{l,i2}
\fmf{phantom_arrow,tension=2}{l,i2}
\fmffreeze
\fmf{fermion,tension=2}{i1,r2}
\fmf{fermion}{i2,i1}
\fmffreeze
\fmfforce{(0.4w,h)}{i3}
\fmf{photon,left=0.5}{i3,i1}
\fmf{phantom_arrow,left=0.5}{i3,i1}
\fmf{photon,right=0.5}{i3,i2}
\fmf{phantom_arrow,right=0.5}{i3,i2}
\fmfdot{i2}
\end{fmfgraph}}
\put(32,4){\mbox{$r$}}
\put(24,31){\mbox{$-k,\rho$}}
\put(3,20){\mbox{$k,\sigma$}}
\put(4,11){\mbox{$p,\mu$}}
\put(23,18){\mbox{$k{+}p{+}r$}}
\end{picture}}
\right)_{\!\!\tfrac{1}{\varepsilon}}
\nonumber
\\*
& \qquad\qquad = \left(
\parbox{40mm}{\begin{picture}(40,20)
\put(0,6){\begin{fmfgraph}(40,10)
\fmfbottom{l,r}
\fmftop{t}
\fmf{fermion}{i1,l}
\fmf{fermion,tension=0.5}{i2,i1}
\fmf{fermion}{r,i2}
\fmffreeze
\fmf{photon,tension=0.5,right=0.5}{t,i1}
\fmf{phantom_arrow,tension=0.5,right=0.5}{t,i1}
\fmf{photon,tension=0.5,left=0.5}{t,i2}
\fmf{phantom_arrow,tension=0.5,left=0.5}{t,i2}
\fmfdot{i2}
\end{fmfgraph}}
\put(34,3){\mbox{$p{+}r$}}
\put(7,15){\mbox{$-k$}}
\put(7,10){\mbox{$\rho$}}
\put(31,10){\mbox{$\sigma$}}
\put(29,15){\mbox{$k$}}
\put(14,3){\mbox{$k{+}p{+}r$}}
\end{picture}}
\quad - \quad 
\parbox{40mm}{\begin{picture}(40,20)
\put(0,6){\begin{fmfgraph}(40,10)
\fmfbottom{l,r}
\fmftop{t}
\fmf{fermion}{i1,l}
\fmf{fermion,tension=0.5}{i2,i1}
\fmf{fermion}{r,i2}
\fmffreeze
\fmf{photon,tension=0.5,right=0.5}{t,i1}
\fmf{phantom_arrow,tension=0.5,right=0.5}{t,i1}
\fmf{photon,tension=0.5,left=0.5}{t,i2}
\fmf{phantom_arrow,tension=0.5,left=0.5}{t,i2}
\fmfdot{i2}
\end{fmfgraph}}
\put(34,3){\mbox{$r$}}
\put(7,15){\mbox{$-k$}}
\put(7,10){\mbox{$\rho$}}
\put(31,10){\mbox{$\sigma$}}
\put(29,15){\mbox{$k$}}
\put(16,3){\mbox{$k{+}r$}}
\end{picture}}
\right)_{\!\!\tfrac{1}{\varepsilon}}~,
\label{W32}
\\
&  p_\mu \left( 
\parbox{40mm}{\begin{picture}(40,36)
\put(0,6){\begin{fmfgraph}(40,30)
\fmfkeep{AFFG}
\fmfright{r1,r2}
\fmfleft{l}
\fmf{fermion,tension=2}{r1,i1}
\fmf{fermion}{i1,i3}
\fmf{phantom}{i3,i2}
\fmf{fermion,tension=2}{i2,r2}
\fmf{photon,tension=1}{l,i3}
\fmf{phantom_arrow,tension=1}{l,i3}
\fmffreeze
\fmf{fermion}{i3,i5,i2}
\fmf{photon}{i4,i1}
\fmf{phantom_arrow}{i4,i1}
\fmf{photon}{i4,i2}
\fmf{phantom_arrow}{i4,i2}
\fmfdot{i5}
\end{fmfgraph}}
\put(30,5){\mbox{$r$}}
\put(31,15){\mbox{$k,\sigma$}}
\put(31,27){\mbox{$-k,\rho$}}
\put(17,11){\mbox{$k{+}r$}}
\put(4,17){\mbox{$p,\mu$}}
\put(11,28){\mbox{$k{+}p{+}r$}}
\end{picture}} 
{}~+~
\parbox{40mm}{\begin{picture}(40,36)
\put(0,6){\begin{fmfgraph}(40,30)
\fmfright{r1,r2}
\fmfleft{l}
\fmf{fermion,tension=2}{r1,i1}
\fmf{fermion}{i1,i3,i2}
\fmf{fermion,tension=2}{i2,r2}
\fmf{photon,tension=1}{l,i3}
\fmf{phantom_arrow,tension=1}{l,i3}
\fmffreeze
\fmf{photon}{i4,i1}
\fmf{phantom_arrow}{i4,i1}
\fmf{photon}{i4,i2}
\fmf{phantom_arrow}{i4,i2}
\fmfdot{i3}
\end{fmfgraph}}
\put(30,5){\mbox{$r$}}
\put(31,15){\mbox{$k,\sigma$}}
\put(31,27){\mbox{$-k,\rho$}}
\put(17,11){\mbox{$k{+}r$}}
\put(4,17){\mbox{$p,\mu$}}
\put(11,28){\mbox{$k{+}p{+}r$}}
\end{picture}} 
{}~+~
\parbox{40mm}{\begin{picture}(40,36)
\put(0,6){\begin{fmfgraph}(40,30)
\fmfkeep{AFFH}
\fmfright{r1,r2}
\fmfleft{l}
\fmf{fermion,tension=2}{r1,i1}
\fmf{fermion}{i3,i2}
\fmf{phantom}{i1,i3}
\fmf{fermion,tension=2}{i2,r2}
\fmf{photon,tension=1}{l,i3}
\fmf{phantom_arrow,tension=1}{l,i3}
\fmffreeze
\fmf{fermion}{i1,i5,i3}
\fmf{photon}{i4,i1}
\fmf{phantom_arrow}{i4,i1}
\fmf{photon}{i4,i2}
\fmf{phantom_arrow}{i4,i2}
\fmfdot{i5}
\end{fmfgraph}}
\put(30,5){\mbox{$r$}}
\put(31,15){\mbox{$k,\sigma$}}
\put(31,27){\mbox{$-k,\rho$}}
\put(17,11){\mbox{$k{+}r$}}
\put(4,17){\mbox{$p,\mu$}}
\put(11,28){\mbox{$k{+}p{+}r$}}
\end{picture}} 
\right)_{\!\!\tfrac{1}{\varepsilon}}
\nonumber
\\*
& \qquad\qquad = \left(
\parbox{40mm}{\begin{picture}(40,20)
\put(0,6){\begin{fmfgraph}(40,10)
\fmfbottom{l,r}
\fmftop{t}
\fmf{fermion}{i1,l}
\fmf{fermion,tension=1}{i2,i3,i1}
\fmf{fermion}{r,i2}
\fmffreeze
\fmf{photon,tension=0.5,right=0.5}{t,i1}
\fmf{phantom_arrow,tension=0.5,right=0.5}{t,i1}
\fmf{photon,tension=0.5,left=0.5}{t,i2}
\fmf{phantom_arrow,tension=0.5,left=0.5}{t,i2}
\fmfdot{i3}
\end{fmfgraph}}
\put(34,3){\mbox{$p{+}r$}}
\put(7,15){\mbox{$-k$}}
\put(7,10){\mbox{$\rho$}}
\put(31,10){\mbox{$\sigma$}}
\put(29,15){\mbox{$k$}}
\put(14,3){\mbox{$k{+}p{+}r$}}
\end{picture}}
\quad - \quad 
\parbox{40mm}{\begin{picture}(40,20)
\put(0,6){\begin{fmfgraph}(40,10)
\fmfbottom{l,r}
\fmftop{t}
\fmf{fermion}{i1,l}
\fmf{fermion,tension=1}{i2,i3,i1}
\fmf{fermion}{r,i2}
\fmffreeze
\fmf{photon,tension=0.5,right=0.5}{t,i1}
\fmf{phantom_arrow,tension=0.5,right=0.5}{t,i1}
\fmf{photon,tension=0.5,left=0.5}{t,i2}
\fmf{phantom_arrow,tension=0.5,left=0.5}{t,i2}
\fmfdot{i3}
\end{fmfgraph}}
\put(34,3){\mbox{$r$}}
\put(7,15){\mbox{$-k$}}
\put(7,10){\mbox{$\rho$}}
\put(32,10){\mbox{$\sigma$}}
\put(29,15){\mbox{$k$}}
\put(16,3){\mbox{$k{+}r$}}
\end{picture}}
\right)_{\!\!\tfrac{1}{\varepsilon}}~,
\label{W33}
\\
&  p_\mu \left( 
\parbox{40mm}{\begin{picture}(40,38)
\put(0,6){\begin{fmfgraph}(40,30)
\fmfright{r1,r2}
\fmfleft{l}
\fmfforce{((0.3w,h))}{i2}
\fmf{fermion,tension=1}{r1,i1}
\fmf{fermion,tension=1}{i1,r2}
\fmf{photon,tension=1}{l,i1}
\fmf{phantom_arrow,tension=1}{l,i1}
\fmffreeze
\fmf{photon,left=0.65}{i2,i1}
\fmf{photon,right=0.65}{i2,i1}
\fmf{phantom_arrow,left=0.65}{i2,i1}
\fmf{phantom_arrow,right=0.65}{i2,i1}
\fmfdot{i1}
\end{fmfgraph}}
\put(30,14){\mbox{$r$}}
\put(4,17){\mbox{$p,\mu$}}
\put(3,30){\mbox{$-k$}}
\put(21,32){\mbox{$k$}}
\end{picture}}
\right)_{\!\!\tfrac{1}{\varepsilon}}
\nonumber
\\*
& \qquad\qquad = \left(
\parbox{40mm}{\begin{picture}(40,20)
\put(0,6){\begin{fmfgraph}(40,15)
\fmfbottom{l,r}
\fmftop{t}
\fmf{fermion}{i,l}
\fmf{fermion}{r,i}
\fmffreeze
\fmf{photon,tension=0.5,right=0.8}{t,i}
\fmf{phantom_arrow,tension=0.5,right=0.8}{t,i}
\fmf{photon,tension=0.5,left=0.8}{t,i}
\fmf{phantom_arrow,tension=0.5,left=0.8}{t,i}
\fmfdot{i}
\end{fmfgraph}}
\put(29,3){\mbox{$p{+}r$}}
\put(11,15){\mbox{$k$}}
\put(12,10){\mbox{$\rho$}}
\put(27,10){\mbox{$\sigma$}}
\put(27,15){\mbox{$-k$}}
\end{picture}}
{}~-~
\parbox{40mm}{\begin{picture}(40,20)
\put(0,6){\begin{fmfgraph}(40,15)
\fmfbottom{l,r}
\fmftop{t}
\fmf{fermion}{i,l}
\fmf{fermion}{r,i}
\fmffreeze
\fmf{photon,tension=0.5,right=0.8}{t,i}
\fmf{phantom_arrow,tension=0.5,right=0.8}{t,i}
\fmf{photon,tension=0.5,left=0.8}{t,i}
\fmf{phantom_arrow,tension=0.5,left=0.8}{t,i}
\fmfdot{i}
\end{fmfgraph}}
\put(34,3){\mbox{$r$}}
\put(11,15){\mbox{$k$}}
\put(12,10){\mbox{$\rho$}}
\put(27,10){\mbox{$\sigma$}}
\put(27,15){\mbox{$-k$}}
\end{picture}}
\right)_{\!\!\tfrac{1}{\varepsilon}}~,
\label{W34}
\\
& p_\mu \left(
\parbox{44mm}{~~\begin{picture}(40,30)
\put(-1,0){\begin{fmfgraph}(40,30)
\fmfright{r1,r2}
\fmfleft{l}
\fmf{fermion,tension=2}{r1,i1}
\fmf{photon}{i4,i1}
\fmf{phantom_arrow,tension=0}{i4,i1}
\fmf{photon}{i4,i3}
\fmf{phantom_arrow,tension=0}{i4,i3}
\fmf{photon}{i5,i2}
\fmf{phantom_arrow,tension=0}{i5,i2}
\fmf{photon}{i5,i3}
\fmf{phantom_arrow,tension=0}{i5,i3}
\fmf{fermion,tension=2}{i2,r2}
\fmf{photon,tension=1}{l,i3}
\fmf{phantom_arrow,tension=1}{l,i3}
\fmffreeze
\fmf{fermion}{i1,i2}
\fmfdot{i3}
\end{fmfgraph}}
\put(33,3){\mbox{$r$}}
\put(32,15){\mbox{$-k$}}
\put(7,6){\mbox{$k{+}r,\nu$}}
\put(12,1){\mbox{$-k{-}r,\rho$}}
\put(2,11){\mbox{$p,\mu$}}
\put(10,26){\mbox{$k{+}p{+}r,\tau$}}
\put(-4,20){\mbox{$-k{-}p{-}r,\sigma$}}
\end{picture}}
\right)_{\!\!\frac{1}{\varepsilon}}
=0~.
\label{W35}
\end{align}
\end{subequations}
The four graphs in (\ref{ffp1}) which involve the four-fermion vertex turn out
to be transversal (contraction with $p_\mu$ yields zero). That zero can
formally be written as the difference of the corresponding graphs in
(\ref{fs1}) containing the four-fermion vertex, with external momenta $p+r$
and $r$, respectively, because the singular part of these graphs is
independent of the external momentum. Thus, (\ref{W3}) proves the
Ward identity (\ref{WI1}) for $(\tau,\ell)=(1,1)$. We stress that this proof
only uses the possibility of a naive factorization (\ref{anreg}) of common
terms in numerator and denominator of the integrand, valid for analytic
regularization. It is not necessary to evaluate the divergent integrals.

In the same way we can prove the Ward identities (\ref{WI2}) and
(\ref{WI3}) without computing the (already very complicated) divergent
integrals. For instance, we have

\begin{subequations}
\begin{align}
&p_\mu \left(
\parbox{42mm}{\begin{picture}(40,28)
\put(0,0){\begin{fmfgraph}(40,25)
\fmfkeep{AAFFA}
\fmfright{r1,r2}
\fmfleft{l1,l2}
\fmf{fermion,tension=2}{r1,i1}
\fmf{fermion,tension=2}{i4,r2}
\fmf{fermion,tension=1}{i1,i2,i3,i4}
\fmf{phantom,tension=1}{i1,i4}
\fmf{photon,tension=1}{l1,i2}
\fmf{phantom_arrow,tension=1}{l1,i2}
\fmf{photon,tension=1}{l2,i3}
\fmf{phantom_arrow,tension=1}{l2,i3}
\fmffreeze
\fmf{photon,tension=1}{i5,i1}
\fmf{phantom_arrow,tension=0}{i5,i1}
\fmf{photon,tension=1}{i5,i4}
\fmf{phantom_arrow,tension=0}{i5,i4}
\fmfdot{i1}
\end{fmfgraph}}
\put(1,4){\mbox{$q,\nu$}}
\put(1,19){\mbox{$p,\mu$}}
\put(29,0){\mbox{$s$}}
\put(28,24){\mbox{$-r$}}
\put(18,2){\mbox{$k$}}
\put(13,21){\mbox{$k{+}p{+}q$}}
\put(30,15){\mbox{$s{-}k,\rho$}}
\put(30,8){\mbox{$k{-}s,\sigma$}}
\put(3,12){\mbox{$k{+}q$}}
\end{picture}}
{}~+~ 
\parbox{42mm}{\begin{picture}(40,28)
\put(-1,0){\fmfreuse{AAFFA}}
\put(1,4){\mbox{$p,\mu$}}
\put(1,19){\mbox{$q,\nu$}}
\put(29,0){\mbox{$s$}}
\put(28,24){\mbox{$-r$}}
\put(18,2){\mbox{$k$}}
\put(13,21){\mbox{$k{+}p{+}q$}}
\put(30,15){\mbox{$s{-}k,\rho$}}
\put(30,8){\mbox{$k{-}s,\sigma$}}
\put(3,12){\mbox{$k{+}p$}}
\end{picture}}
{}~+
\parbox{41mm}{\begin{picture}(40,28)
\put(0,0){\begin{fmfgraph}(40,25)
\fmfright{r1,r2}
\fmfleft{l1,l2}
\fmf{fermion,tension=3}{r1,i1}
\fmf{fermion,tension=3}{i3,r2}
\fmf{fermion,tension=1}{i1,i2,i3}
\fmf{phantom,tension=1}{i1,i3}
\fmf{photon,tension=2}{l1,i1}
\fmf{phantom_arrow,tension=1}{l1,i1}
\fmf{photon,tension=2}{l2,i2}
\fmf{phantom_arrow,tension=1}{l2,i2}
\fmffreeze
\fmf{photon,tension=1}{i5,i1}
\fmf{phantom_arrow,tension=0}{i5,i1}
\fmf{photon,tension=1}{i5,i3}
\fmf{phantom_arrow,tension=0}{i5,i3}
\fmfdot{i1}
\end{fmfgraph}}
\put(1,2){\mbox{$p,\mu$}}
\put(1,19){\mbox{$q,\nu$}}
\put(32,3){\mbox{$s$}}
\put(33,21){\mbox{$-r$}}
\put(13,22){\mbox{$k{+}p{+}q$}}
\put(28,15){\mbox{$s{-}k,\rho$}}
\put(24,8){\mbox{$k{-}s,\sigma$}}
\put(7,12){\mbox{$k{+}p$}}
\end{picture}} \right)_{\!\!\tfrac{1}{\varepsilon}}
\nonumber
\\*
& \qquad\qquad = 
\left(
\parbox{42mm}{\begin{picture}(40,36)
\put(0,6){\begin{fmfgraph}(40,30)
\fmfright{r1,r2}
\fmfleft{l}
\fmf{fermion,tension=2}{r1,i1}
\fmf{fermion}{i1,i3,i2}
\fmf{fermion,tension=2}{i2,r2}
\fmf{photon,tension=1}{l,i3}
\fmf{phantom_arrow,tension=1}{l,i3}
\fmffreeze
\fmf{photon}{i4,i1}
\fmf{phantom_arrow}{i4,i1}
\fmf{photon}{i4,i2}
\fmf{phantom_arrow}{i4,i2}
\fmfdot{i1}
\end{fmfgraph}}
\put(36,9){\mbox{$p{+}s$}}
\put(36,33){\mbox{$-r$}}
\put(31,15){\mbox{$k{-}s,\sigma$}}
\put(31,27){\mbox{$s{-}k,\rho$}}
\put(17,11){\mbox{$k{+}p$}}
\put(4,17){\mbox{$q,\nu$}}
\put(10,28){\mbox{$k{+}p{+}q$}}
\end{picture}} 
\quad - \quad 
\parbox{47mm}{\begin{picture}(40,36)
\put(0,6){\begin{fmfgraph}(40,30)
\fmfright{r1,r2}
\fmfleft{l}
\fmf{fermion,tension=2}{r1,i1}
\fmf{fermion}{i1,i3,i2}
\fmf{fermion,tension=2}{i2,r2}
\fmf{photon,tension=1}{l,i3}
\fmf{phantom_arrow,tension=1}{l,i3}
\fmffreeze
\fmf{photon}{i4,i1}
\fmf{phantom_arrow}{i4,i1}
\fmf{photon}{i4,i2}
\fmf{phantom_arrow}{i4,i2}
\fmfdot{i1}
\end{fmfgraph}}
\put(36,9){\mbox{$s$}}
\put(36,33){\mbox{$-r{-}p$}}
\put(31,15){\mbox{$k{-}s,\sigma$}}
\put(31,27){\mbox{$s{-}k,\rho$}}
\put(21,11){\mbox{$k$}}
\put(4,17){\mbox{$q,\nu$}}
\put(17,28){\mbox{$k{+}q$}}
\end{picture}} 
\right)_{\!\!\tfrac{1}{\varepsilon}}~,
\\
&p_\mu \left(
\parbox{48mm}{\hspace*{-2mm}\begin{picture}(48,28)
\put(0,0){\begin{fmfgraph}(48,25)
\fmfright{r1,r2}
\fmfleft{l1,l2}
\fmf{fermion,tension=3}{r1,i1}
\fmf{fermion,tension=3}{i3,l2}
\fmf{fermion,tension=1}{i1,i2,i3}
\fmf{photon,tension=2}{l1,i2}
\fmf{phantom_arrow,tension=1}{l1,i2}
\fmf{photon,tension=2}{r2,i4}
\fmf{phantom_arrow,tension=1}{r2,i4}
\fmf{phantom,tension=1}{i3,i4,i1}
\fmffreeze
\fmf{photon,tension=1}{i5,i1}
\fmf{phantom_arrow,tension=0}{i5,i1}
\fmf{photon,tension=1}{i5,i4}
\fmf{phantom_arrow,tension=0}{i5,i4}
\fmf{photon,tension=1}{i6,i3}
\fmf{phantom_arrow,tension=0}{i6,i3}
\fmf{photon,tension=1}{i6,i4}
\fmf{phantom_arrow,tension=0}{i6,i4}
\fmfdot{i4}
\end{fmfgraph}}
\put(2,5){\mbox{$p,\mu$}}
\put(2,19){\mbox{$-r$}}
\put(39,4){\mbox{$s$}}
\put(43,21){\mbox{$q,\nu$}}
\put(23,1){\mbox{$-k$}}
\put(16,22){\mbox{$-k{-}q{-}s,\sigma$}}
\put(38,15){\mbox{$k{+}s,\rho$}}
\put(3,12){\mbox{$p{-}k$}}
\end{picture}} \right)_{\!\!\tfrac{1}{\varepsilon}}
\nonumber
\\*
&\qquad\qquad
= \left( 
\parbox{44mm}{~~\begin{picture}(40,30)
\put(-1,0){\begin{fmfgraph}(40,30)
\fmfright{r1,r2}
\fmfleft{l}
\fmf{fermion,tension=2}{r1,i1}
\fmf{photon}{i4,i1}
\fmf{phantom_arrow,tension=0}{i4,i1}
\fmf{photon}{i4,i3}
\fmf{phantom_arrow,tension=0}{i4,i3}
\fmf{photon}{i5,i2}
\fmf{phantom_arrow,tension=0}{i5,i2}
\fmf{photon}{i5,i3}
\fmf{phantom_arrow,tension=0}{i5,i3}
\fmf{fermion,tension=2}{i2,r2}
\fmf{photon,tension=1}{l,i3}
\fmf{phantom_arrow,tension=1}{l,i3}
\fmffreeze
\fmf{fermion}{i1,i2}
\fmfdot{i3}
\end{fmfgraph}}
\put(34,3){\mbox{$p{+}s$}}
\put(34,26){\mbox{$-r$}}
\put(32,15){\mbox{$p{-}k$}}
\put(7,6){\mbox{$k{+}s,\rho$}}
\put(2,11){\mbox{$q,\nu$}}
\put(-4,20){\mbox{$-k{-}q{-}s,\sigma$}}
\end{picture}}
{}~-~
\parbox{48mm}{~~\begin{picture}(40,30)
\put(-1,0){\begin{fmfgraph}(40,30)
\fmfkeep{AFFD}
\fmfright{r1,r2}
\fmfleft{l}
\fmf{fermion,tension=2}{r1,i1}
\fmf{photon}{i4,i1}
\fmf{phantom_arrow,tension=0}{i4,i1}
\fmf{photon}{i4,i3}
\fmf{phantom_arrow,tension=0}{i4,i3}
\fmf{photon}{i5,i2}
\fmf{phantom_arrow,tension=0}{i5,i2}
\fmf{photon}{i5,i3}
\fmf{phantom_arrow,tension=0}{i5,i3}
\fmf{fermion,tension=2}{i2,r2}
\fmf{photon,tension=1}{l,i3}
\fmf{phantom_arrow,tension=1}{l,i3}
\fmffreeze
\fmf{fermion}{i1,i2}
\fmfdot{i3}
\end{fmfgraph}}
\put(34,3){\mbox{$s$}}
\put(34,26){\mbox{$-r{-}p$}}
\put(32,15){\mbox{$-k$}}
\put(7,6){\mbox{$k{+}s,\rho$}}
\put(2,11){\mbox{$q,\nu$}}
\put(-4,20){\mbox{$-k{-}q{-}s,\sigma$}}
\end{picture}}
\right)_{\!\!\tfrac{1}{\varepsilon}}~.
\end{align}
\end{subequations}
Thus, the evaluation of the divergent part of 
$\Gamma^{(1,1)\mu\nu}_{AA\bar{\psi}\psi}(p,q,r,s)$ and 
$\Gamma^{(1,1)\mu\nu\rho}_{AAA\bar{\psi}\psi}(p,q,r,s,t)$ is
compatible with the $\hbar$-renormalizations of $\kappa_i$ and the
additional counterterms required by (\ref{ffp1b}). 

\enlargethispage{5mm}
Let us finally show (\ref{WI0}) for $n=2$, the generalization to
higher $n$ being obvious. We have 
\begin{subequations}
\begin{align}
&p_\mu \left(
- \hspace*{-1.5em} 
\parbox{40mm}{\begin{picture}(40,30)
\put(0,0){\begin{fmfgraph}(40,30)
\fmfsurround{l1,l2,l3}
\fmf{photon,tension=2}{l1,i1}
\fmf{phantom_arrow,tension=1}{l1,i1}
\fmf{photon,tension=2}{l2,i2}
\fmf{phantom_arrow,tension=1}{l2,i2}
\fmf{photon,tension=2}{l3,i3}
\fmf{phantom_arrow,tension=1}{l3,i3}
\fmf{fermion,left=0.7}{i1,i3}
\fmf{fermion,left=0.4}{i3,i2}
\fmf{phantom,left=0.7}{i2,i1}
\fmffreeze
\fmfforce{(0.6w,0.78h)}{i4}
\fmf{fermion,left=0.3}{i2,i4}
\fmf{fermion,left=0.3}{i4,i1}
\fmfdot{i4}
\end{fmfgraph}}
\put(35,17){\mbox{$p,\mu$}}
\put(12,1){\mbox{$q,\nu$}}
\put(6,22){\mbox{$r,\rho$}}
\put(19,26){\mbox{$k{+}r$}}
\put(8,14){\mbox{$k$}}
\put(26,3){\mbox{$k{+}p{+}r$}}
\end{picture}} 
{} ~- 
\parbox{40mm}{\begin{picture}(40,30)
\put(0,0){\begin{fmfgraph}(40,30)
\fmfsurround{l1,l2,l3}
\fmf{photon,tension=2}{l1,i1}
\fmf{phantom_arrow,tension=1}{l1,i1}
\fmf{photon,tension=2}{l2,i2}
\fmf{phantom_arrow,tension=1}{l2,i2}
\fmf{photon,tension=2}{l3,i3}
\fmf{phantom_arrow,tension=1}{l3,i3}
\fmf{fermion,left=0.7}{i1,i3}
\fmf{fermion,left=0.4}{i3,i2}
\fmf{phantom,left=0.7}{i2,i1}
\fmffreeze
\fmfforce{(0.6w,0.78h)}{i4}
\fmf{fermion,left=0.3}{i2,i4}
\fmf{fermion,left=0.3}{i4,i1}
\fmfdot{i4}
\end{fmfgraph}}
\put(35,17){\mbox{$q,\nu$}}
\put(12,1){\mbox{$r,\rho$}}
\put(6,22){\mbox{$p,\mu$}}
\put(19,26){\mbox{$k{+}p{+}r$}}
\put(3,14){\mbox{$k{+}r$}}
\put(26,3){\mbox{$k$}}
\end{picture}} 
{}~ - 
\parbox{40mm}{\begin{picture}(40,30)
\put(0,0){\begin{fmfgraph}(40,30)
\fmfsurround{l1,l2,l3}
\fmf{photon,tension=2}{l1,i1}
\fmf{phantom_arrow,tension=1}{l1,i1}
\fmf{photon,tension=2}{l2,i2}
\fmf{phantom_arrow,tension=1}{l2,i2}
\fmf{photon,tension=2}{l3,i3}
\fmf{phantom_arrow,tension=1}{l3,i3}
\fmf{fermion,left=0.7}{i1,i3}
\fmf{fermion,left=0.4}{i3,i2}
\fmf{phantom,left=0.7}{i2,i1}
\fmffreeze
\fmfforce{(0.6w,0.78h)}{i4}
\fmf{fermion,left=0.3}{i2,i4}
\fmf{fermion,left=0.3}{i4,i1}
\fmfdot{i4}
\end{fmfgraph}}
\put(35,17){\mbox{$q,\nu$}}
\put(12,1){\mbox{$p,\mu$}}
\put(6,22){\mbox{$r,\rho$}}
\put(19,26){\mbox{$k{+}r{+}p$}}
\put(3,14){\mbox{$k{+}p$}}
\put(26,3){\mbox{$k$}}
\end{picture}} 
\right.
\nonumber
\\*
& \hspace*{20em} \left.   
- \hspace*{-1em} 
\parbox{40mm}{\begin{picture}(40,30)
\put(0,0){\begin{fmfgraph}(40,30)
\fmfsurround{l1,l2,l3}
\fmf{photon,tension=2}{l1,i1}
\fmf{phantom_arrow,tension=1}{l1,i1}
\fmf{photon,tension=2}{l2,i2}
\fmf{phantom_arrow,tension=1}{l2,i2}
\fmf{photon,tension=2}{l3,i3}
\fmf{phantom_arrow,tension=1}{l3,i3}
\fmf{fermion,left=0.7}{i1,i3}
\fmf{fermion,left=0.4}{i3,i2}
\fmf{fermion,left=0.7}{i2,i1}
\fmfdot{i1}
\end{fmfgraph}}
\put(35,17){\mbox{$p,\mu$}}
\put(12,1){\mbox{$q,\nu$}}
\put(6,22){\mbox{$r,\rho$}}
\put(19,26){\mbox{$k{+}r$}}
\put(8,14){\mbox{$k$}}
\put(26,3){\mbox{$k{+}p{+}r$}}
\end{picture}} 
\right)_{\!\!\tfrac{1}{\varepsilon}} \!\!\!
=  0\;,
\label{3P1}
\\
& p_\mu \left( - 
\parbox{40mm}{\begin{picture}(40,30)
\put(0,0){\begin{fmfgraph}(40,30)
\fmfsurround{l1,l2,l3}
\fmf{photon,tension=2}{l1,i1}
\fmf{phantom_arrow,tension=1}{l1,i1}
\fmf{photon,tension=2}{l2,i2}
\fmf{phantom_arrow,tension=1}{l2,i2}
\fmf{photon,tension=2}{l3,i3}
\fmf{phantom_arrow,tension=1}{l3,i3}
\fmf{fermion,left=0.7}{i1,i3}
\fmf{fermion,left=0.4}{i3,i2}
\fmf{fermion,left=0.7}{i2,i1}
\fmfdot{i1}
\end{fmfgraph}}
\put(35,17){\mbox{$q,\nu$}}
\put(12,1){\mbox{$r,\rho$}}
\put(6,22){\mbox{$p,\mu$}}
\put(19,26){\mbox{$k{+}p{+}r$}}
\put(3,14){\mbox{$k{+}r$}}
\put(26,3){\mbox{$k$}}
\end{picture}}
{}~-
\parbox{40mm}{\begin{picture}(40,30)
\put(0,0){\begin{fmfgraph}(40,30)
\fmfsurround{l1,l2,l3}
\fmf{photon,tension=2}{l1,i1}
\fmf{phantom_arrow,tension=1}{l1,i1}
\fmf{photon,tension=2}{l2,i2}
\fmf{phantom_arrow,tension=1}{l2,i2}
\fmf{photon,tension=2}{l3,i3}
\fmf{phantom_arrow,tension=1}{l3,i3}
\fmf{fermion,left=0.7}{i1,i3}
\fmf{fermion,left=0.4}{i3,i2}
\fmf{fermion,left=0.7}{i2,i1}
\fmfdot{i1}
\end{fmfgraph}}
\put(35,17){\mbox{$q,\nu$}}
\put(12,1){\mbox{$p,\mu$}}
\put(6,22){\mbox{$r,\rho$}}
\put(19,26){\mbox{$k{+}p{+}r$}}
\put(3,14){\mbox{$k{+}p$}}
\put(26,3){\mbox{$k$}}
\end{picture}}
~- \hspace*{-1.5em} 
\parbox{40mm}{\begin{picture}(40,30)
\put(0,0){\begin{fmfgraph}(40,30)
\fmfsurround{l1,l2,l3}
\fmf{photon,tension=2}{l1,i2}
\fmf{phantom_arrow,tension=2}{l1,i2}
\fmf{photon,tension=2}{l3,i1}
\fmf{phantom_arrow,tension=2}{l3,i1}
\fmf{photon,tension=2}{l2,i1}
\fmf{phantom_arrow,tension=2}{l2,i1}
\fmf{fermion,left=1}{i1,i2}
\fmf{fermion,left=1}{i2,i1}
\fmfdot{i1}
\end{fmfgraph}}
\put(35,17){\mbox{$r,\rho$}}
\put(12,1){\mbox{$q,\nu$}}
\put(4,22){\mbox{$p,\mu$}}
\put(19,26){\mbox{$k$}}
\put(26,3){\mbox{$k{+}r$}}
\end{picture}} 
\right)_{\!\!\tfrac{1}{\varepsilon}} \!\!\!
=  0\;,
\label{3P2}
\\
& p_\mu \left( - ~
\parbox{40mm}{\begin{picture}(40,30)
\put(0,0){\begin{fmfgraph}(40,30)
\fmfsurround{l1,l2,l3}
\fmf{photon,tension=2}{l1,i2}
\fmf{phantom_arrow,tension=2}{l1,i2}
\fmf{photon,tension=2}{l3,i1}
\fmf{phantom_arrow,tension=2}{l3,i1}
\fmf{photon,tension=2}{l2,i1}
\fmf{phantom_arrow,tension=2}{l2,i1}
\fmf{fermion,left=1}{i1,i2}
\fmf{fermion,left=1}{i2,i1}
\fmfdot{i1}
\end{fmfgraph}}
\put(35,17){\mbox{$p,\mu$}}
\put(12,1){\mbox{$q,\nu$}}
\put(4,22){\mbox{$r,\rho$}}
\put(19,26){\mbox{$k$}}
\put(26,3){\mbox{$k{+}p$}}
\end{picture}} 
\quad -  \hspace*{-1em} 
\parbox{42mm}{\begin{picture}(40,30)
\put(0,0){\begin{fmfgraph}(40,30)
\fmfsurround{l1,l2,l3}
\fmf{photon,tension=2}{l1,i1}
\fmf{phantom_arrow,tension=2}{l1,i1}
\fmf{photon,tension=2}{l3,i1}
\fmf{phantom_arrow,tension=2}{l3,i1}
\fmf{photon,tension=2}{l2,i1}
\fmf{phantom_arrow,tension=2}{l2,i1}
\fmffreeze
\fmf{fermion,left}{i1,i1}
\fmfdot{i1}
\end{fmfgraph}}
\put(33,17){\mbox{$p,\mu$}}
\put(12,1){\mbox{$q,\nu$}}
\put(4,22){\mbox{$r,\rho$}}
\put(28,4){\mbox{$k$}}
\end{picture}}
\right)_{\!\!\tfrac{1}{\varepsilon}} \!\!\! =0\,.
\end{align}
\end{subequations}
By exchange $(q,\nu)\leftrightarrow(r,\rho)$ in (\ref{3P1}) and
(\ref{3P2}), all graphs contributing to (\ref{ppp1}) are obtained, which
proves (\ref{WI0}) for $n=2$.

Proceeding analogously one proves (\ref{WI0}) for $n\in\{3,4,5\}$. But
this means that the coefficient of $\frac{1}{\varepsilon}$ in analytic
regularization of the 1-loop Green's functions
$\Gamma^{(1,1)\mu_0\dots\mu_n}_{A\dots A}(p_0,\dots ,p_n)$ is the
Fourier-transformed of a gauge-invariant local field polynomial
\begin{align}
\sum \int d^4x \prod_{i=1}^{n+1} d^4x_i \; 
\theta \;(\partial\dots \partial F)(x_1) \dots 
(\partial\dots \partial F)(x_{n+1}) \;\prod_{j=1}^{n+1} 
\delta^4(x{-}x_j)~,
\end{align}
for an appropriate contraction of Lorentz indices. On the other hand, the
integral has to be of power-counting dimension zero (see footnote \ref{foot}),
which cannot be achieved for $n>2$. This means
\begin{align}
\Gamma^{(1,1)\mu_0\dots\mu_n}_{A\dots A}(p_0,\dots ,p_n) =
\mathcal{O}(1) \qquad \text{for}\quad n > 2~.
\end{align}
Individual graphs contributing to
$\Gamma^{(1,1)\mu_0\dots\mu_n}_{A\dots A}(p_0,\dots ,p_n)$ for $n\in
\{3,4,5\}$ will be divergent, of course.

\section{Discussion}

We have computed or derived all divergent one-loop corrections to Green's
functions of $\theta$-expanded noncommutative QED, up to first order in
$\theta$. Let us summarize the results:
\begin{itemize}
\item[1)] Taking the $\theta$-expansion serious, the model is obviously
  \emph{not renormalizable}. This is the death of all attempts to avoid
  considering the full noncommutative quantum field theory by Seiberg-Witten
  expansion\footnote{It could still be meaningful to consider
    $\theta$-expansions of noncommutative field theories as effective
    actions.}. 
  
  The problem is first of all due to the divergence of the fermion four-point
  function. From a mathematically appealing point of view, fermions are
  elements of an inner product space. Therefore, any \emph{local} field
  monomial can never contain more than two fermion fields, which means that
  divergences in graphs with more than two external fermions cannot be
  renormalized. At order $T$ in $\theta$, graphs with $N_F$ external fermion
  lines are divergent for $3N_F \leq 8{+}4T$. There is no reason why this
  infinite number of divergences could cancel for a quantum field theoretical
  model with a finite number of fields. Anyway, the inclusion of (on
  noncommutative level non-local) fermion number-changing field redefinitions
  does not yield a renormalizable quantum field theory either.

\item[2)] Let us `solve' the above problem in graphs with more than two
  external electrons by ignoring it (to be made precise below). Then
  $\theta$-expanded noncommutative QED is not renormalizable if the
  electrons are massive, with the mass term appearing explicitly in
  the noncommutative Dirac action. This does not exclude a fermion
  mass coming from a Higgs mechanism. It would therefore be important
  to study an Abelian Higgs model.
  
\item[3)] Let us therefore consider $\theta$-expanded \emph{massless}
  noncommutative QED with the divergence in graphs with $n>2$ external
  electrons being ignored. We have proved that in this case our model
  (\ref{YMD}) is multiplicatively one-loop renormalizable---including field
  redefinitions---up to first order in $\theta$. This 
  $\hbar$-dependence of the parameters of the model is given by
  (\ref{hbarren}) and 
\begin{subequations}
\begin{align}
\kappa_1 &= \kappa_{1,0} - \frac{\hbar g_0^2}{(4\pi)^2 \varepsilon}
\Big(
-\frac{13}{48} 
- \frac{1}{6} \kappa_{1,0} + \frac{1}{2} \kappa_{2,0} - \frac{4}{3}
\kappa_{3,0} - \frac{3}{2} \kappa_{4,0} \Big) 
+ \mathcal{O}(\hbar^2) ~,
\\
\kappa_2 &= \kappa_{2,0} -\frac{\hbar g_0^2}{(4\pi)^2 \varepsilon}
\Big( -\frac{1}{12} + \frac{2}{3} \kappa_{1,0} -\frac{2}{3} \kappa_{3,0} 
+ 3 \kappa_{4,0} \Big) + \mathcal{O}(\hbar^2) ~,
\\
\kappa_3 &= \kappa_{3,0}  -\frac{\hbar g_0^2}{(4\pi)^2 \varepsilon}
\Big( \frac{13}{32} - \frac{1}{4} \kappa_{1,0} -\frac{1}{4} \kappa_{2,0} 
+ \frac{3}{4} \kappa_{4,0}\Big) 
+ \mathcal{O}(\hbar^2) ~,
\\
\kappa_4 &= \kappa_{4,0} -\frac{\hbar g_0^2}{(4\pi)^2 \varepsilon}
\Big(\frac{1}{48} - \frac{1}{6} \kappa_{1,0} + \frac{1}{2} \kappa_{2,0} 
+ \frac{2}{3} \kappa_{3,0} - \frac{1}{2} \kappa_{4,0} \Big) 
+ \mathcal{O}(\hbar^2) ~,
\\
\kappa_6 &= \kappa_{6,0}  -\frac{\hbar g_0^2}{(4\pi)^2 \varepsilon}
\Big(\frac{1}{6} + \frac{2}{3} \kappa_{1,0} - \kappa_{2,0} -
\frac{2}{3} \kappa_{3,0} - 3 \kappa_{4,0} - 3 \kappa_{6,0} 
+ 2 \kappa_{7,0} \Big) 
+ \mathcal{O}(\hbar^2) ~,
\\
\kappa_7 &= \kappa_{7,0} -\frac{\hbar g_0^2}{(4\pi)^2 \varepsilon}
\Big(- \frac{1}{2} \kappa_{2,0} - \kappa_{3,0} - \frac{3}{2} \kappa_{4,0} 
+ \frac{1}{2} \kappa_{6,0} - 3 \kappa_{7,0} \Big) 
+ \mathcal{O}(\hbar^2) ~,
\\
\kappa_8 &= \kappa_{8,0} -\frac{\hbar g_0^2}{(4\pi)^2 \varepsilon}
\Big(\frac{1}{16} - \frac{1}{2} \kappa_{1,0} + \kappa_{2,0} 
+ \kappa_{3,0} + \kappa_{4,0} + \kappa_{6,0} + 2 \kappa_{7,0} 
- 6 \kappa_{8,0} \Big) 
+ \mathcal{O}(\hbar^2) \;.
\end{align}
\end{subequations}
But there is no reason why Green's functions with not more than two external
electrons are renormalizable up to the considered order. There could be four
additional divergences which are allowed by gauge symmetry (Ward identity) and
Lorentz symmetry---two in the pure photon sector, one corresponding to a
renormalization of $\theta$ and one in the photon-electron sector. All of
these four divergences are absent! This cannot be accidental!

\end{itemize}

Our task is now to start developing a renormalization scheme for
noncommutative gauge theories implementing these results. The
divergences in graphs with more than two external electrons discussed in
1) tell us that one \emph{must not expand} the noncommutative field
theory in $\theta$. Then only graphs with 
\begin{align}
\hat{N}_B+\frac{3}{2} \hat{N}_F \leq 4 
\end{align}
are divergent, where $\hat{N}_B$ and $\hat{N}_F$ are the numbers of
($\theta$-unexpanded) external gauge bosons and fermions,
respectively\footnote{In other words, the divergence of the fermion
  four-point function should be regarded as an artifact of the
  $\theta$-expansion. Note, however, that similar divergences will
  appear in the $\theta$-unexpanded theory if $\theta$-dependent field
  monomials such as $\int \theta^{\alpha\beta} \hat{F}_{\alpha\beta}
  \star \hat{F}_{\mu\nu} \star \hat{F}^{\mu\nu}$ are included. These
  terms are perfectly compatible with power-counting dimension and
  gauge and Lorentz symmetries!}.  In this way we solve 1)
automatically. One might object immediately that now the famous UV/IR
mixing destroys renormalizability. However, our results 3) tell us to
be more careful, although the problem seems to persist.

Actually the UV/IR mixing \cite{Minwalla:2000px} was due to the distinction of
the Feynman graphs into \emph{planar} and \emph{non-planar} ones. The custom
was to subtract the planar graphs as usual by \emph{local counterterms} and to
keep the non-planar graphs untouched, because non-planar graphs (seem to)
correspond to non-local counterterms \cite{Chepelev:2001hm}. The trouble came
when inserting the non-planar (at first sight finite) graphs as subgraphs into
a bigger graph, which then turned out to be divergent and non-local. It is at
this point where our results propose to modify the subtraction scheme. 
We have seen that
there is a way to \emph{subtract the non-planar graphs} at least partially
\emph{without destroying the symmetries}---using the Seiberg-Witten map in a
crucial way.

It is convenient to represent this idea graphically. Let us draw
Feynman rules for the $\theta$-unexpanded model as double lines. The
Seiberg-Witten differential equation (including all possible field
redefinitions) expands the noncommutative Yang-Mills-Dirac action in
the following way:
\begin{align}
\sum_{\hat{N}_B=2}^4 ~
\parbox{23mm}{\begin{picture}(20,20)
\put(0,0){\begin{fmfgraph}(20,20)
\fmfleft{l1,l2} 
\fmfright{r}
\fmf{dbl_wiggly}{l1,i}
\fmf{dbl_wiggly}{l2,i}
\fmf{dbl_wiggly}{r,i}
\end{fmfgraph}}
\put(5,0){\mbox{\small$1$}}
\put(17,6){\mbox{\small$2$}}
\put(5,17){\mbox{\small$\hat{N}_B$}}
\put(15.5,15){.}
\put(14.6,16.2){.}
\put(13.5,17){.}
\end{picture}} 
= \sum_{n \geq 0} \sum_{N_B\geq 2} ~
\parbox{20mm}{\begin{picture}(20,20)
\put(0,0){\begin{fmfgraph}(20,20)
\fmfsurround{v1,v2,v3,v4,v5,v6} 
\fmf{photon}{v1,i}
\fmf{photon}{v2,i}
\fmf{photon}{v3,i}
\fmf{photon}{v4,i}
\fmf{photon}{v5,i}
\fmf{photon}{v6,i}
\fmfdot{i}
\end{fmfgraph}}
\put(2,6){\mbox{\small$1$}}
\put(7,0){\mbox{\small$2$}}
\put(1,14){\mbox{\small$N_B$}}
\put(10.8,17){.}
\put(9.6,17.6){.}
\put(8.3,18){.}
\end{picture}}\quad , ~~\bullet=\mathcal{O}(\theta^n)~.
\nonumber
\\
\sum_{\hat{N}_B=0}^1 ~
\parbox{23mm}{\begin{picture}(20,20)
\put(0,0){\begin{fmfgraph}(20,20)
\fmfbottom{b1,b2} 
\fmftop{t}
\fmf{dbl_wiggly}{t,i}
\fmf{dbl_plain_arrow,tension=1}{b1,i}
\fmf{dbl_plain_arrow,tension=1}{i,b2}
\end{fmfgraph}}
\put(11,14){\mbox{\small$\hat{N}_B$}}
\end{picture}} 
= \sum_{n \geq 0} \sum_{N_B\geq 0} ~
\parbox{20mm}{\begin{picture}(20,20)
\put(0,0){\begin{fmfgraph}(20,20)
\fmfsurround{v1,v2,v3,v4,v5,v6} 
\fmf{fermion,tension=1}{v4,i,v5}
\fmf{photon}{v3,i}
\fmf{photon}{v6,i}
\fmf{photon}{v1,i}
\fmf{photon}{v2,i}
\fmfdot{i}
\end{fmfgraph}}
\put(12,0){\mbox{\small$1$}}
\put(17,7){\mbox{\small$2$}}
\put(1,14){\mbox{\small$N_B$}}
\put(10.8,17.2){.}
\put(9.6,17.7){.}
\put(8.3,18){.}
\end{picture}}\quad , ~~\bullet=\mathcal{O}(\theta^n)~.
\end{align}

A superficially divergent Green's function $\hat{\Gamma}$ of the
$\theta$-unexpanded theory will then via Seiberg-Witten map be
expressed in terms of Green's functions $\Gamma$ of the
$\theta$-expanded theory.  Renormalization has to proceed order by
order in $\hbar$ (which is the number of loops). In order $\hbar^1$ we
have proved that massless QED is renormalizable up to first order of
$\theta$, because graphs with more than two external fermions are not
expanded. Let us assume that one-loop renormalizability can be
extended to any order of $\theta$.  Under this assumption all
divergences would have been removed for a certain
$\hbar^1$-renormalization of the initial noncommutative
Yang-Mills-Dirac action.

Starting with order $\hbar^2$ there is however a new problem to 
solve. Let us consider the graph
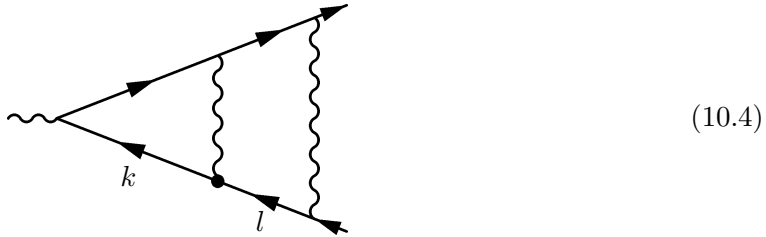
\begin{align}
\parbox{50mm}{\begin{picture}(50,30)
\put(0,0){\begin{fmfgraph}(50,30)
\fmfleft{l}
\fmfright{r1,r2}
\fmf{photon,tension=4}{l,i5}
\fmf{fermion,tension=1}{i1,i3}
\fmf{fermion,tension=1}{i4,i2}
\fmf{fermion,tension=0.6}{i3,i5}
\fmf{fermion,tension=0.6}{i5,i4}
\fmf{fermion,tension=3}{r1,i1}
\fmf{fermion,tension=3}{i2,r2}
\fmffreeze
\fmf{photon}{i1,i2}
\fmf{photon}{i3,i4}
\fmfdot{i3}
\end{fmfgraph}}
\put(15,6){\mbox{$k$}}
\put(33,0){\mbox{$l$}}
\end{picture}}
\label{l=2}
\end{align}
This graph contains an overlapping divergence in $\theta$-expanded
noncommutative QED, because the fermion four-point function is divergent.
In a (renormalizable) quantum field theory on commutative space-time
local counterterms are obtained only if \emph{all} subdivergences are
treated according to the forest formula. Otherwise there are divergent
terms involving logarithms of external momenta and masses.
For the graph (\ref{l=2}), however, we are not
allowed to treat the divergent $l$-integration as a subdivergence in
the forest formula. One has therefore to show that first integrating
over $k$, subtracting then the divergent part, and finally 
integrating over $l$ does not yield divergent terms which contain
logarithms of external momenta---at least on the level of Green's
functions. It seems unlikely that this can work, but one cannot
exclude the possibility of symmetries for the $\theta$-deformed action
which do the job. We had also expected four additional divergences at
order $\hbar^1\theta^1$ which eventually were absent. Thus, one has to
perform the two-loop computation in order to be sure.

\section{Symmetries}

We have mentioned several times possible (additional) symmetries of gauge
theories on $\theta$-deformed space-time. There are clear hints now that such
symmetries exist. Otherwise the absence of the four divergences which are not
reached by field redefinitions cannot be explained\footnote{These symmetries
  could also be reductions of couplings \cite{Zimmermann:1984sx} only.}. A
general idea about these symmetries can be obtained from the mathematical
foundation \cite{Connes:1996gi} of noncommutative geometry. The noncommutative
Dirac action can be written identically as
\begin{align}
\big\langle \hat{\psi} , (\mathrm{i} \setminus\kern-0.8em\partial + 
\setminus\kern-0.6em\hat{A} ) \star \hat{\psi} \big\rangle_{\mathcal{H}} 
= \big\langle U \star \hat{\psi} , 
(\mathrm{i} \setminus\kern-0.8em\partial + U \star 
\setminus\kern-0.6em\hat{A} \star U^* + U \star [\mathrm{i}
\setminus\kern-0.8em\partial ,U^*])
\star (U \star \hat{\psi})
\big\rangle_{\mathcal{H}}  
\end{align}
for $U \in \mathcal{M}$, $U\star U^*=U^* \star U=1$ (see (\ref{MA})).
One is tempted to regard
\begin{align}
\setminus\kern-0.6em\hat{A} & \mapsto 
\setminus\kern-0.6em\hat{A}_U:=U \star \setminus\kern-0.6em\hat{A}
\star U^* + U \star [\mathrm{i} \setminus\kern-0.8em\partial ,U^*]~,
&
\hat{\psi} & \mapsto 
\hat{\psi}_U :=U \star \hat{\psi}~,\qquad
\label{U}
\end{align}
as a gauge transformation. However, the space of all $U$ is very big, for
instance, Lorentz transformations can be implemented via $U$
\cite{Gracia-Bondia:2001ct}. In fact all automorphisms of the algebra
$\mathcal{A}$ are inner \cite{Connes:2001ef}. Thus, $U$ is the candidate for
additional symmetries. However, one has to make sure that the noncommutative
Yang-Mills action is invariant under (\ref{U}). Since we look for an action
invariant under all automorphisms, an action like $\int
\setminus\kern-0.6em\hat{F} \star \setminus\kern-0.6em\hat{F}$ which does not
contain gravity (probably in a different shape) is certainly not the right
choice. There is---at least formally---a natural candidate for a Yang-Mills
action invariant under (\ref{U}), the spectral action
\cite{Chamseddine:1996zu}
\begin{align}
S_\Lambda(\setminus\kern-0.6em\hat{A}) = \mathrm{trace}\, f \Big( 
\frac{ \setminus\kern-0.7em{}D_A^2}{\Lambda^2} \Big)~,
\qquad \setminus\kern-0.7em{}D_A =\mathrm{i} \setminus\kern-0.8em\partial + 
\setminus\kern-0.6em\hat{A} ~,
\label{spectral}
\end{align}
where $f$ is an appropriate cut-off with $f(0)=1$ and $f(x)=0$ for $x\geq 1$.
This is nothing but the weighted sum of the eigenvalues of
$\setminus\kern-0.7em{}D_A^2$ smaller than $\Lambda^2$. The problem is that we
are in a non-compact case so that the eigenvalues of
$\setminus\kern-0.6em{}D_A^2$ are continuous. Recently there has been a lot of
progress on non-compact spectral geometries \cite{Gracia-Bondia:2001ct}.

Our computations provide an indirect support for the spectral action. What we
have computed in Section~\ref{Sec85} is (together with (\ref{ps0})) the
divergent part of the effective action for fermions coupled to an external
photon field, the coupling involving $\theta$. In the commutative case one
recovers the Maxwell action $-\frac{1}{4 g^2} \int F_{\mu\nu} F^{\mu\nu}$ as
the coefficient of that (logarithmic) divergence. This result was rigorously
proved by Langmann \cite{Langmann:2001cv} in the language of regularized
traces for pseudodifferential operators, i.e.\ similar techniques as those
which are used to evaluate the spectral action. There is some rumour that
Langmann's work and the spectral action are equivalent when restricted to the
Yang-Mills part. Whereas in the commutative case gauge and Lorentz symmetries
do not permit a different coefficient of the divergence than the Maxwell
action, we could in the $\theta$-expanded noncommutative case obtain in
principle the additional terms (\ref{YM-ext}). But this does not happen. We
expect therefore that the spectral action forbids these additional terms
trilinear in the noncommutative field strength. Thus, making sense of
(\ref{spectral}), computing the spectral action and identifying the additional
local symmetries of the spectral action is one of the most important next
steps for the renormalization of noncommutative gauge theories.

\acknowledgments

I would like to thank Andreas Bichl, Jesper Grimstrup, Harald Grosse, Lukas
Popp and Manfred Schweda for numerous helpful discussions and a lot of
inspiration which finally resulted in this paper, outcome of our fruitful
collaboration. I am grateful to Jos\'e Gracia-Bond\'{\i}a for advice on the
mathematical background. Finally, the computations of Feynman graphs were
performed using the \emph{Mathematica$^{TM}$} package \cite{Ertl} by Martin
Ertl.

\begin{appendix}

\section{Analytic regularization}
\label{appa}

The one-loop integrals to compute are of the form
\begin{align}
I^{(s+1)}_{\mu_1\dots \mu_n} (p_0,p_1,\dots,p_s;m_0,m_1,\dots, m_s) :=
\lim_{\epsilon\to 0}\int \frac{d^4k}{(2\pi)^4}\;\frac{\prod_{l=1}^n
  k_{\mu_l} }{ \prod_{i=0}^s 
[(k{+}p_i)^2-m_i^2 +\mathrm{i}\epsilon]}~.
\label{Is}
\end{align}
Using Zimmermann's $\epsilon$-trick we replace a propagator 
\begin{align}
\frac{1}{k^2-m^2 + \mathrm{i}\epsilon} & \equiv  
\frac{1}{k_0^2-(\vec{k}\,^2 {+}m^2) + \mathrm{i}\epsilon} \nonumber
\\*
& \mapsto 
\frac{1}{k_0^2-(1{-}\mathrm{i}\epsilon)(\vec{k}\,^2 {+}m^2) } 
= \frac{(\epsilon'{-}\mathrm{i})}{
(\epsilon'{-}\mathrm{i})k_0^2 +
(\epsilon{-}\epsilon'{+}\mathrm{i}{+}\mathrm{i} \epsilon \epsilon')
(\vec{k}\,^2 {+}m^2) } ~.
\end{align}
For $0<\epsilon'<\epsilon$ the denominator of the last expression has a
positive real part so that standard Euclidean integrations techniques using
Schwinger and Feynman parameters can be applied. Analytic regularization
\cite{Speer:gj} consists in the following replacement of the denominator of
the integrand:
\begin{align}
&\frac{(\epsilon'{-}\mathrm{i})^{s+1}}{
\big\{ (\epsilon'{-}\mathrm{i})[k_0^2 + 2 k_0 q_0 + M_0^2]
+ (\epsilon{-}\epsilon' {+}\mathrm{i}{+}\mathrm{i}\epsilon\epsilon')
[\vec{k}^2 + 2\vec{k} \vec{q} + \vec{M}^2]
\big\}^{s+1} }
\nonumber
\\*
& \qquad \mapsto 
\frac{\mu^{2\varepsilon}
(\epsilon'{-}\mathrm{i})^{s+1+\varepsilon}}{
\big\{ (\epsilon'{-}\mathrm{i})[k_0^2 + 2 k_0 q_0 + M_0^2]
+ (\epsilon{-}\epsilon' {+}\mathrm{i}{+}\mathrm{i}\epsilon\epsilon')
[\vec{k}^2 + 2\vec{k} \vec{q} + \vec{M}^2]
\big\}^{s+1+\varepsilon} }~.
\end{align}
Then the integration over the loop momentum $k$ and over the Schwinger
parameter can be performed, with the following result (which now depends
additionally on the renormalization scale $\mu$ and on $\varepsilon$):
\begin{align}
I^{(s+1)}&_{\mu_1\dots \mu_n} (p_0,p_1,\dots,p_s;m_0,m_1,\dots,
m_s;\mu,\varepsilon)    
\label{integral}
\\*
&= \lim_{\epsilon\to 0,~\epsilon'<\epsilon}
\sum_{l=0}^{[n/2]} \frac{(-1)^n \Gamma(s{-}l{-}1{+}\varepsilon) 
\Gamma(s{+}1) \mu^{2\varepsilon}
(\epsilon'{-}\mathrm{i})^{s+\frac{1}{2}-l+\varepsilon}}{
2^l (4\pi)^2 \Gamma(s{+}1{+}\varepsilon)
(\epsilon{-}\epsilon'{+}\mathrm{i}
  {+}\mathrm{i}\epsilon\epsilon')^{\frac{3}{2}}  }
\nonumber
\\*
& \hspace*{1em}
\times \int \! d^s x \; 
T^{n-2l}_{\mu_1\dots \mu_n}\Big(p_0 + \sum_{i=1}^s x_i(p_i{-}p_0)\,,
\epsilon\Big) 
\Big\{ (\epsilon'{-}\mathrm{i}) \sum_{i,j=1}^s x_i (\delta_{ij}{-}x_j)
(p_{i0}{-}p_{00}) (p_{j0}{-}p_{00}) 
\nonumber
\\*
& \hspace*{5em} 
+ (\epsilon{-}\epsilon'{+}\mathrm{i} {+}\mathrm{i}\epsilon\epsilon') 
\Big(M^2_s(x)+ \sum_{i,j=1}^s x_i (\delta_{ij}{-}x_j)
(\vec{p}_{i}{-}\vec{p}_{0}) (\vec{p}_{j}{-}\vec{p}_{0}) \Big)
\Big\}^{l+1-s-\varepsilon} , 
\nonumber
\end{align}
where 
\begin{align}
\int d^s x &= \int_0^1\!\! dx_1 \int_0^{1-x_1}\!\!\!\! dx_2
\; \dots \int_0^{1-\sum_{j=1}^{s-1} x_j} \!\!\!\! dx_s ~,
&
M^2_s(x) &= m_0^2 + \sum_{i=1}^s x_i (m_i^2-m_0^2)~.
\end{align}
Here the completely symmetric tensors $T^k_{\mu_1\dots\mu_n}(p)$ 
are inductively defined by $T^0(p):=1$ and 
\begin{subequations}
\begin{align}
T^0_{\mu_1\dots\mu_n}(p,\epsilon) &:= \sum_{j=2}^n 
g_{\mu_1\mu_j}(\epsilon)\,
T^0_{\mu_2\dots\mu_{j-1}\mu_{j+1}\dots\mu_n}(p,\epsilon)~,
\\
T^k_{\mu_1\dots \mu_n}(p,\epsilon) &:= \frac{1}{k} \sum_{j=1}^n p_{\mu_j} 
T^{k-1}_{\mu_1 \dots \mu_{j-1} \mu_{j+1}\dots \mu_n}(p,\epsilon)~,\qquad 
k\in \{n,n{-}2,\dots\},~ k \geq 1~,
\label{T2}
\\
g_{\mu\nu}(\epsilon) &= \mathrm{diag}(1,-\tfrac{1}{1-\mathrm{i}\epsilon},
-\tfrac{1}{1-\mathrm{i}\epsilon},-\tfrac{1}{1-\mathrm{i}\epsilon})~.
\end{align}
\end{subequations}
In particular, $T^1_\mu(p,\epsilon)=p_\mu$ and
$T^0_{\mu\nu}(p)=g_{\mu\nu}(\epsilon) $. We define 
$T^k_{\mu_1 \dots \mu_n}(p)=0$ for $k<0$ and $k>n$.

The leading terms in $\varepsilon$ of (\ref{integral}) are then given by 
\begin{align}
I^{(s+1)}&_{\mu_1\dots \mu_n} (p_0,p_1,\dots,p_s;m_0,m_1,\dots,
m_s;\mu,\varepsilon)    
\label{div}
\\*
&= \frac{\mathrm{i}}{(4\pi)^2 \varepsilon} 
\sum_{\delta=0}^{[n/2]-s+1} \frac{(-1)^{n-\delta}}{
2^{\delta+s-1} \delta!} 
\int \! d^s x \;
T^{n-2\delta -2s+2}_{\mu_1\dots \mu_n}\Big(
p_0 + \sum_{i=1}^s x_i (p_i{-}p_0)\Big)
\nonumber
\\*[-1ex]
& \hspace*{12em} \times 
\Big\{ \sum_{i,j=1}^s x_i(\delta_{ij}{-}x_j) (p_i{-}p_0)(p_j{-}p_0) 
- M^2_s(x) \Big\}^\delta + \mathcal{O}(1)\;,
\nonumber
\\
\mu^2 \frac{d}{d \mu^2} & \Big(I^{(s+1)}_{\mu_1\dots \mu_n} 
(p_0,p_1,\dots,p_s;m_0,m_1,\dots, m_s;\mu,\varepsilon)\Big)    
\label{scale}
\\*
&= \frac{\mathrm{i}}{(4\pi)^2} 
\sum_{\delta=0}^{[n/2]-s+1} \frac{(-1)^{n-\delta}}{2^{\delta+s-1} \delta!} 
\int \! d^s x 
T^{n-2\delta -2s+2}_{\mu_1\dots \mu_n}\Big(
p_0 + \sum_{i=1}^s x_i (p_i{-}p_0)\Big)
\nonumber
\\*[-1ex]
& \hspace*{12em} \times 
\Big\{ \sum_{i,j=1}^s x_i(\delta_{ij}{-}x_j) (p_i{-}p_0)(p_j{-}p_0) 
- M^2_s(x) \Big\}^\delta + \mathcal{O}(\varepsilon)\;.
\nonumber
\end{align}

We need two important properties of (\ref{div}) and (\ref{scale}): (A) 
Invariance
under the shift of the integration momentum and (B) naive 
factorization of common
terms in numerator and denominator. 
\\[2ex]
(A) To formulate shift invariance, let
$\mathcal{P}^l_n\{\mu_1\dots\mu_n\}$ be any subset of $l$ elements of 
$\{\mu_1,\dots,\mu_n\}$, preserving the order. 
Let $\overline{\mathcal{P}^{n-l}_{n}}\{\mu_1\dots\mu_n\}
=\{\mu_1,\dots,\mu_n\}\setminus \mathcal{P}^l_n\{\mu_1\dots\mu_n\}$ be its
complement. The empty set and the total set are regarded as subsets. 
If $\mathcal{P}^l_n\{\mu_1\dots\mu_n\}=\{\nu_1,\dots \nu_l\}$ let 
$q_{\mathcal{P}^l_n\{\mu_1\dots\mu_n\}}=q_{\nu_1}\cdots
q_{\nu_l}$ for $l>0$ and $q_{\mathcal{P}^0_n\{\mu_1\dots\mu_n\}}=1$.
According to (\ref{Is}), shift invariance means
\begin{align*}
I^{(s+1)}_{\mu_1\dots \mu_n} &(p_0,p_1,\dots,p_s;m_0,m_1,\dots, m_s) 
\nonumber
\\
&= \sum_{\mathcal{P}^l_n} q_{\mathcal{P}^l_n\{\mu_1\dots\mu_n\}} 
I^{(s+1)}_{\overline{\mathcal{P}^{n-l}_n}\{\mu_1\dots \mu_n\}} 
(p_0{+}q,p_1{+}q,\dots,p_s{+}q;m_0,m_1,\dots, m_s) ~.
\end{align*}
For the leading terms in (\ref{div}) and (\ref{scale}) this amounts to verify
\begin{align}
T^\alpha_{\mu_1\dots\mu_n}(p)=  
\sum_{\mathcal{P}^l_n,\;0\leq l \leq \alpha} 
(-1)^{l}\, q_{\mathcal{P}^l_n\{\mu_1\dots\mu_n\}} \,
T^{\alpha-l}_{\overline{\mathcal{P}^{n-l}_n}\{\mu_1\dots \mu_n\}}
(p{+}q) ~.
\label{Ta}
\end{align}
Eq.\ (\ref{Ta}) is obvious for $\alpha \in \{0,1\}$ and any $n$ and follows
from (\ref{T2}) by induction in $\alpha$. Assuming it holds for $\alpha{-}1$
and $n{-}1$ we have
\begin{align*}
T^\alpha_{\mu_1\dots\mu_n}(p) &= 
\frac{1}{\alpha} 
\sum_{j=1}^n (p_{\mu_j} {+} q_{\mu_j}) 
\sum_{\mathcal{P}^l_{n-1}} (-1)^{l} 
q_{\mathcal{P}^l_{n-1}\{\mu_1\dots\mu_{j-1}\mu_j\dots\mu_n\}} 
T^{\alpha-1-l}_{\overline{\mathcal{P}^{n-1-l}_{n-1}}
\{\mu_1\dots\mu_{j-1}\mu_j\dots\mu_n\}} (p{+}q)
\\
& - 
\frac{1}{\alpha} 
\sum_{j=1}^n q_{\mu_j}
\sum_{\mathcal{P}^l_{n-1}} (-1)^{l} 
q_{\mathcal{P}^l_{n-1}\{\mu_1\dots\mu_{j-1}\mu_j\dots\mu_n\}} 
T^{\alpha-1-l}_{\overline{\mathcal{P}^{n-1-l}_{n-1}}
\{\mu_1\dots\mu_{j-1}\mu_j\dots\mu_n\}} (p{+}q)
\\
&=  
\sum_{\mathcal{P}^l_{n},\;l<n} \frac{\alpha-l}{\alpha} 
(-1)^{l} 
q_{\mathcal{P}^l_{n}\{\mu_1\dots\mu_n\}} 
T^{\alpha-l}_{\overline{\mathcal{P}^{n-l}_{n}}
\{\mu_1\dots\mu_n\}} (p{+}q)
\\
& + \sum_{\mathcal{P}^{l+1}_{n},\;l<n} \frac{l+1}{\alpha} 
(-1)^{l+1} 
q_{\mathcal{P}^{l+1}_{n}\{\mu_1\dots\mu_n\}} 
T^{\alpha-l-1}_{\overline{\mathcal{P}^{n-l-1}_{n}}
\{\mu_1\dots\mu_n\}} (p{+}q)~.
\end{align*}
After a shift in $l$ we confirm (\ref{Ta}).
\\[2ex]
(B)
Naive factorization means according to (\ref{Is}), using the shift 
invariance,  
\begin{align}
I^{(s)}_{\mu_1\dots \mu_n} &(p_0{-}p_s,p_1{-}p_s,\dots,p_{s-1}{-}p_s;
m_0,m_1,\dots, m_{s-1}) 
\nonumber
\\*
&= g^{\nu_1\nu_2} 
I^{(s+1)}_{\mu_1\dots \mu_n\nu_1\nu_2} (p_0{-}p_s,p_1{-}p_s,\dots,
p_{s-1}{-}p_s,0;m_0,m_1,\dots, m_s) 
\nonumber
\\*
& - m_s^2 
I^{(s+1)}_{\mu_1\dots \mu_n} (p_0{-}p_s,p_1{-}p_s,\dots,p_{s-1}{-}p_s,0;
m_0,m_1,\dots, m_s) ~.
\label{factor}
\end{align}
For the leading terms in (\ref{div}) and (\ref{scale}) this amounts to 
verify, when inserting the identity
\begin{align*}
g^{\mu_{n-1}\mu_n} T^k_{\mu_1\dots\mu_n}(p) &= 
(k{+}n{+}2) T^k_{\mu_1\dots\mu_{n-2}}(p) 
+ p^2 T^{k-2}_{\mu_1\dots\mu_{n-2}}(p) ~,
\end{align*}
the following equation
\begin{align}
&\int d^{s-1}x 
\sum_{\delta=0}^{[n/2]-s+2} \frac{(-1)^{n-\delta}}{
2^{\delta+s-2} \delta!} 
T^{n-2\delta -2s+4}_{\mu_1\dots \mu_n}\Big(
(p_0{-}p_s) + \sum_{i=1}^{s-1} x_i (p_i{-}p_0)\Big)
\nonumber
\\*[-1ex]
&\hspace*{10em} \times 
\Big\{ \sum_{i,j=1}^{s-1} x_i(\delta_{ij}{-}x_j) (p_i{-}p_0)(p_j{-}p_0) 
- M^2_{s-1}(x) \Big\}^\delta 
\nonumber
\\
&=\int d^s x 
\sum_{\delta=0}^{[n/2]-s+2} \frac{(-1)^{n-\delta}}{
2^{\delta+s-2} \delta!} 
(n{-}\delta {-}s{+}4) T^{n-2\delta -2s+4}_{\mu_1\dots \mu_n}
\Big((p_0{-}p_s) + \sum_{i=1}^s x_i (p_i{-}p_0)\Big)
\nonumber
\\*[-1ex]
& \hspace*{10em} \times 
\Big\{ \sum_{i,j=1}^s x_i(\delta_{ij}{-}x_j) (p_i{-}p_0)(p_j{-}p_0) 
- M^2_s(x) \Big\}^\delta 
\nonumber
\\
&+ \int d^s x 
\sum_{\delta=0}^{[n/2]-s+1} \frac{(-1)^{n-\delta}}{
2^{\delta+s-1} \delta!} 
\Big(\Big((p_0{-}p_s) + \sum_{i=1}^s x_i (p_i{-}p_0)\Big)^2 
-m_s^2\Big)
\nonumber
\\*[-1ex]
& \hspace*{2em} \times 
T^{n-2\delta -2s+2}_{\mu_1\dots \mu_n}
\Big((p_0{-}p_s) + \sum_{i=1}^s x_i (p_i{-}p_0)\Big)
\Big\{ \sum_{i,j=1}^s x_i(\delta_{ij}{-}x_j) (p_i{-}p_0)(p_j{-}p_0) 
- M^2_s(x) \Big\}^\delta .
\label{appell}
\end{align}
We have verified (\ref{appell}) for the values of $s\in\{1,2,3\}$ and $0\leq n
\leq 2s$ relevant for this paper by explicit calculation. Unfortunately we do
not have a general proof. The integration over $x_s$ on the rhs.\ of
(\ref{appell}) leads to Appell hypergeometric functions which we did not
succeed to treat.

\end{appendix}

\end{fmffile}

\end{document}